\documentclass[aps,pra,a4paper,reprint,twocolumn,floatfix,superscriptaddress,dvipsnames,svgnames,notitlepage,table,nofootinbib,longbibliography,amsmath,amssymb,showkeys]{revtex4-2}

\usepackage[english]{babel}
\usepackage[T1]{fontenc}
\usepackage{times}
\usepackage[utf8]{inputenc}  
\usepackage{mlmodern}         
\usepackage{amsmath, amssymb} 
\usepackage{graphicx}
\usepackage{arydshln}
\usepackage{natbib}
\usepackage{bm}
\usepackage{tabularx}
\usepackage{braket}
\usepackage{xcolor}
\usepackage{amsfonts}
\usepackage{stmaryrd}
\usepackage{comment}
\usepackage{dirtytalk}
\usepackage[pdfusetitle]{hyperref}
\hypersetup{pdflang={English},colorlinks=true,linkcolor=RoyalBlue,citecolor=RoyalBlue,urlcolor=RoyalBlue,breaklinks=true}
\usepackage[nameinlink,capitalize,capitalise]{cleveref}
\usepackage{qcircuit}

\usepackage{pdfpages}
\makeatletter
\AtBeginDocument{\let\LS@rot\@undefined}
\makeatother

\usepackage[autostyle,english=american]{csquotes}
\MakeOuterQuote{"}

\usepackage[normalem]{ulem}
\usepackage{url}
\usepackage{bbold}
\usepackage{realboxes}

\usepackage{enumitem}
\usepackage{makecell} 
\usepackage{multirow}
\usepackage{tikz}
\usepackage{siunitx}
\usepackage{booktabs}
\usepackage{algpseudocode}
\usepackage[ruled,vlined]{algorithm2e}
\usepackage[page]{appendix}
\usepackage[font=scriptsize]{subcaption}
\usepackage[font=scriptsize, justification=raggedright]{caption}
\usepackage{placeins}
\usepackage{lipsum, babel}

\newcommand{\second}[0]{s}

\usepackage{mathtools}
\usepackage{dsfont}

\newcommand{\eat}[1]{}

\usepackage{listings}

\definecolor{dkgreen}{RGB}{0, 150, 150}
\definecolor{bkgd}{RGB}{240,242,246}
\definecolor{ceruleanblue}{rgb}{0.16, 0.32, 0.75}
\definecolor{orange-red}{rgb}{1.0, 0.27, 0.0}
\definecolor{anotherblue}{RGB}{37,92,243}
\definecolor{blackblue}{RGB}{46,60,85}
\definecolor{goldyellow}{RGB}{199,146,12}
\lstdefinestyle{altstyle2}{
    backgroundcolor=\color{bkgd},
    basicstyle=\ttfamily\footnotesize\color{blackblue},
    breakatwhitespace=false,
    breaklines=true,
    captionpos=b,   
    commentstyle=\color{goldyellow},
    keepspaces=true,
    keywordstyle=\color{orange-red},
    language=Python,
    numbersep=5pt,
    numberstyle=\tiny\color{ceruleanblue},
    showspaces=false,
    showstringspaces=false,
    showtabs=false,
    stringstyle=\color{anotherblue},
    tabsize=2
}
\graphicspath{{img/}}

\begin{document}

\title{Partially Fault-Tolerant Quantum Computation for Megaquop Applications}

\author{Ming-Zhi Chung}
\affiliation{QunaSys Inc.,  Tokyo, Japan}

\author{Ali H. Z. Kavaki}
\affiliation{1QB Information Technologies (1QBit), BC, Canada}

\author{Artur Scherer}
\thanks{{\vskip-10pt}{\hskip-8pt}Corresponding author: \href{mailto:artur.scherer@1qbit.com}{artur.scherer@1qbit.com}\\}
\affiliation{1QB Information Technologies (1QBit), BC, Canada}

\author{Abdullah Khalid}
\affiliation{1QB Information Technologies (1QBit), BC, Canada}

\author{Xiangzhou Kong}
\affiliation{1QB Information Technologies (1QBit), BC, Canada}

\author{\mbox{Toru Kawakubo}}
\affiliation{QunaSys Inc.,  Tokyo, Japan}

\author{Namit Anand}
\affiliation{HPE Quantum, Emergent Machine Intelligence, HPE Labs, CA, USA}

\author{Gebremedhin A Dagnew}
\affiliation{1QB Information Technologies (1QBit), BC, Canada}

\author{Zachary Webb}
\affiliation{1QB Information Technologies (1QBit), BC, Canada}

\author{Allyson Silva}
\affiliation{1QB Information Technologies (1QBit), BC, Canada}

\author{Gaurav Gyawali}
\affiliation{HPE Quantum, Emergent Machine Intelligence, HPE Labs, CA, USA}

\author{\mbox{Tennin Yan}}
\affiliation{QunaSys Inc.,  Tokyo, Japan}

\author{Keisuke Fujii}
\affiliation{Fujitsu Quantum Computing Joint Research Division, Center for Quantum Information and Quantum Biology, Osaka University, Japan}
\affiliation{Graduate School of Engineering Science, Osaka University, Japan}
\affiliation{Center for Quantum Information and Quantum Biology, Osaka University, Japan} \affiliation{RIKEN Center for Quantum Computing (RQC), Wako Saitama, Japan}

\author{Alan Ho}
\affiliation{Qolab, WI, USA}

\author{Masoud Mohseni}
\affiliation{HPE Quantum, Emergent Machine Intelligence, HPE Labs, CA, USA}

\author{Pooya Ronagh}
\affiliation{1QB Information Technologies (1QBit), BC, Canada}
\affiliation{Department of Physics \& Astronomy, University of Waterloo, ON, Canada}
\affiliation{Perimeter Institute for Theoretical Physics, ON, Canada}
\affiliation{Institute for Quantum Computing, University of Waterloo, ON, Canada}

\author{John Martinis}
\affiliation{Qolab, WI, USA}

\date{\today}

\begin{abstract}

Partially fault-tolerant quantum computing (FTQC) has recently emerged as a promising approach for the execution of megaquop-scale circuits with millions of logical operations. In this work, we demonstrate the strengths and the limitations of this approach by conducting quantum resource estimation (QRE) of the space--time-efficient analog rotation (STAR) architecture using realistic hardware specifications for superconducting processors, and compare it against the QRE of the full FTQC architecture. We show how the performance of the STAR architecture's protocols is affected by hardware improvements. We also reduce the space requirements for partial FTQC by developing a procedure leveraging code growth to decrease the size of a factory producing analog rotation states. 
Our results reveal a non-trivial dependence of the optimal pre-growth code distance on the rotation angle with respect to post-growth infidelity.
Further, we analyze space--time trade-offs between the factory size and the error-mitigation overhead, and observe that in an application-agnostic setting, there is a Goldilocks zone for circuits in the regime of roughly $10^5$--$10^6$ small-angle rotation gates. We show that quantum simulation of 2D Fermi--Hubbard model systems is a particularly well-suited application for the STAR architecture, requiring only hundreds of thousands of physical qubits and runtimes on the order of minutes for modest system sizes. Due to its favourable algorithmic scaling to larger system sizes, utility-scale simulation of the 2D Fermi--Hubbard model could potentially be attained using partial FTQC.

\end{abstract}

\maketitle


\section{Introduction}
\label{sec:introduction}

Achieving a utility-scale quantum advantage for practical applications in quantum chemistry, condensed-matter physics, and materials science, with surface-code error-corrected fault-tolerant quantum computers typically requires more than a million physical qubits, as was demonstrated by recent quantum resource estimation (QRE) studies presented in Ref.~\cite{mohseni2025buildquantumsupercomputer}, and consistent with other estimates~\cite{reiher2017elucidating,kivlichan2020improved,lee2021even,yoshioka2024hunting,Low_2025}. 
Even if physical-level error rates far below the fault-tolerance threshold  become attainable, existing fault-tolerant quantum computing (FTQC) schemes are expected to incur considerable overheads in both space (i.e., the number of required physical qubits) and time (i.e., the overall runtime for executing a given quantum algorithm). On the other hand, fault-tolerant architectures based on quantum low-density parity-check (qLDPC) codes~\cite{yoder2025tourgrossmodularquantum,webster2026pinnaclearchitecturereducingcost} offer a promising avenue toward drastically reducing the number of physical qubits required per logical qubit and substantially lowering the space overhead compared to architectures based on the standard surface codes. Unfortunately, they also introduce  other significant challenges, including complex, long-range connectivity requirements, lower operational thresholds, and high-overhead real-time decoding.  

Recent demonstrations of quantum error correction (QEC) on actual hardware~\cite{google_2024_qec_below_threshold, quera_2025_fault_tolerant} have opened a new era of quantum information processing beyond the NISQ era. If quantum hardware comprising only tens or hundreds of thousands of physical qubits could be harnessed to solve practically relevant problems prior to reaching utility-scale FTQC, it would help sustain a healthy cycle of hardware advancement. Demonstrating a practical quantum advantage on \mbox{``early FTQC''} (EFTQC) hardware platforms, such as the error-corrected ``megaquop machines'' capable of reliably executing quantum circuits with a million quantum operations, thus targeting a logical per-gate error rate of approximately $10^{-6}$, has been proposed as a promising avenue for bridging the gap between the NISQ and FTQC eras~\cite{preskill2025beyondNISQ,Katabarwa_2024}. 
Even if achieving practical quantum advantage, or even quantum utility, on EFTQC hardware remains uncertain, initial EFTQC demonstrations may help identify key scalability challenges and inform the most effective strategies for scaling towards fully fault-tolerant architectures.

Innovations in the EFTQC era proceed along two complementary directions: architectural advances and algorithmic developments. Architectural innovations focus on error-corrected schemes that minimize resource overhead.
A promising strategy to reducing the substantial space and time overheads of FTQC is adopting partially fault-tolerant compilation of quantum algorithms. Recently, increasing attention has been given to compilation schemes that combine error-corrected Clifford gates with space--time-efficient analog 
\mbox{rotations~\cite{akahoshi2024partially,toshio2025practical,ismail2025transversal}}. This approach entirely omits the costly magic state distillation procedures that are typically required for a reliable implementation of non-Clifford gates.  
It also avoids decompositions of arbitrary-angle rotation gates into the Clifford$+T$ gate set via the well-established \mbox{Solovay--Kitaev} algorithm or alternative techniques~\cite{selinger2015efficient,ross2015optimal} designed for the same purpose, which  typically result in a large number of $T$ gates. 
Instead, arbitrary-angle rotation gates are directly executed through analog rotation and ancilla state injection followed by error 
detection and post-selection in a repeat-until-success (RUS) fashion. Referred to as the \mbox{\em space--time-efficient analog rotation} (STAR) 
quantum computing architecture in Refs.~\cite{akahoshi2024partially,toshio2025practical,ismail2025transversal}, 
this framework has recently gained substantial attention  as a promising methodology suitable for implementations on EFTQC hardware platforms. 

On the algorithmic front, researchers are exploring methods such as post-VQE and pre-FTQC algorithms. As a post-VQE algorithm, quantum selected configuration interaction (QSCI)~\cite{kanno2023quantum} avoids the estimation of expectation values, which often undermines the accuracy of VQE, and instead uses bit strings sampled from a quantum computer to restrict the subspace in which the ground-state energy is calculated on a classical computer~\cite{robledo2024chemistry}.
As pre-FTQC algorithms, various quantum phase estimation (QPE) schemes have been proposed to operate within the resource constraints of partially error-corrected architectures; see Refs.~\cite{Lin:2021rwb, Wang:2022gxu, Ding:2024qvu, Kiss:2024sep, Wan:2021non, Blunt:2023gqs, Ding:2022xue, Wang:2023ruq, Wang:2022qjk, ding2023even, Chung:2024dyf}. These approaches 
are suitable for being implemented on EFTQC devices, and therefore they are commonly referred to as ``EFTQC algorithms''.

Particularly well-known examples include variants of statistical phase estimation (SPE) \cite{Lin:2021rwb, Wang:2022gxu} and quantum complex exponential least squares (QCELS) \cite{ding2023even}. 
Compared with standard QPE, these methods trade execution time for higher-fidelity circuit execution by subdividing the phase estimation procedure into a series of estimations obtained from shallower circuits that can be executed sequentially or in parallel. An earlier study~\cite{chung2024contrasting} demonstrates that SPE can outperform VQE on partially error-corrected devices when the physical error rate is on the order of $10^{-4}$. In this paper, we benchmark the performance of the QCELS algorithm implemented on partially error-corrected STAR architectures under realistic hardware-quality specifications, using the quantum computation of molecular electronic spectra and quantum simulations of the 2D Fermi--Hubbard model as representative high-utility practical applications.   

It is worth emphasizing that partial FTQC approaches are not necessarily limited to the EFTQC era. They comprise a novel framework that is useful for problems that naturally comprise small-angle rotations (hence incurring lower overhead), such as Hamiltonian simulation. 
In this sense, given the right application, compiling quantum algorithms to partial FTQC framework can be useful even when we have access to hardware compatible with full-FTQC hardware. This is a function of the relevant space--time trade-offs (physical qubits and physical runtime). An analogy can be made here from classical simulation techniques such as approximate stabilizer rank methods. These rely on quasiprobability decomposition techniques to treat small-angle rotations as opposed to na\"ively compiling them via many $T$ gates, thereby offering an improved scaling. Our goal is to take inspiration from these approaches and think about partial FTQC as a novel quantum compilation scheme, instead of a limitation of the quantum hardware available in the near-term. The work done here takes the first steps to show that partial FTQC compilation is useful for tasks such as Fermi--Hubbard simulation, perhaps even offering a viable pathway to utility-scale quantum computing (USQC).

The main building blocks of the STAR architecture framework comprise: (i) a baseline architecture for efficiently implementing logical Clifford operations, which we here assume to be based on the rotated surface code; (ii) a direct
implementation of logical analog rotation gates $R_Z(\theta)$ by first preparing the associated rotation resource states $|m_\theta\rangle$ and then consuming these states using quantum gate teleportation implemented via lattice surgery; (iii) error mitigation schemes to suppress control and stochastic errors incurred in prepared resource states; and (iv) optimal compilation to take advantage of
locality and parallelism of non-Clifford operations. See Refs.~\cite{akahoshi2024partially,toshio2025practical} for comprehensive studies of this framework. 

A typical STAR architecture thus implements error-corrected Clifford gates via lattice surgery on the planar surface code, while arbitrary-angle rotation gates are realized by 
consuming rotation resource states prepared directly through analog operations within surface code patches. Prior work focused on improving the preparation of such states, whose success probability decreases rapidly as the code distance of the target patch increases, which can result in large resource state factories,  especially in the case of utility-scale applications involving deep circuits. 
Various methods have been proposed to prepare rotation resource states in a  STAR architecture. Ref.~\cite{akahoshi2024partially} suggested encoding the rotation gate  into a minimal surface code of distance $d=2$, which can detect a single error, before expanding it into a larger patch. Because this method requires patch growth even for small distances, the resulting infidelity remains high. Another work~\cite{choi2023pftqc} mitigated this challenge by performing analog rotations with physical rotation gates and applying post-selection to the prepared states, eliminating the need for early growth. While this approach 
yields improved fidelity for the resource states, their success rates remain very low. Reference~\cite{toshio2025practical} further advanced this approach by introducing a transversal multi-rotation protocol alongside optimal post-selection. 
Finally, while our work and the aforementioned studies focus on superconducting systems with fixed qubit connectivity, a recent analysis presented in Ref.~\cite{ismail2025transversal} explored a transversal STAR architecture for neutral-atom  hardware. By removing nearest-neighbor connectivity constraints, their approach allows analog rotations to be implemented directly via transversal gates. 

This work aims to characterize the performance of resource state preparation protocols through emulations with realistic noise models exceeding the level of detail in prior works.  In particular, using realistic assumptions specifying the quality of superconducting quantum hardware, we demonstrate how the protocols' performance is affected by hardware improvements. Our benchmarking studies of the sensitivity of partial FTQC protocols to hardware improvements here closely follow the methodology presented in Ref.~\cite{mohseni2025buildquantumsupercomputer}. 
In particular, to characterize the quality of hardware, we use two comprehensive sets of values specifying the various superconducting hardware parameters, which we refer to as \lq\lq{}target\rq\rq{} and  \lq\lq{}desired\rq\rq{}; see \Cref{tab:physical_params}. While the target hardware specifications are used here as a promising near-term goal, the desired hardware specifications represent a degree of quality that would yield a noise model with an error suppression rate twice as good as that of the target hardware.

A primary focus of this work is a detailed analysis of the growth of rotation resource state patches to larger target code distances. Boosting the size of the code hosting the rotation resource state becomes necessary when  large code distances are used for the logical qubit patches within the core processor in order to protect the quantum computation sufficiently well to keep it below an allowable target logical error rate. This is because resource states consumed during quantum gate teleportation must have a code distance that matches that of the computational logical qubit patches in the core processor.
The requirement for growing the resource states arises because the success probability of rotation resource state preparation decreases significantly at larger code distances, which in turn leads to a substantial increase in the required size of the rotation resource state factory. For this reason, resource states are initially prepared in a code patch with a sufficiently small code distance to ensure an acceptably high preparation success probability, and are subsequently grown to the larger target code distance required for executing deep circuits that are typically necessary for utility-scale quantum computation. 
However, since the resource state growth protocol is not fault tolerant, it introduces additional errors that significantly increase the logical error rate of the final target patch, thereby resulting in a large error mitigation overhead. 
Our simulation studies demonstrate how success probabilities and logical error rates scale with respect to the initial and final code distances of the rotation resource-state code patches under varying hardware-quality specifications of \Cref{tab:physical_params}.  

Our findings for the performance of resource
state preparation and growth protocols allow us to analyze trade-offs between rotation resource state factory size and error-mitigation overheads. To demonstrate such trade-offs, we have performed detailed QRE studies using medium-size circuits in the megaquop regime and beyond. 
For our QRE studies, we examined two representative high-utility applications: quantum computations of molecular electronic spectra—using small active spaces of the $p$-benzyne molecule as a concrete example—and quantum simulations of the Fermi--Hubbard model. Our analyses of the resource requirements for the STAR architecture also include a direct comparison against the resource estimates for   
FTQC architectures in executing the associated logical quantum circuits. The resource estimates for   
FTQC architectures reported in this work were generated using the TopQAD toolkit~\cite{1qbit2024topqad}, which  is based on architecture design optimizations proposed in Ref.~\cite{silva2024optimizing}. 
These case studies enable us to assess the practical limitations of STAR schemes. 

We complement our QRE studies with an analysis on how 
the overhead of generating resource states (i.e., factory size) in a STAR architecture scales with increasing circuit size. 
STAR architectures aim to provide a near-term reduction in the cost of preparing the resource states required for universal computation, which currently represents one of the dominant overheads in large-scale quantum applications.
It is therefore instructive to compare the factory-size requirements of the partially fault-tolerant STAR scheme with those of fully fault-tolerant implementations on EFTQC platforms, especially in light of recent advances in more-efficient $T$ state preparation techniques, such as magic-state cultivation~\cite{gidney2024magicstatecultivationgrowing}.
We find that for quantum circuits containing fewer than roughly $10^5$ small-angle rotation gates, fault-tolerant schemes based on magic-state cultivation require significantly smaller factories than the STAR approach. However, cultivation alone, without subsequent growth and distillation of magic states, becomes infeasible beyond this scale. For deeper circuits containing approximately $10^5$--$10^6$ small-angle rotation gates, the STAR scheme outperforms fully fault-tolerant approaches based on magic-state distillation. For circuits exceeding 
$10^6$ small-angle rotation gates, however, the overheads associated with error mitigation quickly becomes prohibitive, rendering the STAR scheme impractical at larger scales.

This paper is organized as follows. In \Cref{sec:Preliminaries}, we review the STAR architecture framework and the phase estimation algorithms suitable for EFTQC implementations 
considered in this work.   In \Cref{sec:simulation_results}, we report the results of our numerical analyses on the performance of the rotation resource state preparation and subsequent growth under varying hardware specifications. In \Cref{sec:QRE-studies}, we present our resource estimation 
studies for quantum computation of electronic spectra and the Fermi--Hubbard model. Finally, \Cref{sec:QRE-studies} provides the main conclusion of our studies and outlines our perspectives and promising directions for  future research. Complementary technical details are presented in appendices \ref{sec:partial-FTQC_appendix}--\ref{app:appendix-QCELS}. In appendix~\ref{sec:partial-FTQC_appendix}, the transversal multi-rotation protocol and the methodology to simulate this protocol are thoroughly discussed. Appendix~\ref{app:improved trotter} derives an improved Trotter error bound for simulations of the $p$-benzyne Hamiltonian. Appendix~\ref{app:appendix-QCELS} provides a review of the  QCELS algorithm and its error budget equation.

\begin{table}[t]
    \centering
    \begin{tabular}{ccc}
        \hline\hline
       \hspace{0.7cm} 
        \multirow{2}{*}{\shortstack[c]{  \textbf{Hardware Parameter}}}\hspace{1cm} 
       & 
        \multirow{2}{*}{\shortstack[c]{\textbf{Target}  }}  
       \hspace{1cm} 
       &
       \multirow{2}{*}{\shortstack[c]{\textbf{Desired}  }}  
       \hspace{1cm}\\
        &&\\ 
        \hline
        &&\\ 
        \hspace{0.5cm} $T_1$, $T_2$ times\hspace{1cm} &  \SI{200}{\micro\second} \hspace{1cm} &  \SI{340}{\micro\second} \hspace{1cm}\\
         \hspace{0.5cm}Single-qubit gate error\hspace{1cm}  &  0.0002 \hspace{1cm}  & 0.00012\hspace{1cm}\\
         \hspace{0.5cm}Two-qubit gate error\hspace{1cm}  &  0.0005 \hspace{1cm} & 0.00029 \hspace{1cm}\\
         \hspace{0.5cm}State preparation error\hspace{1cm}  &  0.01 \hspace{1cm} & 0.00588 \hspace{1cm}\\
         \hspace{0.5cm}Measurement error\hspace{1cm}  &  0.005 \hspace{1cm} & 0.00294 \hspace{1cm}\\
         \hspace{0.5cm}Reset error\hspace{1cm}  &  0.005 \hspace{1cm} & 0.00294 \hspace{1cm}\\
         \hspace{0.5cm}Single-qubit gate time\hspace{1cm}  & 25 ns\hspace{1cm}  & 25 ns \hspace{1cm}\\
         \hspace{0.5cm}Two-qubit gate time\hspace{1cm}  &  25 ns \hspace{1cm} & 25 ns \hspace{1cm}\\
         \hspace{0.5cm}State preparation time\hspace{1cm}  & \SI{1}{\micro\second} \hspace{1cm}  &  \SI{1}{\micro\second} \hspace{1cm}\\
         \hspace{0.5cm}Measurement time\hspace{1cm}  &  100 ns \hspace{1cm} & 100 ns \hspace{1cm}\\
         \hspace{0.5cm}Reset time\hspace{1cm}  &  100 ns \hspace{1cm} & 100 ns \hspace{1cm}\\
          \cmidrule{1-3}
         \hspace{0.5cm}Error suppression rate $\Lambda$\hspace{1cm}  &  9.3 \hspace{1cm} & 18 \hspace{1cm}\\&&\\ 
         \hline\hline
    \end{tabular}
    \caption{Hardware specifications for two sets of parameters specifying the quality of a quantum processor based on superconducting qubits: {\em target} and {\em desired} hardware. This table was reproduced from data reported in Ref.~\cite{mohseni2025buildquantumsupercomputer}.  
    The target set represents values envisioned to be a promising near-term goal; the desired set represents synthetically generated hardware specifications corresponding to a noise model with about twice the error suppression rate, $\Lambda$, of the target hardware extracted from the exponential suppression law $\mu d^2 \Lambda^{-(d+1)/2}$ for a quantum memory based on the rotated surface code of distance $d$. For further details on the choice of this error suppression model, see Ref.~\cite{silva2024optimizing}.}
    \label{tab:physical_params}
\end{table}

\section{Preliminaries}
\label{sec:Preliminaries}
\subsection{STAR: A space--time-efficient analog-rotation \\
partially fault-tolerant quantum computing architecture}
\label{sec: STAR intro}

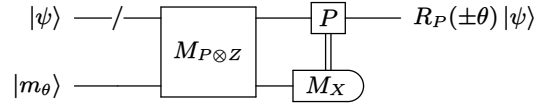
\begin{figure}[t]
    \centering
    \centerline{
    \Qcircuit @C=1.5em @R=1.5em {
    \lstick{\ket{\psi}} & \push{/} \qw & \multigate{1}{M_{P \otimes Z}} & \gate{P} & \qw & & R_{P}(\pm\theta)\ket{\psi} \\
    \lstick{\ket{m_{\theta}}} & \qw & \ghost{M_{P \otimes Z}} & \measureD{M_X} \cwx[-1] & & &
}
}
    \caption{Gate teleportation circuit for implementing the Pauli-product rotation $R_P(\theta)=\exp(-i\theta P/2)$, where $P$ is any multi-qubit tensor product of Pauli operators,  by consuming the ancillary rotation resource state $|m_{\theta}\rangle$~\cite{toshio2024practical}. Here, $M_{P\otimes Z}$ denotes a multi-Pauli measurement of the operator $P\otimes Z$. If the ancilla measurement yields $-1$, then a correction by $P$ is required, which should be done in software. The state $\ket{\psi}$ undergoes a rotation of either $\theta$ or $-\theta$ depending on the outcome of the multi-Pauli measurement.}
    \label{fig: STAR injection}
\end{figure}
One major aspect of standard fault-tolerant schemes preventing their current instantiation is the overhead from magic-state distillation, a key step in implementing $T$ gates for universal quantum computation. To circumvent this overhead, a partially fault-tolerant architecture has been proposed that avoids magic-state distillation entirely by adopting a Clifford$+\theta$ gate set~\cite{akahoshi2024partially},
where the $\theta$ gate indicates an arbitrary phase rotation gate $R_Z(\theta)= e^{-i \theta Z/2}$. Similar to $T$ gates, phase rotations can be implemented via quantum gate teleportation, using Clifford gates and specific resource states
\begin{align}
|m_\theta \rangle \equiv R_Z(-2\theta) |+\rangle  = \cos \theta |+\rangle + i \sin \theta |-\rangle ,
\end{align}
as illustrated in \Cref{fig: STAR injection}. 

Since gate teleportation can under-rotate by $2\theta$ depending on measurement outcomes, the procedure to apply a specific rotation employs a repeat-until-success (RUS) approach. 
This succeeds on average within two attempts, and hence the additional overhead is modest. 
Unlike magic-state distillation, 
the arbitrary angle ancilla states are not distilled to have arbitrarily small error, but instead various error reduction techniques are performed to produce states with small finite error. 
This approach allows us to use fully error-corrected Clifford and high-fidelity phase rotation gates with arbitrary angles.

\begin{figure*}[tb]
    \centering
    \begin{minipage}{0.25\textwidth}
        \centering
        \includegraphics[width=\linewidth]{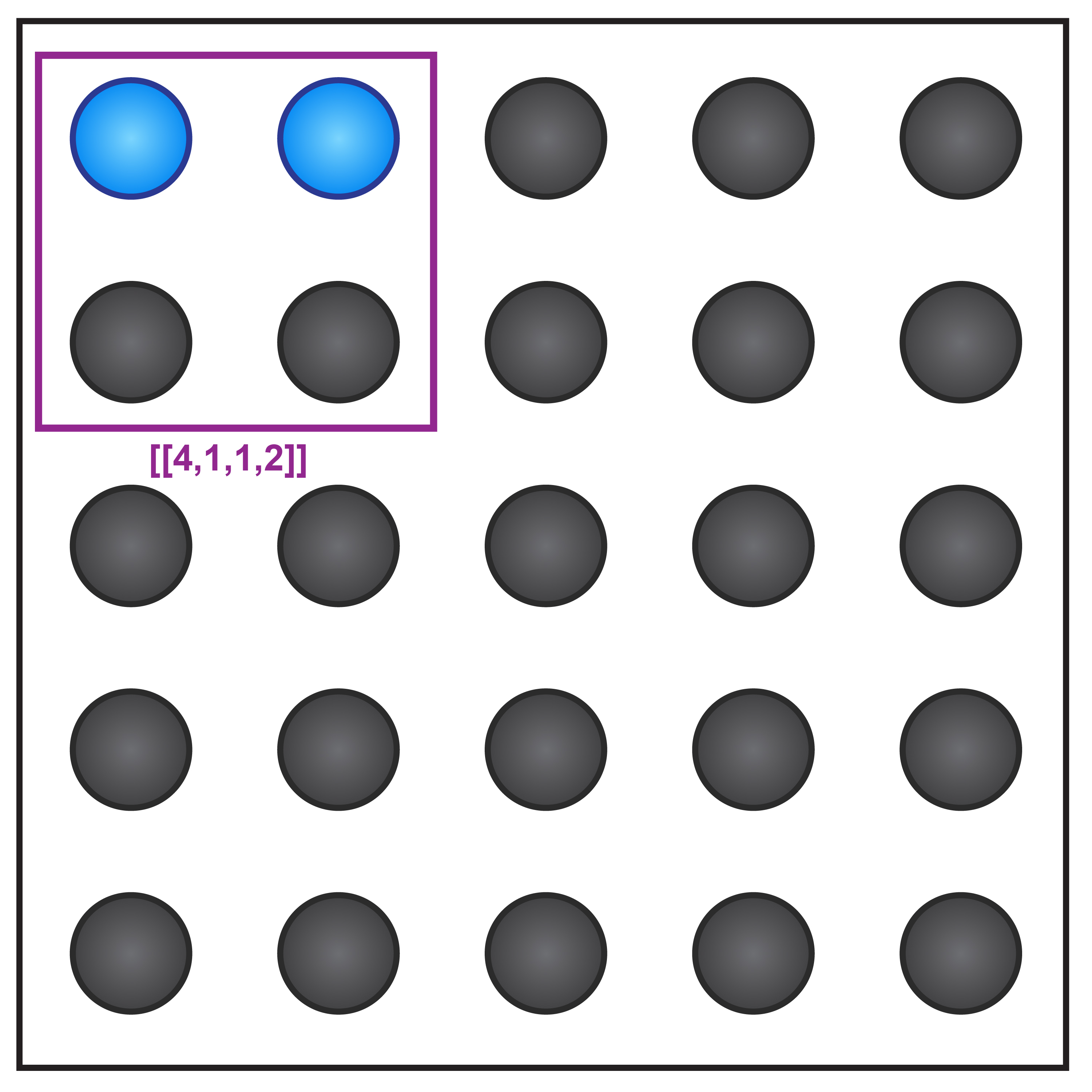}
        \caption*{(a)}
    \end{minipage}
    \hspace{2em}
    \begin{minipage}{0.25\textwidth}
        \centering
        \includegraphics[width=\linewidth]{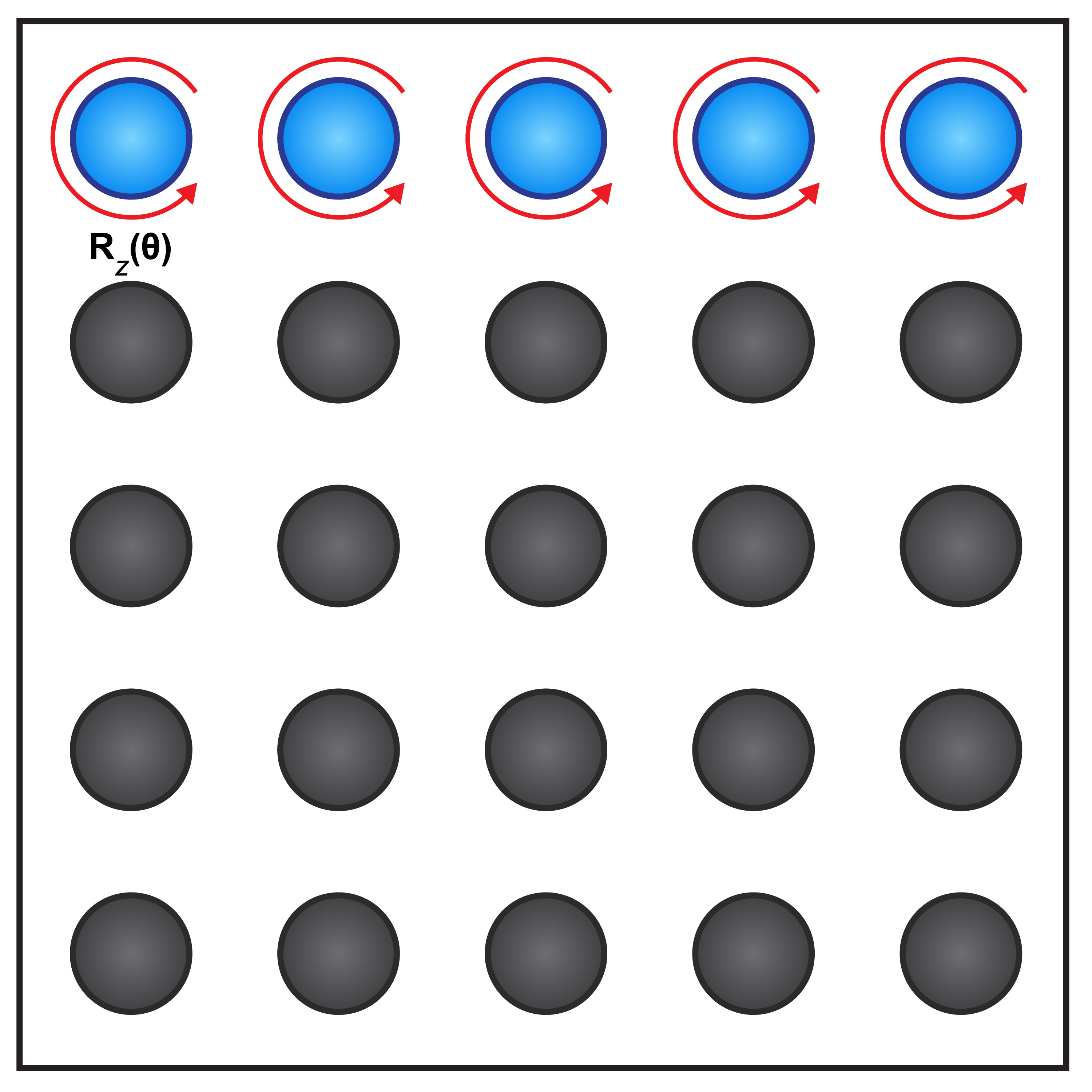}
        \caption*{(b)}
    \end{minipage}
    \hspace{2em}
    \begin{minipage}{0.25\textwidth}
        \centering
        \includegraphics[width=\linewidth]{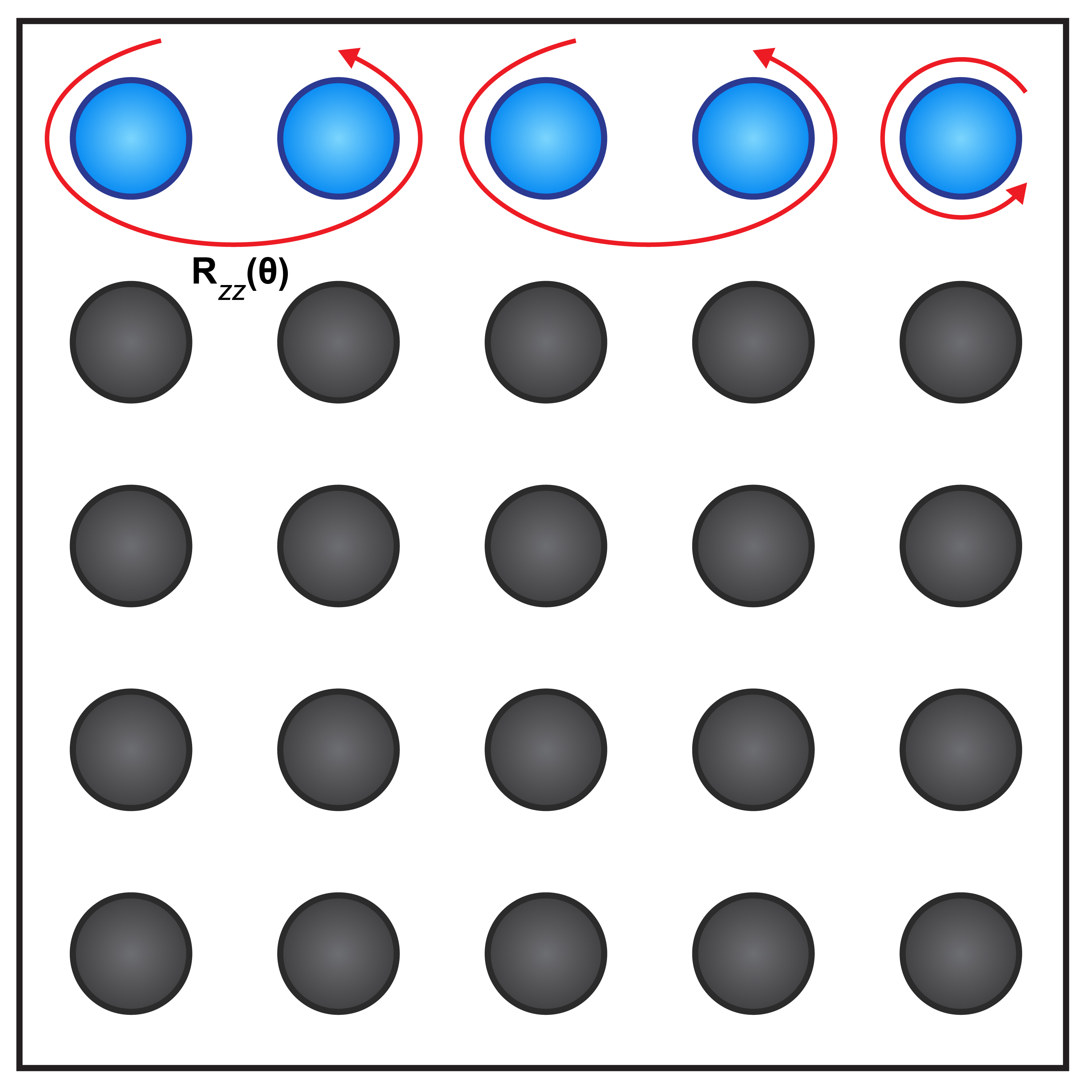}
        \caption*{(c)}
    \end{minipage}

    \caption{Schematic illustration of three different rotation state preparation protocols for a rotated surface code patch with code distance $d = 5$. 
    A transversal rotation protocol applies physical rotation gates to a row of data qubits, typically the top row, followed by error detection. This effectively rotates the logical patch by an angle determined by the individual gate rotations. Three representative methods are illustrated: (a) A small subsystem code, $[[4, 1, 1, 2]]$, is first prepared and error-detected, then expanded into a larger patch to serve as a resource state~\cite{Akahoshi2024PFTQC}. (b) A transversal scheme where each data qubit in the top row is rotated independently~\cite{choi2023pftqc}. (c) A variation of the transversal approach, using a combination of two-qubit rotations and a single one-qubit rotation to implement the logical operation. The need for a lone one-qubit gate in (c) arises due to the use of odd-distance surface codes~\cite{Toshio2025PFTQC}. For clarity, error detection procedures are not shown in the figures. Note that in (b) and (c), error detection follows the rotation gates, while in (a) it is part of the initial code preparation.}
    \label{fig: star rotation}
\end{figure*}

\begin{figure}
    \centering
    \includegraphics[width=0.73\linewidth]{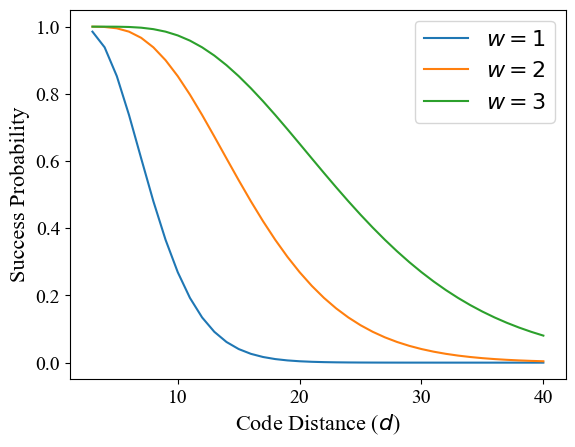}
    \caption{Success probability of the multi-Pauli rotation protocol in the noise-free limit, for 
   a target rotation angle $\theta_* = 10^{-3}$,  
   and for various weights $m$ of the Pauli-rotation generator (as shown in the legend).
   }
    \label{fig:STAR success}
\end{figure}

The method of ancilla state preparation heavily affects their fidelity and required resource overhead, and thus the logical error rate of the applied gates and size of the resulting circuit. \Cref{fig: star rotation} summarizes the history of the advancement of ancilla state preparation. Starting from \Cref{fig: star rotation}(a), which was first introduced in Ref.~\cite{akahoshi2024partially}, the method utilizes a $[[4,1,1,2]]$ subsystem code to prepare the state. This protocol provides high success rate, but the logical error rate per rotation gate is relatively high at $P_L \approx 4p_{\text{phys}}/15$ (where $p_{\text{phys}}$ is the error rate of a physical single-qubit rotation). In contrast, a protocol based on \Cref{fig: star rotation}(b) was developed~\cite{choi2023fault} in which the ancilla state infidelity scales as $d(\theta_*)^{2-2/d}$ with respect to the target rotation angle $\theta_*$ and code distance $d$. Unfortunately, the success rate of this protocol is prohibitively low once the code distance reaches approximately 10, which makes practical use of this method unlikely. The work presented in Ref.~\cite{toshio2024practical} improved upon the protocol in Ref.~\cite{choi2023fault} by adopting transversal multi-Pauli rotation gates as in \Cref{fig: star rotation}(c) and a more efficient post-selection method. In the noise-free limit, the success rate is higher if the rotation generator of individual rotation gates has a larger weight and a lower code distance (see Fig.~6(a) of Ref.~\cite{toshio2024practical}). We leave the quantitative explanation of how the success rate depends on the code distance in Appendix~\ref{sec:partial-FTQC_appendix}, and pictorially demonstrate how the success rate vary with the code distance in \Cref{fig:STAR success}. In the presence of noise, these preparations can cause over-rotation errors, which are mitigated by probabilistic coherent error cancellation that reduces the noise strength from $k|\theta_*|p_{\text{phys}}$ to $k(\theta_*)^{2-2/k}p_{\text{phys}}$ with $k$ being the number of multi-Pauli rotation gates. Due to the RUS procedure combining multiple rotations, the final logical error rate of analog rotations implemented in this manner is given by:
\begin{eqnarray}
    p_{\text{analog}} = \alpha_{\text{\tiny{RUS}}} |\theta_*| p_{\text{phys}},
    \label{eq:STAR error model}
\end{eqnarray}
with $\alpha_{\text{\tiny{RUS}}} \approx 0.4k$ for fixed weight and $\alpha_{\text{\tiny{RUS}}} \approx 1.5$ for all three protocols in \Cref{fig: star rotation} being used in a hybrid manner, as estimated in Ref.~\cite{toshio2024practical}.

The property that 
the logical error rate of an analog rotation is proportional to the angle $\theta_*$
is especially beneficial for quantum simulations, where time-evolution operators are used extensively. 
In EFTQC, the small overhead of Trotterized time evolution often leads to smaller resource requirements than more sophisticated approaches using block encoding and the quantum singular value transformation (QSVT)~\cite{childs2018toward}. 
Trotterization entails many small-angle rotations, each of which fault-tolerant protocols decompose into a long sequence of $H$, $S$, and $T$ gates and many applications of costly magic-state distillation. 
In contrast, partially fault-tolerant architectures directly implement these small-angle rotations, dramatically reducing the time and space overhead.  We will state specific overhead reductions in \Cref{sec:factory_size_requirements}.

While these constructions can apply arbitrary rotations without large overheads, they are specifically not fault-tolerant and the errors from each rotation quickly combine to hinder algorithm performance. 
When high-precision estimation of physical observable expectation values is required, one can employ probabilistic error cancellation (PEC)~\cite{temme_2017_error_mitigation, van2023probabilistic} to mitigate the error. As we expect the noise caused by these arbitrary rotations to be biased towards Pauli-$Z$ errors, PEC inserts $Z$ gates with some probability after each of the noisy $R_Z$ gates to cancel the effect of noise. This results in a sampling overhead of:
\begin{eqnarray}
    \gamma_{\text{\tiny{PEC}}}^2 \approx \exp(4\alpha_{\text{\tiny{RUS}}}\theta_{\text{tot}}p_{\text{phys}}),
\end{eqnarray}
which we control by minimizing the total rotation angle $\theta_{\text{tot}}$ of a circuit.

\subsection{Phase estimation algorithms for EFTQC architectures}\label{sec: SPE budget}

\begin{table}[t]
    \centering
    \begin{tabular}{|c|c|c|}
         \hline
        Algorithm & $T_{\text{max}}$ & $T_{\text{total}}$ \\
        \hline
        \hline
        LT22 SPE \cite{Lin:2021rwb} & ${\widetilde{\mathcal{O}}}(\epsilon^{-1})$ & $\widetilde{\mathcal{O}}(\epsilon^{-1} p_0^{-2})$ \\
        \hline
        \hline
        Gaussian SPE \cite{Wang:2022gxu, Chung:2024dyf} & $\widetilde{\mathcal{O}}
        (\Delta_{\text{True}}^{-1})$ & 
$\widetilde{\mathcal{O}}(p_0^{-2}\epsilon^{-2} \Delta_{\text{True}})$
\\
\hline
\hline
QCELS \cite{ding2023even}& $\widetilde{\mathcal{O}}(\delta\; \epsilon^{-1})$ & $\widetilde{\mathcal{O}}(\delta^{-1}\epsilon^{-1})$\\
\hline
    \end{tabular}
    \caption{Complexities of $T_{\text{max}}$and $T_{\text{total}}$ with respect to the energy estimation accuracy $\epsilon$, normalization factor $\tau$, lower bound of ground state  overlap, $p_0$,  and energy gap lower bound $\Delta_{\text{True}}$.}
    \label{tab:complexity}
\end{table}

In this subsection, we review the complexities and error budgets for the aforementioned EFTQC algorithms. The two major performance indicators of QPE algorithms are the maximal evolution time $T_{\text{max}}$, which corresponds to the runtime required to execute the deepest circuit of the algorithm, and 
the total evolution time $T_{\text{total}}$, which represents the 
total algorithm runtime.
 In a fault-tolerant setting, one would typically prioritize optimizing $T_{\text{total}}$ to reach the Heisenberg limit $T_{\text{total}}\sim\mathcal{O}(\epsilon^{-1})$ and have a minimal total runtime. However, in the EFTQC era where we lack sufficient physical resources to reach fault-tolerance, it can be more desirable to use an algorithm with a smaller $T_{\text{max}}$ complexity since it could be run on a smaller quantum device.

The evolution time complexities of SPE and QCELS are summarized in \Cref{tab:complexity}, which shows varying trade-offs between  $T_{\text{max}}$ and $T_{\text{total}}$. Among these algorithms, Gaussian SPE \cite{Wang:2022gxu} minimizes $T_{\text{max}}$ by using a Gaussian filter whose standard deviation is controlled by the spectral gap of the Hamiltonian. This results in \mbox{$T_{\text{max}}\sim \mathcal{O}(\Delta^{-1})$,} which is usually much smaller than $\mathcal{O}(\epsilon^{-1})$. On the other hand, $T_{\text{max}}$ in QCELS \cite{ding2023even} scales as $\mathcal{O}(\delta\epsilon^{-1})$, where $\delta$ is a constant that scales with the ground state overlap $p_0$ as $\delta\sim \mathcal{O}(\sqrt{1-p_0})$. This means that $T_{\text{max}}$ can be arbitrarily small when the input state is the exact ground state. At the same time, the $T_{\text{total}}$ scaling of $\mathcal{O}(\delta^{-1}\epsilon^{-1})$ implies that we can retain Heisenberg scaling even when $T_{\text{max}}$ is small. This motivates us to select QCELS as our target EFTQC algorithm to study in the subsequent resource estimations.

\section{Simulation of the transversal multi-rotation protocol and code growth}
\label{sec:simulation_results}
\begin{figure*}[tb]
    \centering
  \begin{tabular}{l @{\hskip 0.5in} l}
  (a) Success rate as a function of physical error rate & (b) Logical infidelity as a function of physical error rate\\
\includegraphics[width=0.377\linewidth]{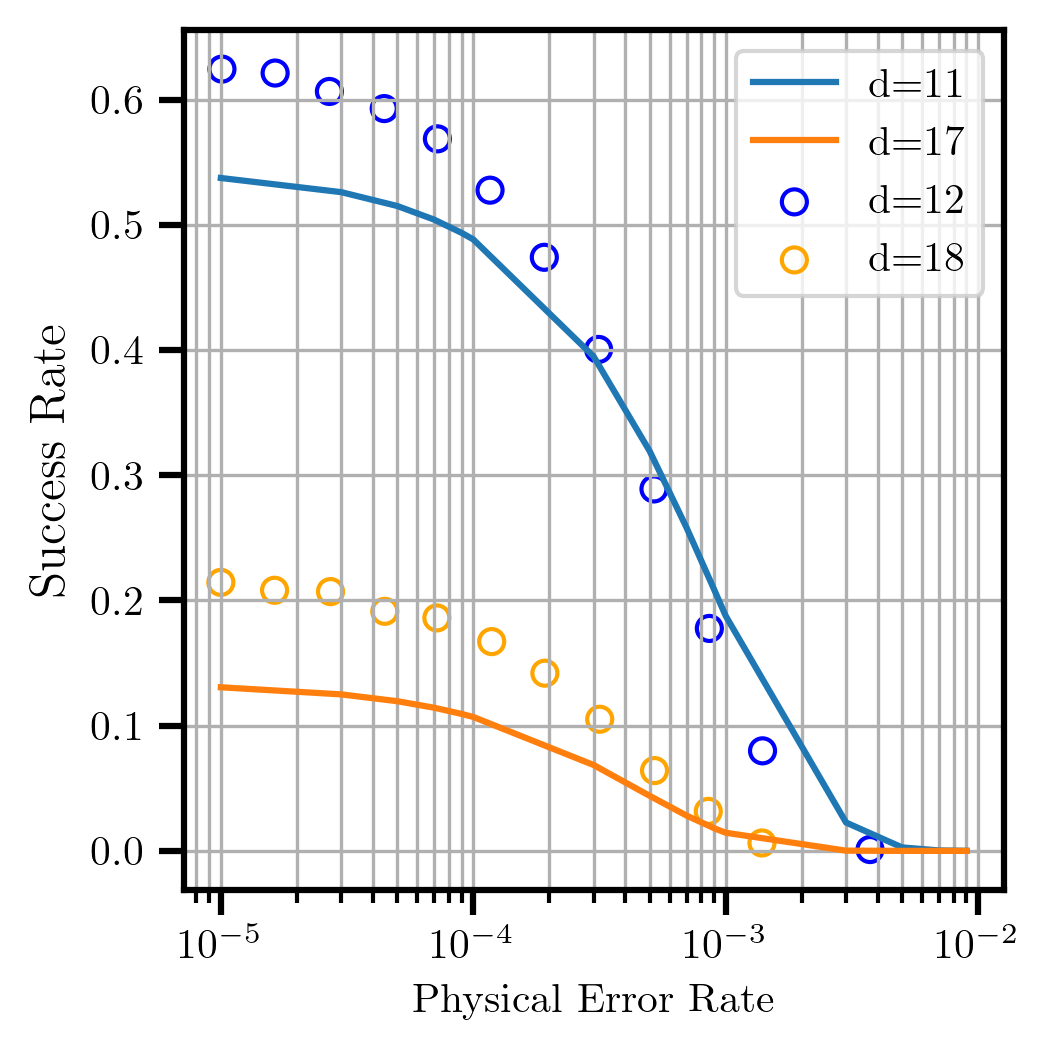} &
   \includegraphics[width=0.377\linewidth]{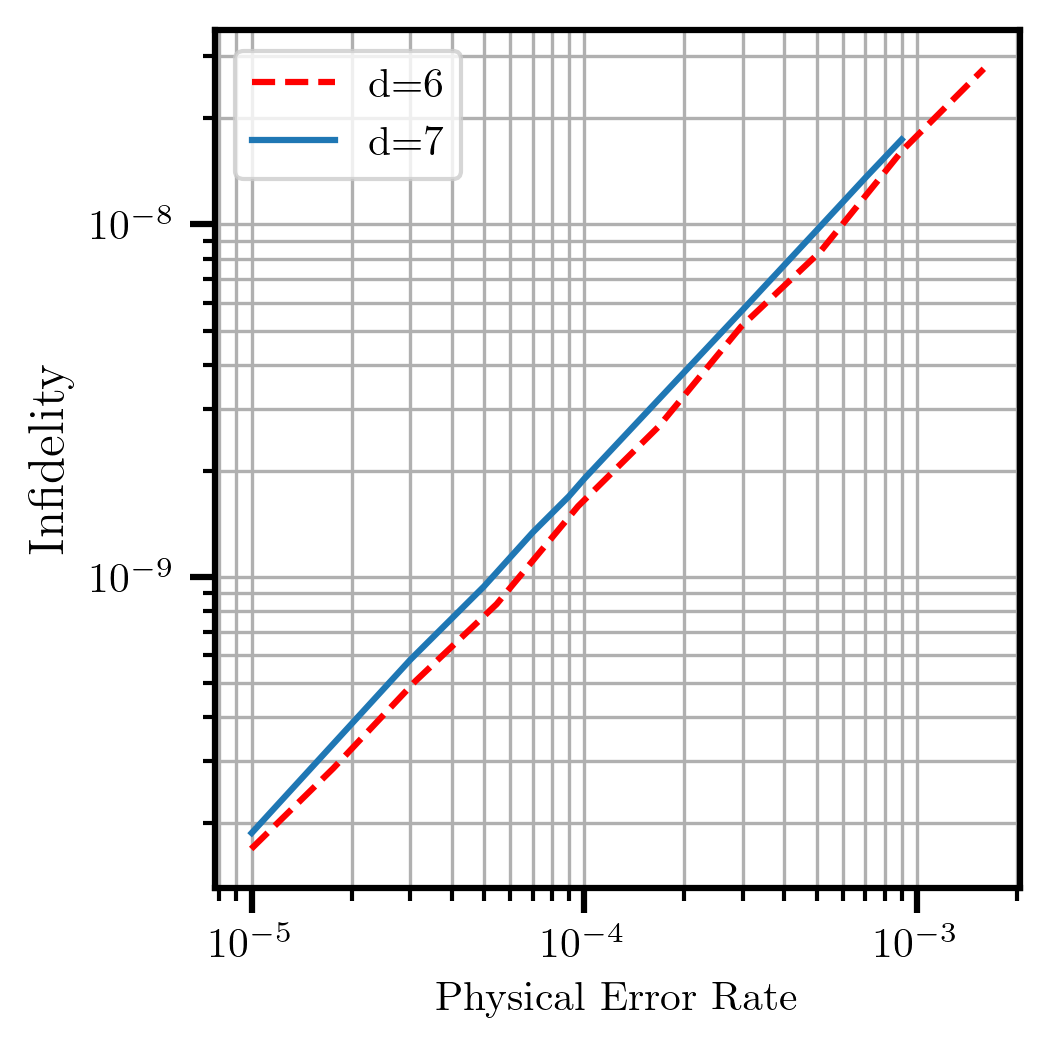}
    \end{tabular}
    \caption{Success rate of the transversal rotation state preparation protocol, and 
    logical infidelity of the resource states prepared by this protocol, as a function of the physical error rate. The results of our simulation (shown as solid lines) are compared with those presented in Ref.~\cite{Toshio2025PFTQC} (shown as circles and dashed line) for $\theta_{*}=10^{-3}$ and $w=2$. (a) Success rates for $d = 11, 17$ (solid lines) obtained from our simulations. For comparison,  success rates are shown 
    for $d = 12, 18$ (circles) from Fig.~9 of Ref.~\cite{Toshio2025PFTQC}). (b) Logical infidelity from our simulation for $d = 7$ 
    in comparison to the result for $d = 6$ presented in Fig. 7 of Ref.~\cite{Toshio2025PFTQC}). The discrepancies between the two results are primarily due to differences in code type and distance. This work uses a rotated surface code of odd distances, whereas the cited paper uses an unrotated surface code of even distances. Additionally, our use of odd distances requires an additional single qubit physical rotation along with the two-qubit rotations.}
    \label{fig:comparison}
\end{figure*}

We have simulated the transversal rotation protocol under a number of different scenarios. For all our simulations, we leveraged the Clifford simulator Stim~\cite{gidney2021stim} and the decoder PyMatching~\cite{higgott2025sparse}. As a first step, we validated our simulation framework against the results presented in prior work of Ref.~\cite{Toshio2025PFTQC}. We simulated the transversal rotation protocol for the rotated surface code with odd distances and the uniform depolarizing error model presented in Appendix C of Ref.~\cite{Toshio2025PFTQC}. In these simulations, post-selection in the second and third rounds of syndrome measurement involves all the syndromes and not only those in the top three rows. In \Cref{fig:comparison}, our simulation results are compared with those for the unrotated surface code of even distances presented in Ref.~\cite{Toshio2025PFTQC}. We observe good consistency between our results and those reported in prior work. The minor discrepancies can be explained by differences in the QEC codes 
used and other specific implementation details.

\subsection{Sensitivity of resource state preparation \\to hardware improvements}
\label{sec:sensitivity-hardware-improvements}
\begin{figure}[tb]
    \centering
    \includegraphics[width=0.43\textwidth]{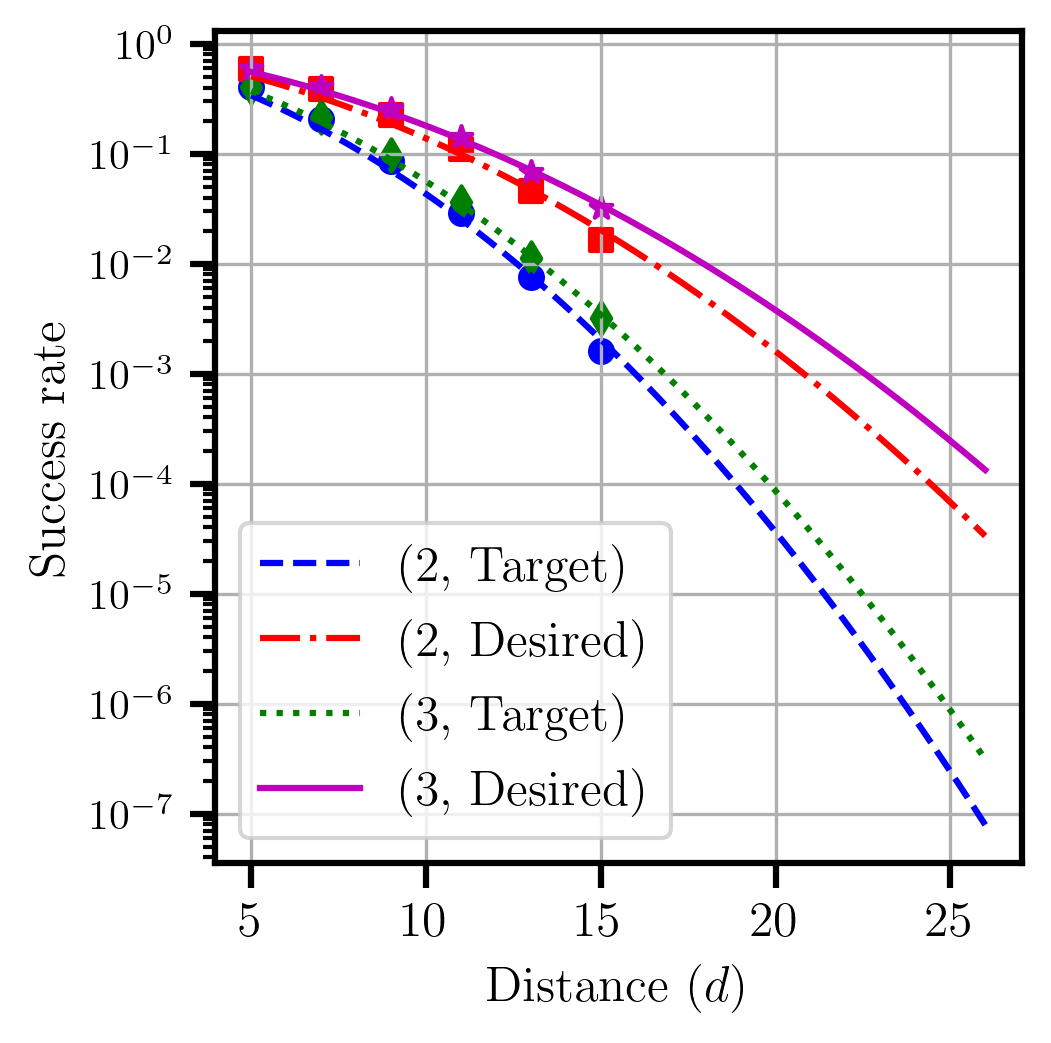}
    \caption{Success rate of the transversal multi-rotation resource state preparation protocol as a function of the code distance, for the optimal post-selection method with $\theta_{*} = 10^{-4}$ and $w=2,3$ (indicated in the legend), and for 
    target and desired hardware parameter specifications. 
    Circles represent the simulation results, while dashed lines indicate fits. We find that the success rate is better for $w=3$ than for $w=2$.}
    \label{fig:success-rate}
\end{figure}

For our resource estimates we simulated the protocol on a rotated surface code with odd distance for the target and desired hardware parameter specifications of \Cref{tab:physical_params}. This is done for two choices of the weight of the physical rotations, $w=2$ and $w=3$. However, if $w$ does not divide $d$, then a smaller weight rotation is used at the end of the row, like in the examples shown in \Cref{fig:post-select-regime} in Appendix~\ref{sec:partial-FTQC_appendix}. The success rate for the parameter sets is shown in \Cref{fig:success-rate}. To find a regression model, we assume that all detectors in the protocol are equally likely to trigger due to physical errors. There are $d^2 + 6d$ detectors that are post-selected on. If each detector triggers with probability $q$, then the success rate can be estimated by $(1 - q)^{d^2 + 6d}$. We find that this model fits the simulated data well.

\begin{figure}[tb]
    \centering
\includegraphics[width=0.43\textwidth]{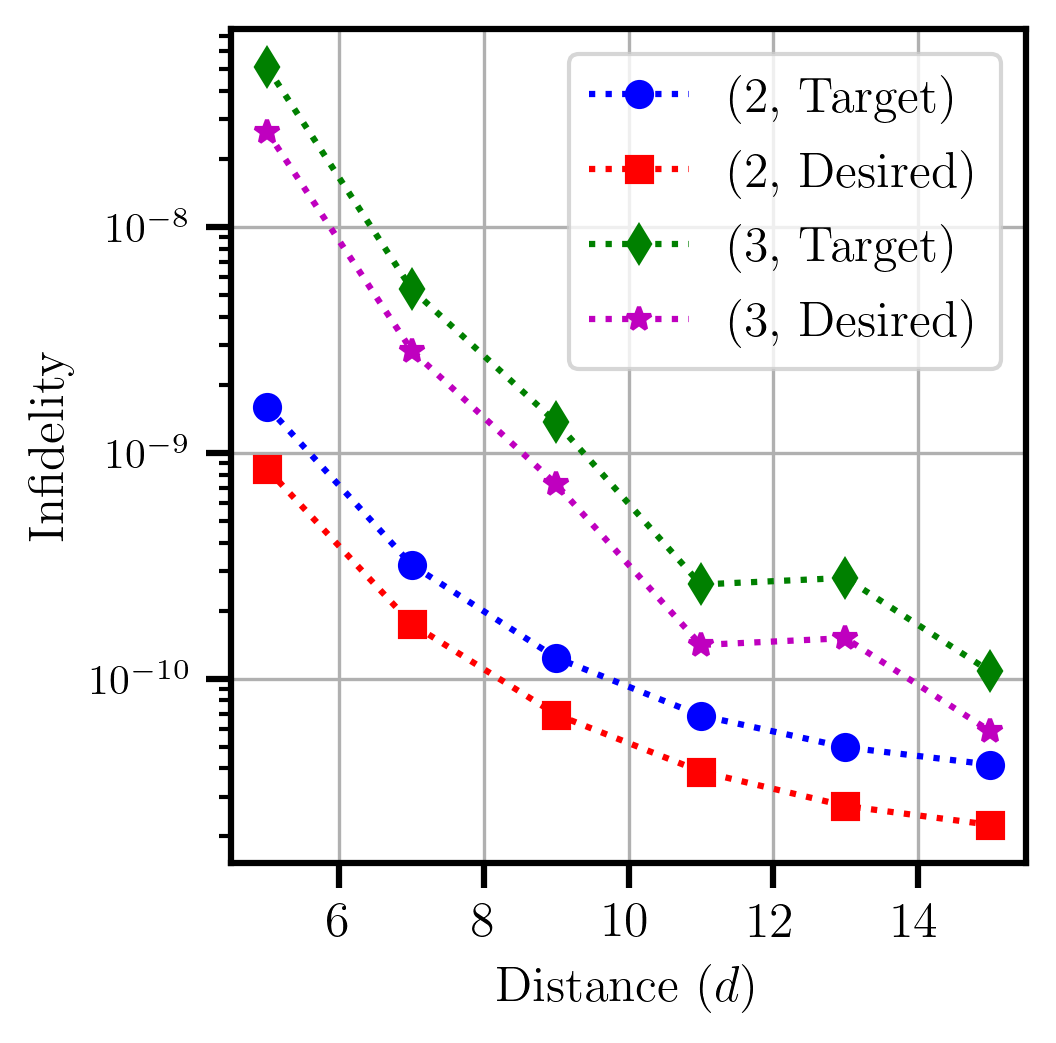}
    \caption{Infidelity of resource states prepared using the transversal multi-rotation protocol as a function of code distance, for the optimal post-selection method with $\theta_{*} = 10^{-4}$ and $w=2,3$ (indicated in the legend), and for 
    target and desired hardware parameter specifications.  The irregularities in the $w=3$ plots are due to  the different weight of the final physical rotation.}
    \label{fig:infidelity}
\end{figure}

Finally, we evaluated infidelity for the transversal rotation protocol, as shown in \Cref{fig:infidelity}. The irregularities in the infidelity plots prevent us from finding a fitting model. Note that while $w=3$ has better success rates than $w=2$, it also has worse infidelity. This is because of the longer rotation circuit for $w=3$ (see \Cref{fig:circuits}), which introduce more errors.

The success rate falls quite fast with distance. Therefore, it is not possible to directly use the resource state created using this protocol in the overall computation. For example, if we set the tolerance on the success rate at one percent, then for the target, $d=13$ is the largest state we can create using this protocol, which is smaller than what is required. Therefore, the resource state is first created on a small patch, and then grown to the distance required by the core processor. This is discussed in the next section.

\subsection{Growing rotation resource states}
\label{sec:growth_protocol}

\begin{figure}[tb]
    \centering    \includegraphics[width=0.83\linewidth]{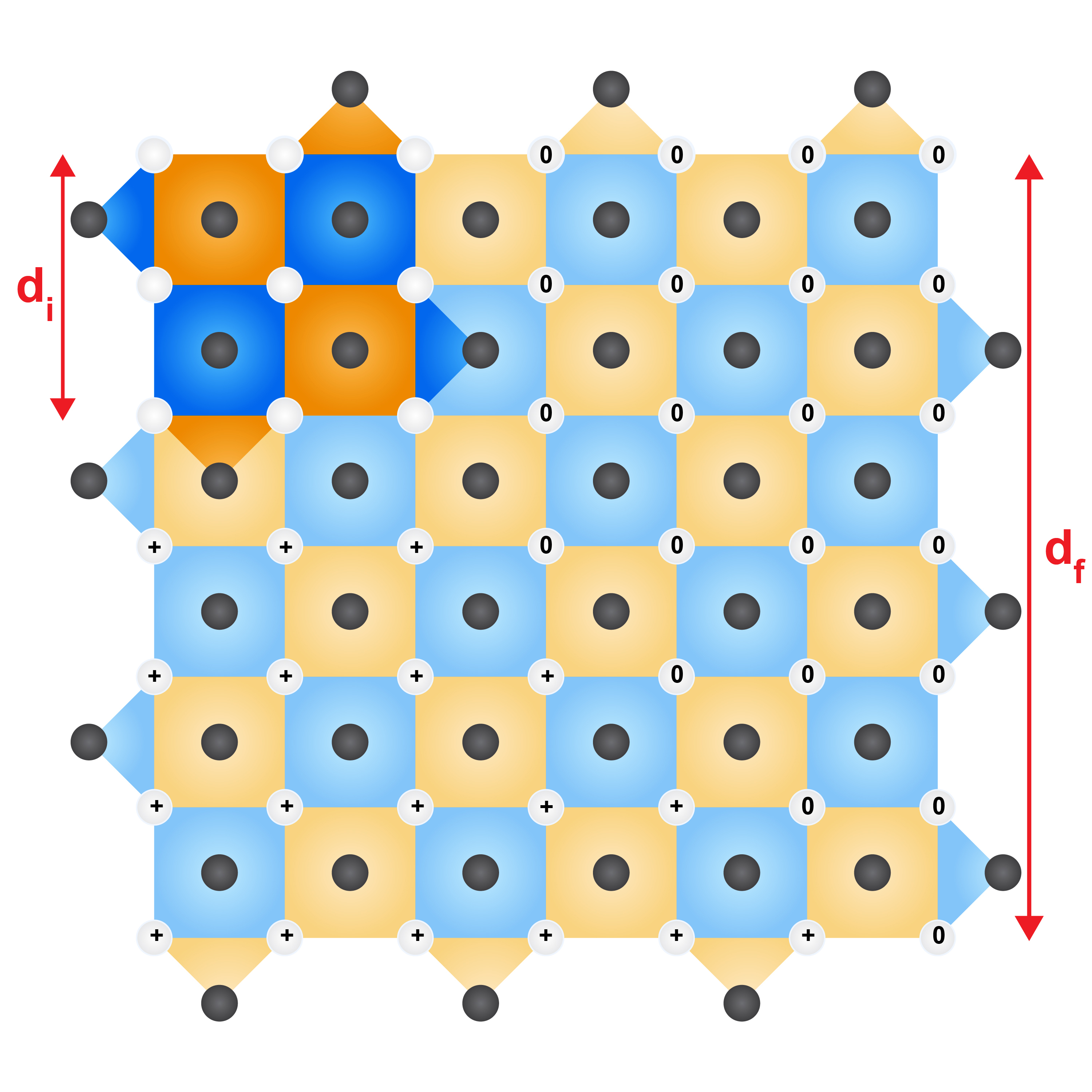}
    \caption{Schematic illustration of rotated surface code growth. A logical state encoded in a small surface code patch of initial code distance $d_i$ can be grown to a larger surface code patch of final target code distance $d_f$ according to the protocol proposed in Ref.~\cite{Fowler2018LowOQ}. Here, the brightly coloured patch with distance $d_i=3$ in the top left corner  indicates the initial state, while the pale coloured zone indicates the target code patch of distance  
    $d_f=7$.  White circles represent data qubits, while  grey circles indicate syndrome qubits. Blue faces measure the $Z$-type stabilizers and orange 
    faces measure the $X$-type stabilizers. 
    The $X_L$ operator extends along the left boundary, while the $Z_L$ operator extends across the top boundary. To grow the state of the initial patch, first the data qubits in the expansion zone are prepared in either the $\ket{0}$ or the $\ket{+}$ states, as shown. This is followed by $\mathcal{O}(d_f)$ rounds of error correction.}
    \label{fig:growth_scheme}
\end{figure}

Growing the size of the code hosting the rotation resource state is necessary when large code distances are required for logical qubit patches in the core processor. This is because the success rate of the transversal multi-rotation protocol decreases significantly at larger code distances, as shown in \Cref{fig:success-rate}.
The growth protocol is used to grow resource states on small distance surface codes to the required distance while injecting minimal error into the logical state. \Cref{fig:growth_scheme} shows the essential details of the growth protocol~\cite{Fowler2018LowOQ}. A logical state encoded in a patch of distance $d_i$ is grown into a patch of distance $d_f$. First, the data qubits in the expansion zone are prepared in either $\ket{0}$ or $\ket{+}$ as shown. The qubits along the left edge are prepared in the $\ket{+}$ state to extend the $X_L$ operator along the entire left edge of the larger patch. Similarly, the qubits along the top edge are prepared in the $\ket{0}$ state to extend the $Z_L$ operator across the entire top edge of the larger patch. The value of the other qubits is chosen to minimize the injected error. After this step, $\mathcal{O}(d_f)$ rounds of error correction must be performed to resolve the values of stabilizers in the expansion zone. This completes the growth of the logical state.

\begin{figure*}[tb]
  \centering
  \begin{tabular}{l @{\hskip 0.1in} l}
  \text{(a) Direct growth for target hardware specifications} & \text{(b) Direct growth for desired hardware  specifications}\\
\includegraphics[width=0.39\linewidth]{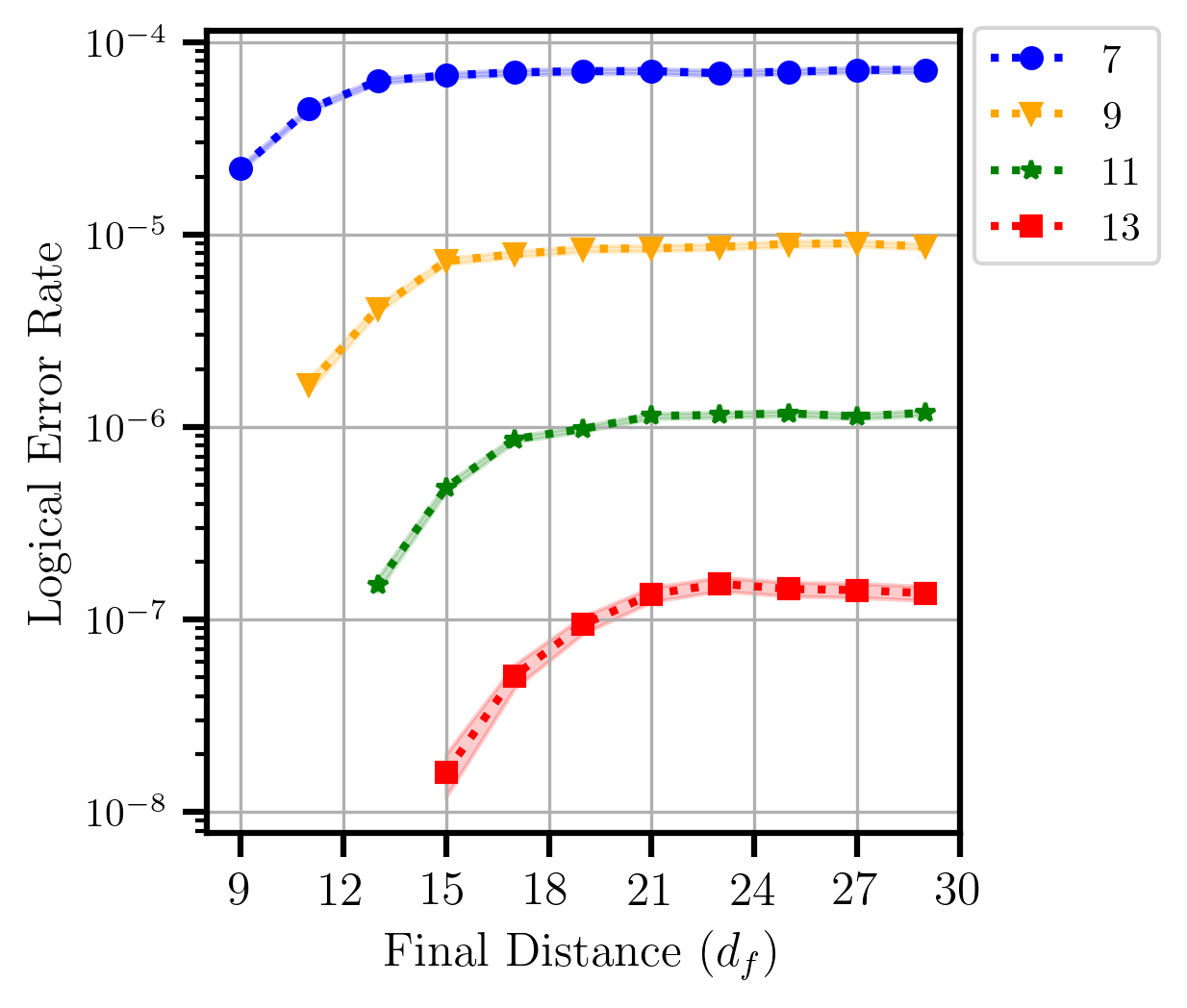} & \includegraphics[width=0.39\linewidth]{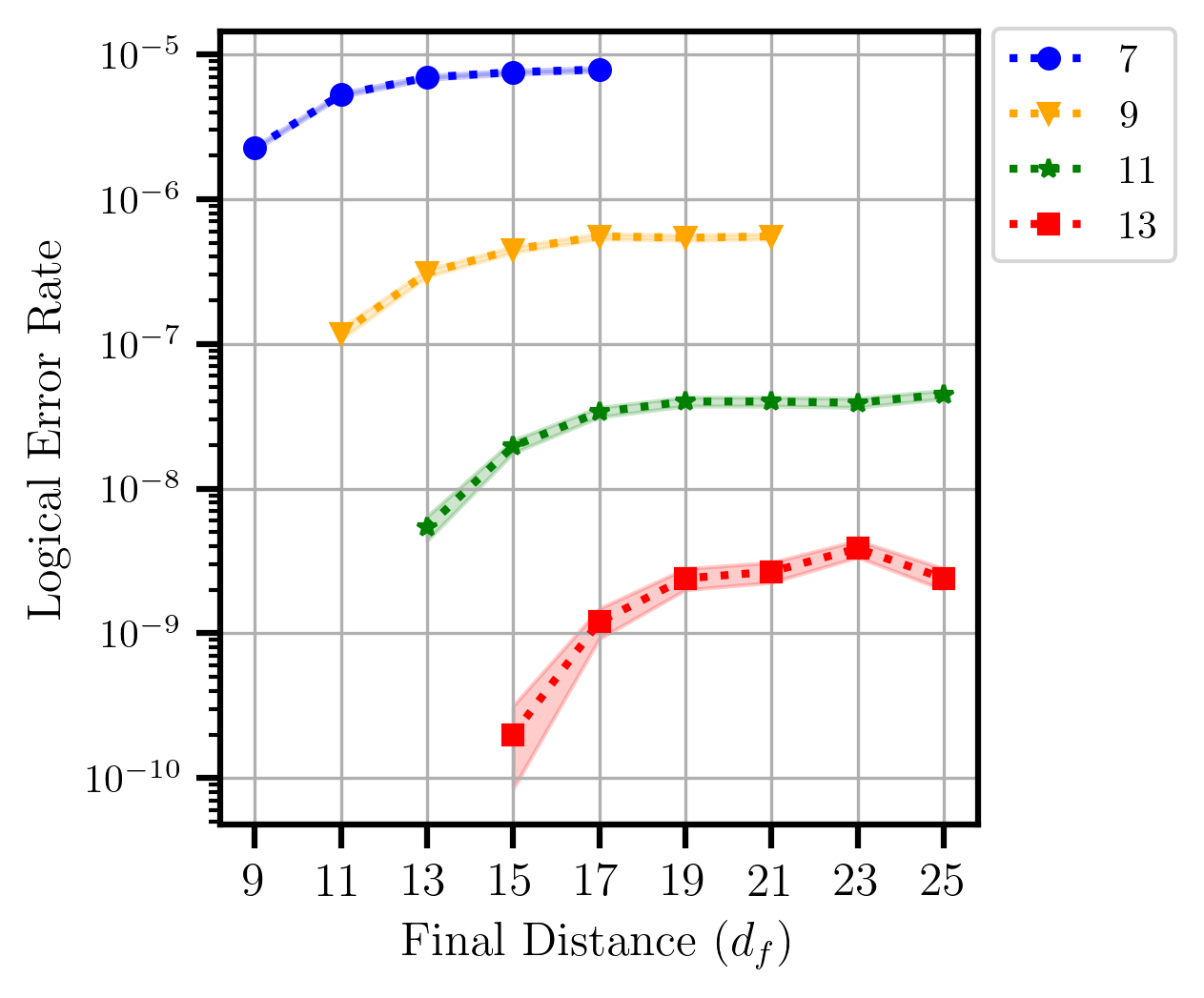}\\
&\\
    \text{(c) Stepwise growth for target hardware  specifications} & 
     \text{(d) Stepwise growth for desired hardware  specifications}\\ \includegraphics[width=0.497\linewidth]{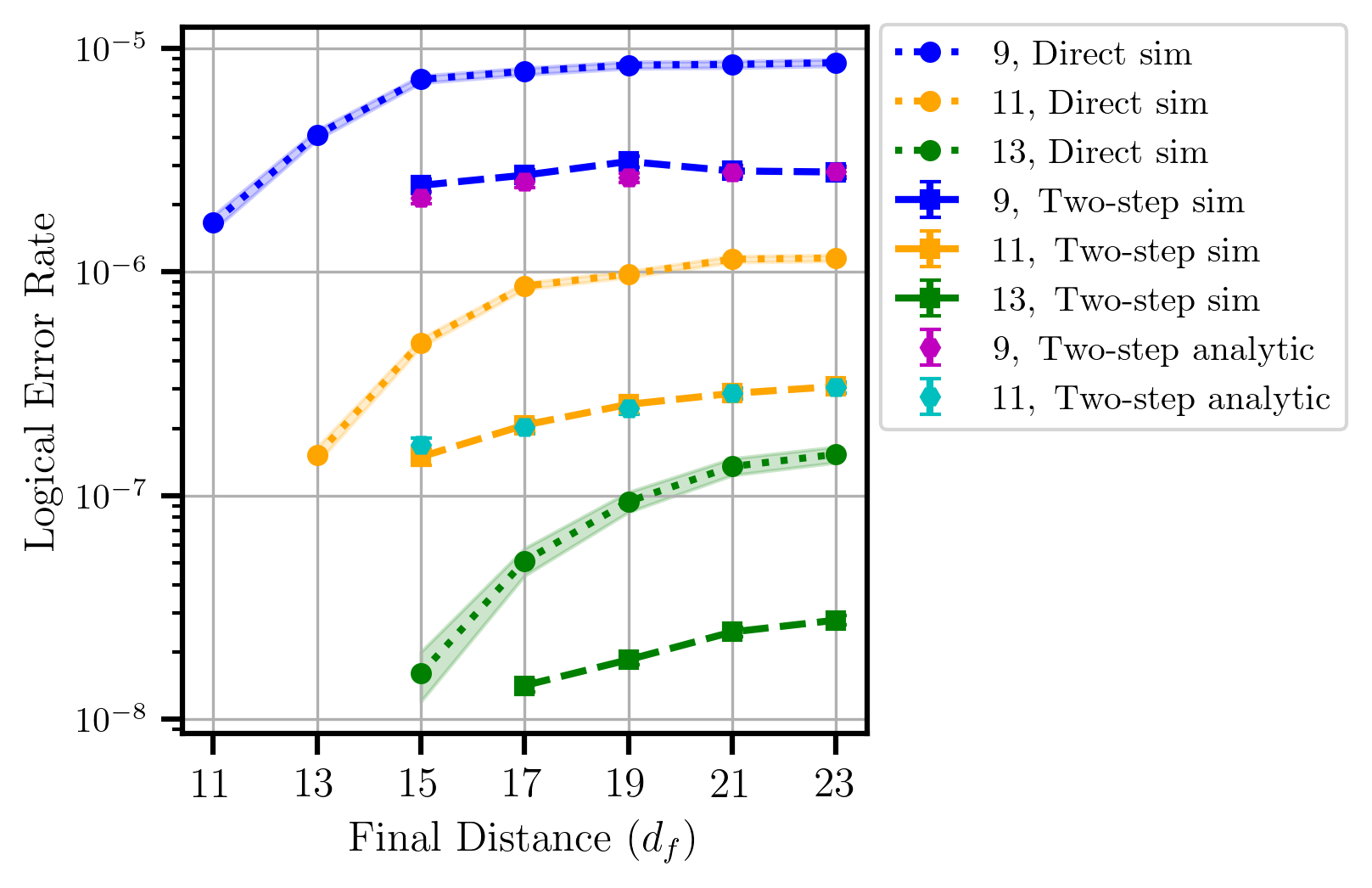} & \includegraphics[width=0.497\linewidth]{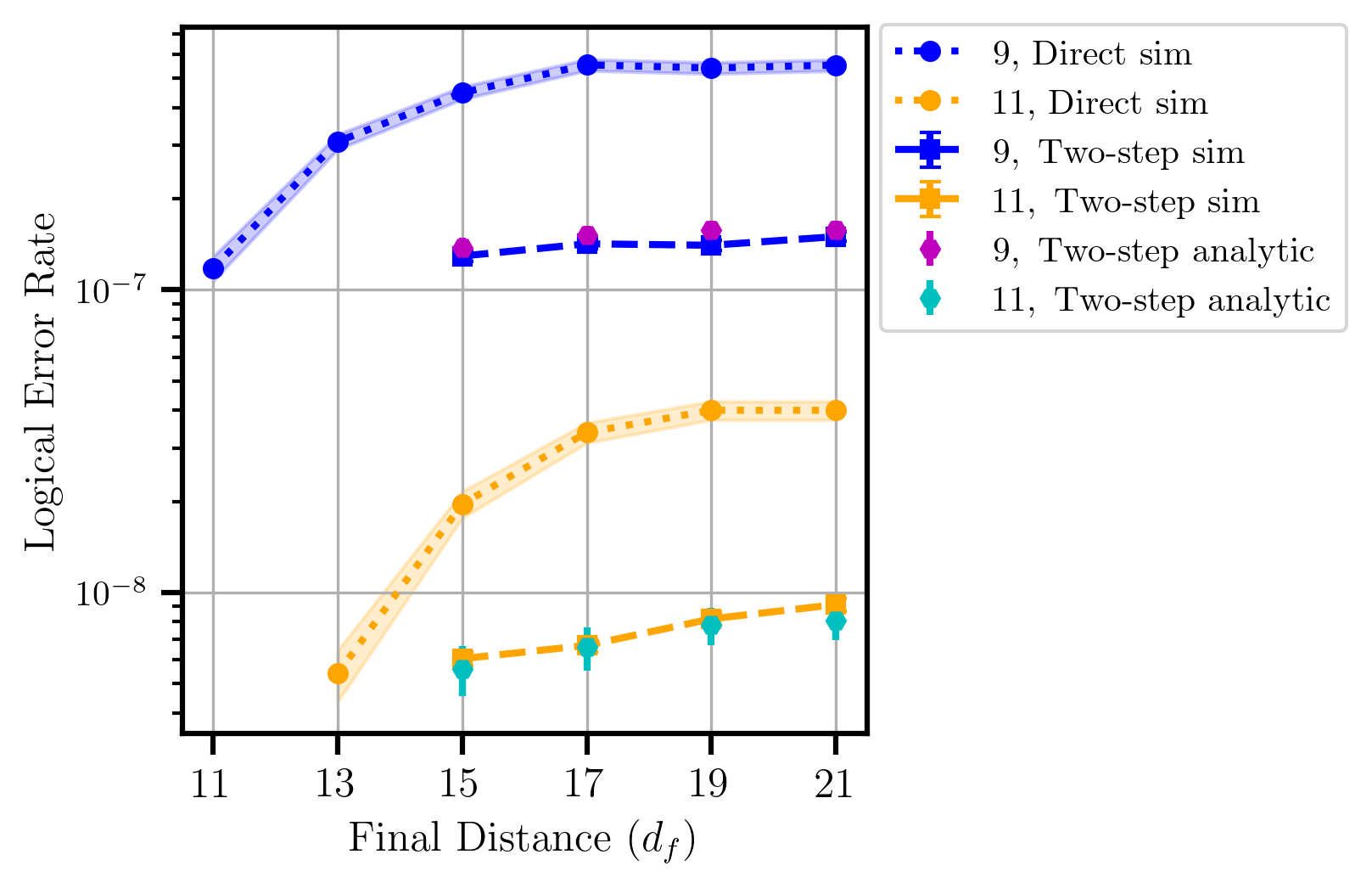}
  \end{tabular}
  \caption{Logical error rate (LER) of the direct-growth and two-step growth protocols as a function of the final target distance $d_f$, for various initial distances $d_i$ as indicated in the legend. Simulation results are shown for (a) direct growth for target and (b) desired hardware parameter specifications. We introduce a two-step growth procedure, where the initial patch at distance $d_i$ is first expanded to $d_i + 2$ and then to the final distance $d_f$. Panels (c) and (d) show a comparison between two-step and direct growth for target specifications (c) and desired hardware parameters (d). Additionally, we compare the simulation results of two-step growth with the analytic inference, $\text{LER}(d_i \rightarrow d_i+2) + \text{LER}(d_i+2 \rightarrow d_f)$.
}
  \label{fig:growth_infidelity}
\end{figure*}

For our application, we want to prepare a resource state of distance $d_f$ in one logical cycle, which consists of $d_f$ rounds of error correction. The transversal rotation protocol takes $4$ rounds to execute. This leaves us $d_f-4$ rounds to grow the state. We simulate the growth protocol to estimate the logical error rate. We first noiselessly prepare a patch of distance $d_i$ in the logical $\ket{+}_L$ state. Then we perform $d_f-4$ rounds of error correction, in which target or desired noise is added. The results for various choices of $d_i$ are shown in \Cref{fig:growth_infidelity}. As the growth protocol is not fault tolerant, that is, it is not possible to grow a state without injecting more errors into it, the logical error rate increases. For each $d_i$ we grow until the logical error rate stabilizes to some value.  We find that larger initial patches result in lower error rates, even when reaching similar final distances.

In addition to the direct growth protocol, where an initial patch of distance $d_i$ is expanded directly to the final target distance $d_f$, we introduce a two-step growth protocol. In this method, the initial patch is first grown to an intermediate distance of $d_i+2$ before reaching the final target distance $d_f$.This two-step approach is motivated by the sensitivity of the logical error rate (LER) to the initial patch distance. As shown in \Cref{fig:growth_infidelity}(a) and (b), increasing the initial distance from $d_i$ to $d_i+2$ typically reduces the LER by nearly an order of magnitude. However, this method introduces a time trade-off because the intermediate step requires at least $d_i+2$ additional stabilization rounds. Consequently, adding further intermediate steps (i.e., multiple-step growth) would continue to compound this round overhead. We hypothesize that the total error of the two-step protocol can be estimated analytically by simply summing the LERs of the two individual steps, expressed as \begin{equation}
\text{LER}(d_i \rightarrow d_i+2) + \text{LER}(d_i+2 \rightarrow d_f).\end{equation}
To verify this assumption, we compared this analytical inference against our simulation results. As shown in \Cref{fig:growth_infidelity}(c) and (d), the simulations confirm our hypothesis and align closely with the analytical prediction. 

\begin{figure}[tb]
  \centering
  \includegraphics[width=0.43\textwidth]{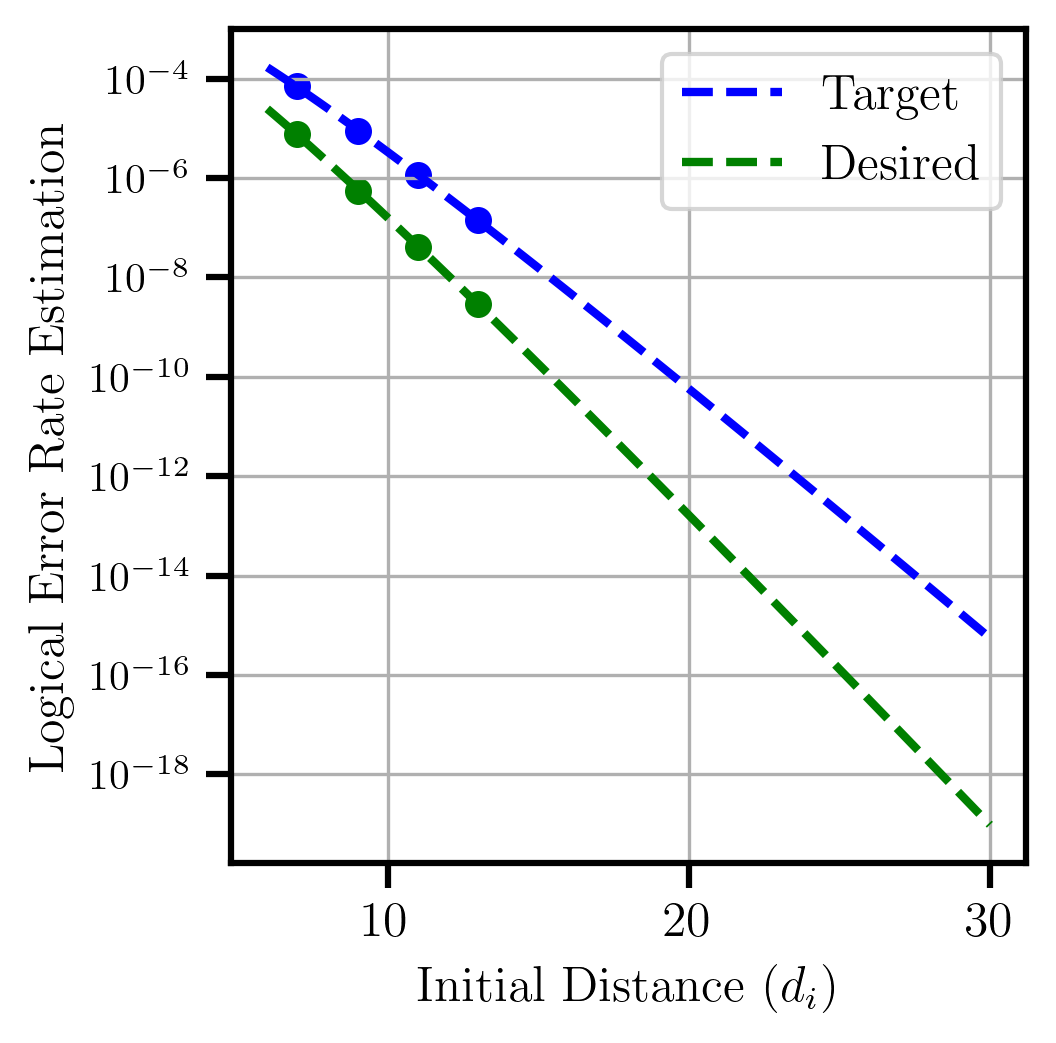}
  \caption{Asymptotic logical error rates for the growth protocol as a function of initial distance. The dots are the asymptotic values from simulation, while the dashed lines are fits according to the error model given in~\Cref{eq:growth_model}. Simulation results are shown for target (blue colour) and desired (green colour) hardware parameter specifications, as indicated in the legend.}
  \label{fig:asym_prediction}
\end{figure}

It is not feasible to simulate the growth protocol for the values of $d_i$ and $d_f$ used in the resource estimates. To estimate these values, we fit the asymptotic values of the logical error rate of each $d_i$ against $d_f$. These asymptotic values provide fairly tight upper bounds of the logical error rate for any choice of $d_i$ and $d_f$. We find that the model
\begin{equation}
P_\text{growth}(d_i) = \mu d_i^2 \Lambda^{-\frac{d_i + 1}{2}}, \label{eq:growth_model}
\end{equation}
where $\mu$ and $\Lambda$ are fitting parameters, provides good fits. This can be observed in \Cref{fig:asym_prediction}.

The final fidelity $F$ of a logical rotation resource state, initially prepared in a code patch of distance $d_i$, and subsequently grown to a patch of target distance $d_f$, is obtained as 
\begin{equation}
    F = F_\text{resource} \times F_\text{growth}.
    \label{eq: fidelity def}
\end{equation}
This fidelity is used in our QRE studies presented in \Cref{sec:QRE-studies}.

\subsection{Rotation gate error evaluation}
\label{sec:gate_error}
With the simulation data in the previous sections, we start discussing the rotation gate error. A single rotation gate consists of multiple RUS trials so the infidelity of the rotation state preparation and the growth error accumulates over multiple RUS trials. In \cite{toshio2024practical}, it is shown that the coherent over-rotation error dominates the worst case gate error evaluation and thus probabilistic coherent error cancellation is applied to cancel off the coherent error term, at the cost of doubling the infidelity. This can be seen from comparing Eqs.~(12) and (21) in Ref.~\cite{toshio2024practical}. 

Here, we take the effect of error accumulation into account by setting the worst-case error rate $\varepsilon_\diamond(\mathcal{E}^c_{\theta_*})$ in Eq.~(23) of Ref.~\cite{toshio2024practical} to be \mbox{$2(1 - F_{\text{resource}}(\theta_*, d_i))\equiv 2 \mathcal{I}_{\text{rot}}(\theta_*, d_i)$}. In addition, we also need to take into account the infidelity resulting from the code patch growth. Hence, the total logical error of  implementing a logical $R_Z(\theta_*)$ gate amounts to: 
\begin{equation}
\small
\begin{split}
    P_L{(\theta_*)} 
    &= \sum_{K=1}^{\infty}\frac{1}{2^K} \sum_{n=1}^K\!\left( 2\mathcal{I}_{\text{rot}}(\omega(2^{n-1}\theta_*), d_i) +  \mathcal{I}_{\text{growth}}(d_i, d_f)\right)\\
    & =\left[\sum_{K=1}^{\infty}\frac{1}{2^K} \!\sum_{n=1}^K\!\left( 2\mathcal{I}_{\text{rot}}(\omega(2^{n-1}\theta_*), d_i) \right)\!\right] \! +2 \mathcal{I}_{\text{growth}}(d_i, d_f)\\
    & \sim \mathcal{O}(|\theta_*|)\! + 2 \mathcal{I}_{\text{growth}}(d_i, d_f).
\end{split}
\end{equation}
Here, $\omega(\phi)$ denotes that the magnitude of the rotation angle $\phi$ must be constrained within $\pi/8$. Thus, whenever $\phi$ exceeds $\pi/8$, one should replace it with an $S$ gate and a rotation angle of $|\pi/4 - \phi|$. The next term with $\mathcal{I}_{\text{growth}}$ denotes the error due to growth infidelity. It has no $n$ dependency so its coefficient sums to a factor of 2. The rotation infidelity has $n$ dependency and the final scaling of the error due to resource state preparation scales linearly with respect to the rotation angle magnitude. This is the same scaling as in \Cref{eq:STAR error model}, which was first estimated in \cite{toshio2024practical}.

\begin{figure*}[tb]
  \centering
    \begin{tabular}{l}
  \text{(a) Logical error rate when performing a rotation with analog rotation resource states prepared at distance $d$ without growth}\\
  \\
    \includegraphics[width=0.9\linewidth]{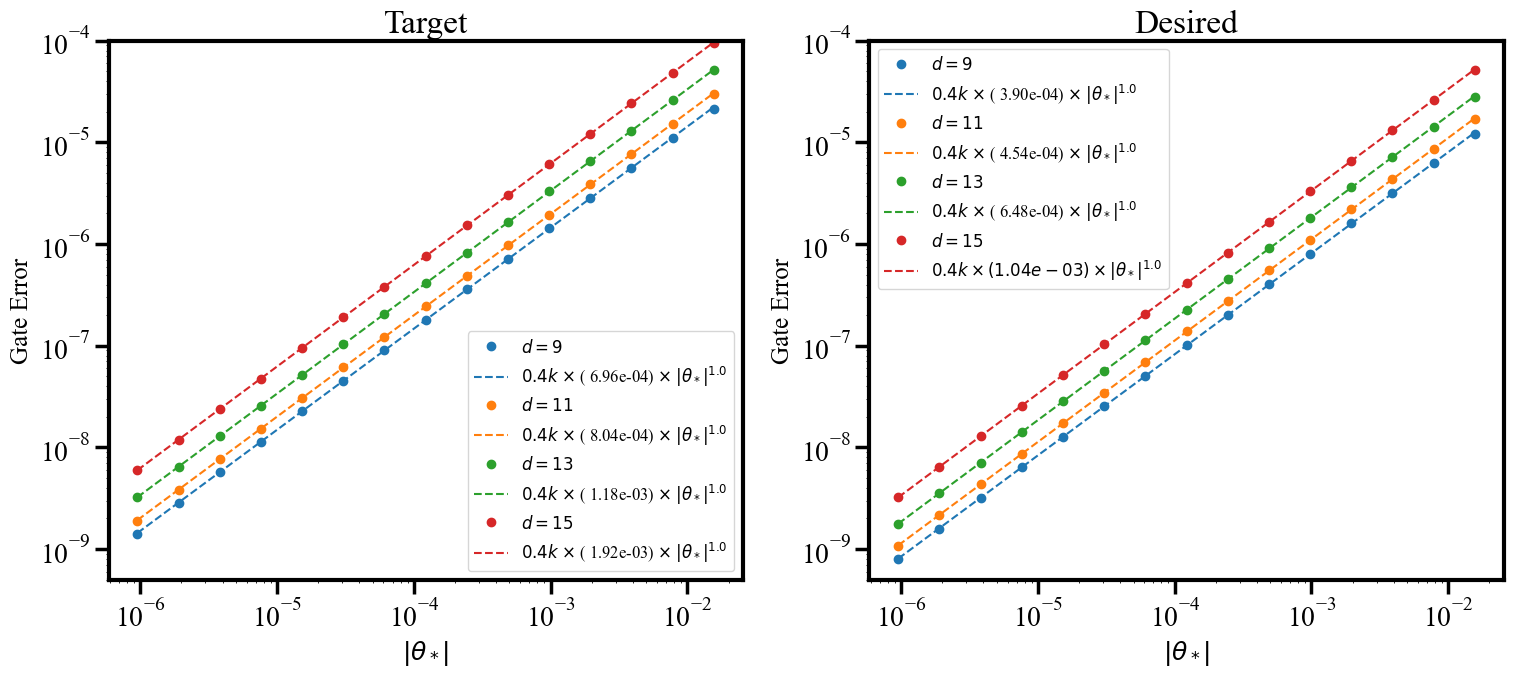}\\
    \\
    \text{(b) Logical error rate when analog rotation resource states are initially prepared at distance $d_i$ and subsequently grown to $d_f$}\\
    \\
    \includegraphics[width=0.9\linewidth]{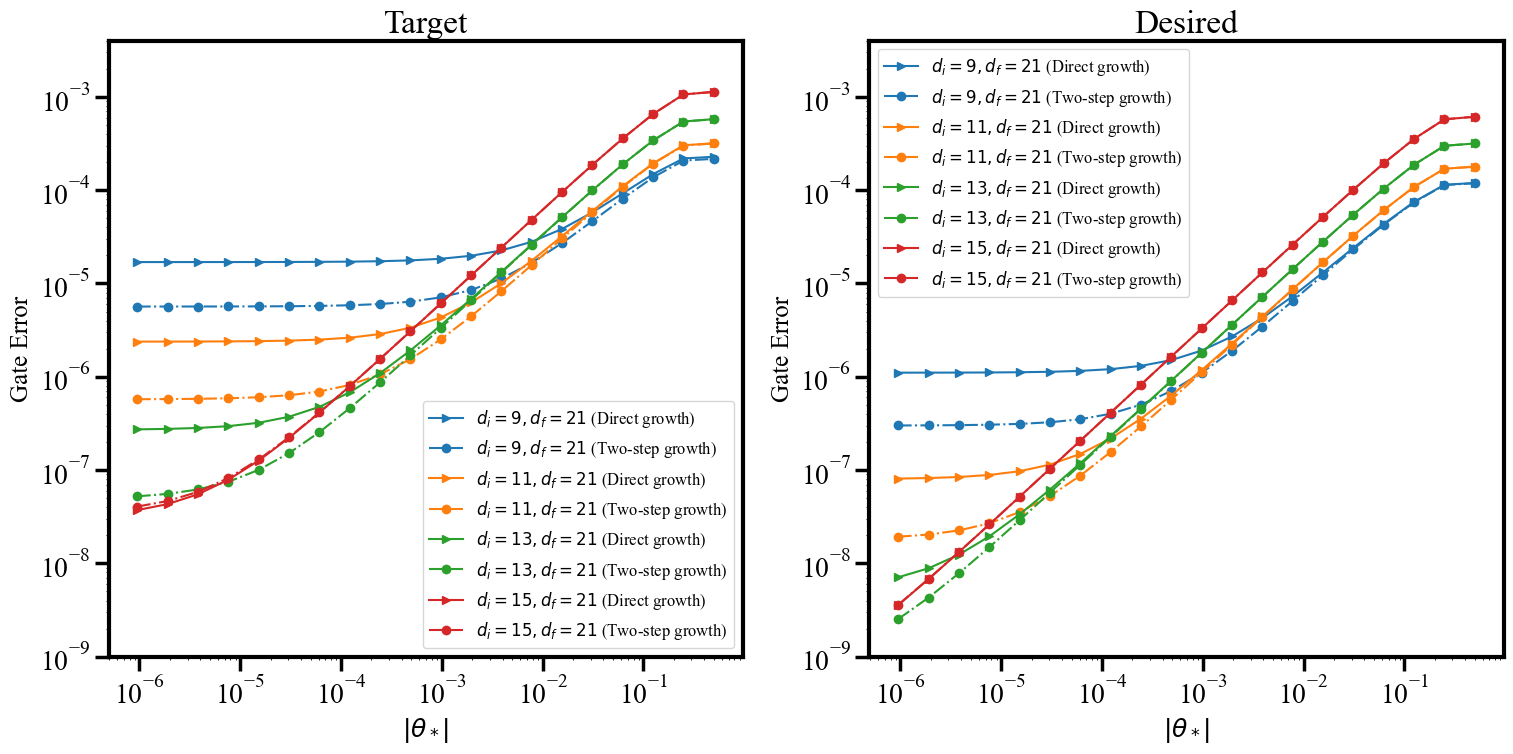}
    \end{tabular}
  \caption{Logical error rate of analog $R_Z(\theta_*)$ rotation as a function of rotation angle $\theta_*$ 
  for (a) resource state preparation in a code patch of distance $d$ without growth; and (b) resource state preparation with growth for various  $(d_i, d_f)$ combinations. Here, $k = \lceil d/2\rceil$ in the case 
  without growth, and $k = \lceil d_i / 2\rceil$ in the case with growth. For each fixed angle $\theta_*$, there exists a best choice for the combination of the initial and final code distance, $(d_i, d_f)$. 
  } 
  \label{fig:rot_gate_error}
\end{figure*}

We hereby show the estimated $R_Z$ rotation error $P_L(\theta_*, d_i, d_f)$ in \Cref{fig:rot_gate_error}. We explicitly show the fitting equation of the error model in the legends of \Cref{fig:rot_gate_error}(a), which indeed is scaling as $\mathcal{O}(\theta_*)$. The factor of $0.4 k$ is separated because it is the $\alpha_{\text{RUS}}$ factor proposed in \cite{toshio2024practical}. Keeping it as an explicit factor helps us compare the PEC overhead in subsequent resource estimation studies. 

\section{Resource estimation studies}
\label{sec:QRE-studies}

In this section, we present detailed resource estimation studies for the ground state energy estimation (GSEE) problem. We consider two representative high-utility applications: quantum computation of electronic spectra of molecules and quantum simulations of the Fermi--Hubbard model. 
Our resource estimations consist of the physical qubit count and the physical wall-clock time for executing the associated quantum circuits. For our QRE studies, we use the success rate and infidelity estimations presented in~\Cref{sec:simulation_results}.


\subsection{Runtime estimation}
Following the discussion in ~\Cref{sec: SPE budget}, we set up the error budget equation for the QCELS algorithm as in Ref.~\cite{Akahoshi:2024yme}. The optimal algorithmic parameters are obtained by minimizing the total number of Trotter steps needed to execute the algorithm while also satisfying the constraint:
\begin{equation}
 \epsilon_{\text{Trotter}} + \epsilon_{\text{QCELS}} = \epsilon_{\text{target}},
 \label{eq:SPE_budget}
\end{equation}
where $\epsilon_{\text{Trotter}}$ is the error due to Trotter decomposition, $\epsilon_{\text{QCELS}}$ is the error caused by the QCELS algorithm, and $\epsilon_{\text{target}}$ is the target energy error we can tolerate. The resulting runtime is given by
\begin{eqnarray}
    T_{\text{total}} = N_s\sum_{j=1}^J\sum_{n=0}^N \gamma_{n,j}^2 \left\lceil n  \frac{2^{j-J}\delta}{\epsilon_{\text{QCELS}}} \sqrt{\frac{w}{\epsilon_{\text{Trotter}}}} \;\right\rceil T_{\text{1-Trotter}}.
    \label{eq:QCELS runtime estimation.}
\end{eqnarray}
Derivation and parameter definitions of this equation are summarized in Appendix~\ref{app: QCELS scaling}. 

The simulation results in~\Cref{sec:simulation_results} affect the runtime via the PEC overhead factor $\gamma_{n, j}^2$ and the execution time of 1-Trotter step $T_{1-\text{Trotter}}$.  These are both dependent on the initial and final code distances, $d_i$ and $d_f$, of the growth protocol discussed in~\Cref{sec:growth_protocol}. As we require the total logical error of the algorithm to be smaller than 1\%, we can find the required final code distance by multiplying the logical space-time volume of the circuit by the logical error rate of memory for the surface code and ensuring the product is below our target error rate:
\begin{equation}
    N_{\text{patch}}\times (N_{\text{clock}} \mathcal{C})\times\mu d_f\Lambda^{-\frac{d_f+1}{2}} \leq \epsilon_{\text{log err}} = 0.01.
    \label{eq:df_det}
\end{equation}
In this equation, $N_{\text{patch}}$ is the total number of logical qubit patches used by the core processor for the algorithm, including data and bus qubits.  Detailed discussions of logical qubit layouts will occur in subsequent subsections, as they are dependent on algorithm requirements.  $\mathcal{C}$ is the circuit depth of a single clock cycle. $N_{\text{clock}}$ is the length of the deepest circuit in the QCELS algorithm on the given architecture in terms of clock cycles.  It depends on the logical patch layout, resource state factory layout, growth protocol and lattice surgery scheduling, all of which need to be implemented in a way that ensures a continuous supply of resource states are available for the algorithm.
Note that \Cref{eq:df_det} only affects the errors from idling and lattice surgery. The errors caused by rotation state preparation are left out from this budget as they will be mitigated by PEC.

The factor $\gamma_{n, j}^2$ in \Cref{eq:QCELS runtime estimation.} describes the PEC overhead and relates to the gate error discussed in \Cref{sec:gate_error} via:
\begin{equation}
    \gamma_{n,j}^2 = \exp\left(4 N_{n, j}\sum_{1-\text{trotter}} P_L(\theta_*)\right),
    \label{PEC overhead}
\end{equation}
where $N_{n, j}$ is the number of Trotter slices needed to evolve the Hamiltonian for the $(n,j)$-th step of the QCELS algorithm, and the summation is over the rotation gate errors in a single Trotter slice.

\subsection{Electronic-structure quantum computations: 
ground-state energy estimation for small molecular active spaces}
\label{sec: p-benzyne estimate}

\begin{table*}[!htbp]
    \centering
    \setlength{\tabcolsep}{1pt} 
    {\scriptsize
    \begin{tabular*}{\textwidth}{|c|c|c|c|c|@{\extracolsep{\fill}}ccccccc@{\extracolsep{\fill}}ccccccc@{\extracolsep{\fill}}}
        \hline
        \hline
        \multicolumn{19}{c}{}\\
        \multicolumn{19}{l}{\normalsize (a) Partially Fault-Tolerant STAR Architecture}\\
        \multicolumn{19}{c}{}\\
       \multicolumn{5}{c}{}&&&&&&&&\\
       \multicolumn{2}{c}{} & \multicolumn{3}{c}{\bf Logical Resources} &  \multicolumn{2}{c}{}&
       \multicolumn{6}{c}{\bf Target Hardware  } & 
       \multicolumn{6}{c}{\bf Desired Hardware  } \\
        \cmidrule{3-5} 
        \cmidrule{6-12} 
        \cmidrule{13-19}
       \multicolumn{1}{c}{\multirow{2}{*}{\shortstack[c]{Alg.}}} & 
       \multicolumn{1}{c}{\multirow{2}{*}{\shortstack[c]{$N_{\text{orb}}$}}} & 
       \multicolumn{1}{c}{\multirow{2}{*}{\shortstack[c]{\# Log. \\qubits }}} & 
       \multicolumn{1}{c}{\multirow{2}{*}{\shortstack[c]{\# Rot. \\ gates}}} & 
       \multicolumn{1}{c}{\multirow{2}{*}{\shortstack[c]{$N_\text{max}$}}} & 
       \multirow{3}{*}{\shortstack[c]{\# Phys.\\ qubits}} & 
        \multirow{3}{*}{\shortstack[c]{Phys. \\ time}} & 
        \multirow{3}{*}{\shortstack[c]{PEC free \\ time}} & 
        \multirow{3}{*}{\shortstack[c]{Max.\\ single- \\ shot time }} &  
        \multirow{3}{*}{\shortstack[c]{$d_i$}} &  
        \multirow{3}{*}{\shortstack[c]{$d_f$}} &  
        \multirow{3}{*}{\shortstack[c]{\# Prep.\\ patches}} &  
        \multirow{3}{*}{\shortstack[c]{\# Phys.\\ qubits}} & 
        \multirow{3}{*}{\shortstack[c]{Phys. \\ time}} & 
        \multirow{3}{*}{\shortstack[c]{PEC free \\ time}} & 
        \multirow{3}{*}{\shortstack[c]{Max.\\ single- \\ shot time}}&
        \multirow{3}{*}{\shortstack[c]{$d_i$}} &
        \multirow{3}{*}{\shortstack[c]{$d_f$}} &
        \multirow{3}{*}{\shortstack[c]{\# Prep.\\ patches}}\\
        \multicolumn{5}{c}{}&&&&&&&&\\
        \multicolumn{5}{c}{}&&&&&&&&\\
        \hline\hline
        \multirow{6}{*}{\rotatebox[origin=c]{90}{\tiny{QCELS, empir.}}} &&&&&&&&&&&&&&&&\\
        & 6
        & 12 (39) & 9.8$\times$10\textsuperscript{3} & 40
        & 2.9$\times$10\textsuperscript{4} & 6.5 min & 2.6 min & 0.1 sec & 11 & 17 & 12
        & 2.1$\times$10\textsuperscript{4} & 3.5 min & 2.3 min & 0.1 sec  & 11 & 15 &  7
        \\
        & 14
        & 28 (87)
        & 6.5$\times$10\textsuperscript{5}  & 93
        & 9.9$\times$10\textsuperscript{4} & 10.9 yrs & 3.8 hrs & 10.5 sec  & 11 & 23 & 7
        & 5.4$\times$10\textsuperscript{4} & 7.3 days & 2.8 hrs & 7.8 sec  & 11 & 17 &  6
        \\
        & 18
        & 36 (111) 
        & 2.3$\times$10\textsuperscript{6} & 120 
        & 1.5$\times$10\textsuperscript{5} & $10^{10}$ days & 14.4 hrs & 40.7 sec & 11 & 25 & 7
        & 8.4$\times$10\textsuperscript{4} & 17.8 yrs & 10.9 hrs & 30.9 sec & 11 & 19 & 6
        \\
        & 26
        & 52 (159) 
        & 1.5$\times$10\textsuperscript{7}  & 173 
        & 2.5$\times$10\textsuperscript{5} & $10^{52}$ days & 4.4 days & 4.9 min & 13 & 27 & 13
        & 1.5$\times$10\textsuperscript{5} & $10^{15}$ days & 3.4 days & 3.8 min & 13 & 21 & 9
        \\
        &&&&&&&&&&&&\\
        \hline\hline
        \multirow{6}{*}{\rotatebox[origin=c]{90}{\tiny{QCELS, theoret.}}} &&&&&&&&&&&&&&&&\\
        & 6
        & 12 (39) & 2.6$\times$10\textsuperscript{2} & 1 
        & 2.3$\times$10\textsuperscript{4} & 17.7 hrs & 17.2 hrs & 3.0 ms & 11 & 15 & 13
        & 2.1$\times$10\textsuperscript{4} & 17.5 hrs & 17.2 hrs & 3.0 ms & 11 & 15 & 7
        \\
        & 14
        & 28 (87) & 7.0$\times$10\textsuperscript{3} & 1 
        & 7.1$\times$10\textsuperscript{4} & 41.3 days & 32.4 days & 0.1 sec & 11 & 19 & 11
        & 4.2$\times$10\textsuperscript{4} & 28.1 days & 25.6 days & 0.1 sec & 11 & 15 & 7
        \\
        & 18
        & 36 (111) 
        & 3.8$\times$10\textsuperscript{4} & 2 
        & 1.1$\times$10\textsuperscript{5} & 136.5 days & 108.7 days & 0.6 sec & 11 & 21 & 10
        & 5.3$\times$10\textsuperscript{4} & 85.8 days & 77.6 days & 0.4 sec & 11 & 15 & 7
        \\
        & 26
        & 52 (159) 
        & 1.8$\times$10\textsuperscript{5} & 2 
        & 1.8$\times$10\textsuperscript{4} & 3.8 yrs & 1.7 yrs & 2.9 sec & 11 & 23 & 7
        & 9.5$\times$10\textsuperscript{4} & 1.7 yrs & 1.3 yrs & 2.2 sec & 11 & 17 & 6
        \\
        &&&&&&&&&&&&\\
        \hline\hline
    \end{tabular*}

\begin{tabular*}{\textwidth}{|c|c|cc|@{\extracolsep{\fill}}ccc ccc@{\extracolsep{\fill}}ccc ccc}
    \multicolumn{16}{c}{}\\
    \multicolumn{16}{l}{\normalsize (b) Fault-Tolerant Architecture}\\
    \multicolumn{16}{c}{}\\
    \multicolumn{4}{c}{} & \multicolumn{6}{c}{\bf Time-optimal} & \multicolumn{6}{c}{\bf Space-optimal} \\
    \cmidrule{5-10}\cmidrule{11-16}
    \multicolumn{2}{c}{} &
    \multicolumn{2}{c}{\bf Logical Resources} &
    \multicolumn{3}{c}{\bf Target Hardware} &
    \multicolumn{3}{c}{\bf Desired Hardware} &
    \multicolumn{3}{c}{\bf Target Hardware} &
    \multicolumn{3}{c}{\bf Desired Hardware} \\
    \cmidrule{3-4}
    \cmidrule{5-7}\cmidrule{8-10}
    \cmidrule{11-13}\cmidrule{14-16}
    \multicolumn{1}{c}{\multirow{2}{*}{\shortstack[c]{Alg.}}} &
    \multicolumn{1}{c}{\multirow{2}{*}{\shortstack[c]{$N_{\text{orb}}$}}} &
    \multicolumn{1}{c}{\multirow{2}{*}{\shortstack[c]{\# Log.\\ qubits}}} &
    \multicolumn{1}{c}{\multirow{2}{*}{\shortstack[c]{\# $T$\\ gates}}} &
    \multirow{2}{*}{\shortstack[c]{\# Phys.\\ qubits}} &
    \multirow{2}{*}{\shortstack[c]{Phys.\\ time}} &
    \multirow{2}{*}{\shortstack[c]{QEC code\\ distances}} &
    \multirow{2}{*}{\shortstack[c]{\# Phys.\\ qubits}} &
    \multirow{2}{*}{\shortstack[c]{Phys.\\ time}} &
    \multirow{2}{*}{\shortstack[c]{QEC code\\ distances}} &
    \multirow{2}{*}{\shortstack[c]{\# Phys.\\ qubits}} &
    \multirow{2}{*}{\shortstack[c]{Phys.\\ time}} &
    \multirow{2}{*}{\shortstack[c]{QEC code\\ distances}} &
    \multirow{2}{*}{\shortstack[c]{\# Phys.\\ qubits}} &
    \multirow{2}{*}{\shortstack[c]{Phys.\\ time}} &
    \multirow{2}{*}{\shortstack[c]{QEC code\\ distances}}\\
    \multicolumn{4}{c}{}&&&&&\\
    \hline\hline

    \parbox[t]{4mm}{\multirow{5}{*}{\rotatebox[origin=c]{90}{{DF Qubitiz.}}}} &
    &&&&&&&\\ 
    & 6 & 291 & 4.9$\times$10\textsuperscript{6}
       & 1.4$\times$10\textsuperscript{6} & 42.9 sec & 23 | 25
       & 1.0$\times$10\textsuperscript{6} & 42.9 sec & 17 | 25
       & 1.0$\times$10\textsuperscript{6} & 7.9 min & 23 | 27
       & 8.5$\times$10\textsuperscript{5} & 5.8 min & 17 | 25 \\
    & 18 & 502 & 2.3$\times$10\textsuperscript{9}
       & 3.6$\times$10\textsuperscript{6} & 6.9 hrs & 11, 29 | 31
       & 2.5$\times$10\textsuperscript{6} & 6.9 hrs & 23 | 31
       & 2.6$\times$10\textsuperscript{6} & 2.9 days & 11, 29 | 33
       & 2.1$\times$10\textsuperscript{6} & 2.6 days & 23 | 31 \\
    & 26 & 668 & 9.0$\times$10\textsuperscript{9}
       & 4.7$\times$10\textsuperscript{6} & 1.2 days & 11, 31 | 33
       & 3.5$\times$10\textsuperscript{6} & 1.2 days & 23 | 33
       & 3.6$\times$10\textsuperscript{6} & 19.9 days & 13, 31 | 35
       & 3.2$\times$10\textsuperscript{6} & 10.1 days & 23 | 33 \\
    &&&&&&&\\ 
    \hline\hline
\end{tabular*}
    }
\caption{Comparison of resource requirements of partially fault-tolerant and fully fault-tolerant implementations of the electronic-structure quantum circuits associated with GSEE for \mbox{$p$-benzyne}, for qualitatively accurate computation within a target accuracy \mbox{1.6 mHa}, using  a circuit-level error budget of 0.01. In both cases, we report estimates for the physical wall-clock times and the number of physical qubits required for implementations of the corresponding QPE algorithms for electronic spectra associated with various molecular active spaces with sizes specified by the number of orbitals $N_{\text{orb}}$, using the chemical basis set $\mathtt{6-31G}$ to represent the spin orbitals. The data for $N_{\text{orb}}=6, 14, 18, 26$ correspond to active space selections HL$\pm 2, 6, 8, 12$ (using HL$\pm n$ to denote ``HOMO$-n$ and LUMO$+n$''). 
Physical resources are reported for two  hardware specifications, namely, \lq\lq{}target\rq\rq{} and \lq\lq{}desired\rq\rq{} hardware, as summarized in \Cref{tab:physical_params}. 
Furthermore, for both the STAR architecture and the fully fault-tolerant architecture, 
we completely ignored the impacts of real-time decoder latencies, that is, in both cases the classical reaction time associated with real-time decoding tasks was assumed to be zero.
(a) Physical resource requirements for executing the QCELS algorithm on a partially fault-tolerant STAR architecture. Logical qubit requirements are formatted as $n_1(n_2)$, where $n_1$ specifies system qubits and $n_2$ represents the total logical count within the core processor. The \# Rot. gates and \# Prep. patches columns display the number of rotation gates in the deepest circuit and the number of logical patches allocated for analog rotation resource state preparation, respectively. Circuit execution metrics include "Phys. time", which represents the total runtime including the PEC sampling overhead,  "PEC-free time", which represents the circuit execution time including algorithmic sampling but excluding the PEC sampling overhead, and 
"Max.\ single-shot time", which 
refers to the execution time for a single, deepest circuit in the algorithm suite. Additionally, we report 
the initial and final code distances during the rotation resource state production process: $d_i$ represents the code distance of the patches in which the resource states are initially prepared, and $d_f$ represents the target final code distance to which the prepared resource state patches are subsequently grown. Resource estimates 
are categorized by "empirical" and "theoretical" selections for the QCELS algorithm's $\delta$ parameter, as detailed in Appendix~\ref{app: QCELS scaling}. 
The reported logical qubit counts include bus patches of the core processor. 
(b) Resource estimates generated using the  TopQAD toolkit~\cite{1qbit2024topqad} for QPE implementation on a fully fault-tolerant architecture. Here, the QPE circuits are based on the double-factorized qubitization algorithm (DF Qubitization) of von Burg et al.~\cite{vonBurg2021quantum}. Logical qubit counts refer solely to the number of qubits in the logical circuit, that is, they exclude architecture considerations.
In addition to physical qubit counts and wall-clock execution times, we also report the QEC code distances that are required for running the corresponding circuits fault-tolerantly. For example,  \mbox{[11, 29 | 31]} means that the code distances $d=11$ and $29$ are required for the first and second magic state distillation levels, respectively, while the QEC code distance $d=31$ is needed to encode the logical qubits of the core processor. These choices are determined by the architecture's assembler~\cite{1qbit2024topqad} based on optimizations of the various trade-offs between the space and time costs proposed in Ref.~\cite{silva2024optimizing}. 
} 
\label{tab: p-benzyne qre}
\end{table*}

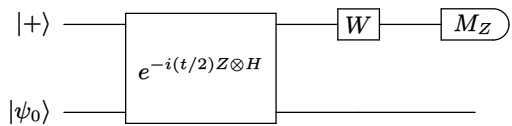
\begin{figure}[t]
    \centering
\centerline{
\Qcircuit @C=2.5em @R=2.5em {
        \lstick{\ket{+}} & \multigate{1}{e^{-i(t/2){Z} \otimes H}} & \gate{W} & \measureD{M_Z} \\
        \lstick{\ket{\psi_0}} & \ghost{e^{-i(t/2)Z \otimes H}} & \qw & \qw
    }}
    \caption{Single-shot quantum circuit used for computing $\langle e^{-iHt}\rangle$ in implementing the QCELS algorithm for GSEE for $p$-benzyne. $W = H$ for estimating $\text{Re}(\langle e^{-iHt}\rangle)$, and $W = SH$ for estimating $\text{Im}(\langle e^{-iHt}\rangle)$.}
    \label{fig:short_hadamard}
\end{figure}

\begin{figure}[t]
    \centering
\includegraphics[width=0.98\linewidth]{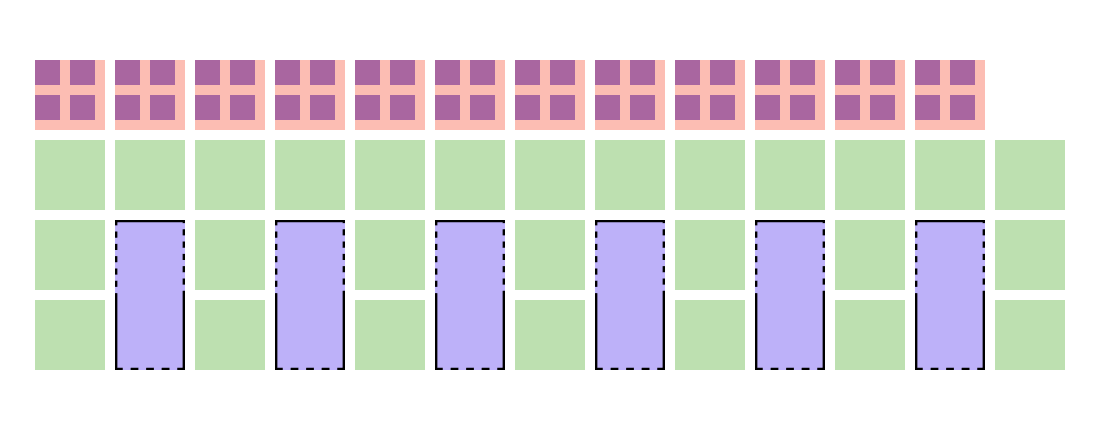}
    \caption{Logical microarchitecture layout to implement the Hadamard test quantum circuit used in QCELS algorithm with empirically inferred $\delta$ parameter $\delta$ in GSEE for $p$-benzyne using an active space with $N_\text{orb} = 6$ orbitals.  The pink-coloured patches represent the rotation resource states along with the core processor's data qubits in violet colour and bus qubits in green colour. 
    }
    \label{fig:layout_partial-FTQC_p_benzyne}
\end{figure}

In this section, we analyze resource requirements for the GSEE problem for $p$-benzyne using various active space sizes.  Different active spaces correspond to different models for the $p$-benzyne molecule, with varying overheads and approximation errors.  The basis set and active spaces for our simulations are chosen to be the same as in Ref.~\cite{mohseni2025buildquantumsupercomputer}: the chemical basis set $\mathtt{6-31G}$ is used to represent the spin orbitals, while the various molecular active spaces for the \mbox{$p$-benzyne} molecule are specified by HL$\pm 2, 6, 8, 12$ (using HL$\pm n$ to denote ``HOMO$-n$ and LUMO$+n$'', where 
``HOMO'' refers to the highest occupied molecular orbital, and ``LUMO'' stands for the lowest unoccupied molecular orbital). The circuit we use for estimating $\langle e^{-iHt}\rangle$ is given by \Cref{fig:short_hadamard}, which has the advantage of halving the circuit depth compared to the standard Hadamard test circuit. We assume that the circuit is implemented on the core processor with logical qubit layout presented in \Cref{fig:layout_partial-FTQC_p_benzyne}. 

Due to the non-locality of the molecular Hamiltonian, the algorithm interacts many different pairs of logical qubits and in order to simplify scheduling of bus qubit use we execute one rotation gate at a time.  Additionally, this layout provides access to all measurement bases for each encoded qubit, so the algorithm only requires a single clock cycle to utilize an analog rotation resource state for gate teleportation. Thus, the total number of clock cycles needed for one Trotter step is the number of rotation gates multiplied by the average number of needed RUS steps (e.g., 2).

Now, we estimate the physical resources needed to compute the ground state (GS) energy of $p$-benzyne with QCELS on a device that implements the STAR architecture. The active-space sizes are chosen to be the same as in Table IV of Ref.~\cite{mohseni2025buildquantumsupercomputer} and we perform the estimation with the improved Trotter error as described in  Appendix~\ref{app:improved trotter}.

In this resource estimation, we picked two QCELS hyperparameter sets, one with the empirical value of $\delta = 0.06$ and the other being the theoretically permitted choice of $\delta = 0.001$ for perfect overlap of the input state with the true ground state. Discussion of the choice of $\delta$ is summarized in Appendix~\ref{app: delta choice}.

The choice of $\delta$ leads to different maximal circuit depths, and correspondingly different final code distances $d_f$. We determine the code distance $d_f$ by solving ~\Cref{eq:df_det}. As we can see in \Cref{tab: p-benzyne qre}, the final code distances mostly exceed 15, leading to a very low probability of success in the creation of resource states as described in \Cref{fig:success-rate} and an an unfeasible amount of overhead for the circuit. To bypass this problem, we use the 2-step growth protocol developed in \Cref{sec:growth_protocol} to improve the success rate. While smaller initial code distances can lead to significantly smaller resource state factory, it comes at the price of a worse PEC overhead due to a higher growth infidelity. Thus, we determine the initial code distance by scanning over $d_i = 9, \;11, \; 13$ and selecting the one that results in the minimal execution time.

As shown in \Cref{tab: p-benzyne qre}, the PEC overhead becomes significant starting from $N_{\text{orb}} \geq 18$, and we would like to determine the maximal active space for which the overhead remains feasible. The PEC overhead of the deepest circuit in QCELS can be approximated as 
\begin{eqnarray}
    \gamma_{\text{max}}^2 \approx \exp\left(4 \alpha _{\text{\tiny RUS}}pT_{\text{max}}\lVert H\rVert_1/2 + 4\times 2 N_{\text{max}}\mathcal{I}_{\text{growth}}\right).
\end{eqnarray}
Here, $p$ could refer to $p_{\text{phys}}$ as in Ref.~\cite{toshio2024practical, Akahoshi:2024yme}, but we can instead choose to have it refer to the fitting coefficient in the legend of \Cref{fig:rot_gate_error} (a). We find a point where
\begin{equation}
    \alpha_{\text{\tiny{RUS}}} p T_{\text{max}}\lVert H\rVert_1/2 + 8 N_{\text{max}}\mathcal{I}_{\text{growth}} \leq 1.
\end{equation}
We temporarily ignore the $\mathcal{I}_{\text{growth}}$ for simplicity and obtain the loose upper bound of:
\begin{equation}
    \lVert H\rVert_1 \leq \frac{2}{\alpha_{\text{\tiny{RUS}}}pT_{\text{max}}},
    \label{eq: h_norm_bound}
\end{equation}
where all the parameters can be simply determined by the error budget equation and \Cref{fig:rot_gate_error} (a). With the error budget equation solution with $\delta=0.06$ giving $T_{\text{max}} \approx 56.25$, we obtain the bounds:
\begin{equation}
    \left. \lVert H\rVert_1 \right|_{\text{target}} \leq 24.7, \;\left. \lVert H\rVert_1 \right|_{\text{desired}} \leq 32.6.
\end{equation}
With $\lVert H\rVert_1 \approx 6.6$ for $N_{\text{orb}} = 6$ and $\lVert H\rVert_1 \approx 45.1$ for \mbox{$N_{\text{orb}} = 14$,} we can understand why the physical time is several orders of magnitude larger than the PEC-free runtime.

One assumption of these resource estimates is that a new resource state is produced every clock cycle. These resource states are created on an area of the architecture called the resource state factory, as shown in \Cref{fig:layout_partial-FTQC_p_benzyne}. The factory consists of a number of patches of size $d_f$, each of which is divided into $\lfloor d_f/d_i \rfloor^2$ patches of size $d_i$. On each small patch, an analog rotation state preparation can be attempted every four stabilization rounds. As soon as a shot succeeds on a small patch, the logical state is grown into its associated large patch in $d_i + 2 + d_f$ rounds. The resource state is then available to be consumed at the start of the next logical cycle, but may be stored in memory if an excess are produced in a given time frame. The size of these factories, as listed in \Cref{tab: p-benzyne qre}, is determined by simulating a factory for $10^5$ steps, incorporating the non-deterministic preparation process as well as the time used to grow, store, and consume the resource states. In these simulations, we use the success probability associated with $\theta_* = 5 \times 10^{-3}$, which is the largest angle we allow in the execution of the algorithm.

\subsection{Ground-state energy estimation for the Fermi--Hubbard model} \label{sec: 2D FH}

\begin{table*}[!htbp]
\centering
{\scriptsize
\centering
\scriptsize 
\setlength{\tabcolsep}{1pt} 

\begin{tabular*}{\textwidth}{|c|c|c|ccc|@{\extracolsep{\fill}}ccccccc|ccccccc|}
        \hline
        \hline
        \multicolumn{19}{c}{}\\
        \multicolumn{19}{l}{\normalsize (a) Partially Fault-Tolerant STAR Architecture}\\
        \multicolumn{19}{c}{}\\
        \multicolumn{3}{c}{} & \multicolumn{3}{c}{\bf Logical Resources} & \multicolumn{7}{c}{\bf Target Hardware} & \multicolumn{7}{c}{\bf Desired Hardware} \\
        \cmidrule{4-6} \cmidrule{7-13} \cmidrule{14-20}
        \multicolumn{1}{c}{\multirow{3}{*}{\shortstack[c]{$U/\tau $}}} & 
        \multicolumn{1}{c}{\multirow{3}{*}{\shortstack[c]{$L$}}} & 
        \multicolumn{1}{c}{\multirow{3}{*}{\shortstack[c]{$\epsilon$}}} & 
        \multicolumn{1}{c}{\multirow{3}{*}{\shortstack[c]{\# Log.\\ qubits}}} &
        \multicolumn{1}{c}{\multirow{3}{*}{\shortstack[c]{Max.\\ Trotter}}} & 
        \multicolumn{1}{c}{\multirow{3}{*}{\shortstack[c]{\# Rot.\\gates}}} & 
        \multicolumn{1}{c}{\multirow{3}{*}{\shortstack[c]{\# Phys.\\ qubits}}} & 
        \multicolumn{1}{c}{\multirow{3}{*}{\shortstack[c]{\# Phys.\\ time}}} & 
        \multicolumn{1}{c}{\multirow{3}{*}{\shortstack[c]{PEC \\free\\ time}}} & 
        \multicolumn{1}{c}{\multirow{3}{*}{\shortstack[c]{1 Trott. \\ clocks}}} & 
        \multicolumn{1}{c}{\multirow{3}{*}{\shortstack[c]{Max.\\ single- \\ shot time}}} & 
        \multicolumn{1}{c}{\multirow{3}{*}{\shortstack[c]{$d_i$}}} & 
        \multicolumn{1}{c}{\multirow{3}{*}{\shortstack[c]{$d_f$}}} & 
        \multicolumn{1}{c}{\multirow{3}{*}{\shortstack[c]{\# Phys.\\ qubits}}} & 
        \multicolumn{1}{c}{\multirow{3}{*}{\shortstack[c]{\# Phys.\\ time}}} & 
        \multicolumn{1}{c}{\multirow{3}{*}{\shortstack[c]{PEC \\free\\ time}}} & 
        \multicolumn{1}{c}{\multirow{3}{*}{\shortstack[c]{1 Trotter \\ clocks}}} & 
        \multicolumn{1}{c}{\multirow{3}{*}{\shortstack[c]{Max.\\ single- \\ shot time}}} & 
        \multicolumn{1}{c}{\multirow{3}{*}{\shortstack[c]{$d_i$}}} & 
        \multicolumn{1}{c}{\multirow{3}{*}{\shortstack[c]{$d_f$}}} \\
        \multicolumn{6}{c}{}&&&&&\\
        \multicolumn{6}{c}{}&&&&&\\
        \hline
        \hline
        \multirow{4}{*}{4} 
        & 4 & 0.08 & 65 & 150 & 3.4$\times$10\textsuperscript{4} & 
        4.7$\times$10\textsuperscript{4} & 21.3 min & 12.1 min & 314 & 561 ms & 9 & 19 & 
        2.9$\times$10\textsuperscript{4} & 12.4 min & 11.5 min & 338 & 532 ms & 9 & 15 \\
        & 6 & 0.18 & 145 & 66 & 3.6$\times$10\textsuperscript{4} & 
        1.0$\times$10\textsuperscript{5} & 12.4 min & 6.9 min & 382 & 300 ms & 9 & 19 & 
        6.5$\times$10\textsuperscript{4} & 7.0 min & 6.5 min & 416 & 288 ms & 9 & 15 \\
        & 8 & 0.32 & 257 & 37 & 3.8$\times$10\textsuperscript{4} & 
        2.3$\times$10\textsuperscript{5} & 8.6 min & 5.0 min & 442 & 205 ms & 9 & 21 & 
        1.2$\times$10\textsuperscript{5} & 4.8 min & 4.4 min & 481 & 186 ms & 9 & 15 \\
        & 10 & 0.50 & 401 & 24  & 3.9$\times$10\textsuperscript{4} & 
        3.5$\times$10\textsuperscript{5} & 6.7 min & 3.8 min & 492 & 148 ms & 9 & 21 & 
        1.8$\times$10\textsuperscript{5} & 3.7 min & 3.4 min & 528 & 133 ms & 9 & 15 \\
        \hline
        \multirow{4}{*}{8} 
        & 4 & 0.08 & 65 & 225 & 5.0$\times$10\textsuperscript{4} & 
        5.7$\times$10\textsuperscript{4} & 46.1 min & 19.1 min & 309 & 875 ms & 9 & 21 & 
        2.9$\times$10\textsuperscript{4} & 18.6 min & 17.0 min & 333 & 785 ms & 9 & 15 \\
        & 6 & 0.18 & 145 & 100 & 5.5$\times$10\textsuperscript{4} & 
        1.3$\times$10\textsuperscript{5} & 24.3 min & 10.5 min & 378 & 476 ms & 9 & 21 & 
        6.5$\times$10\textsuperscript{4} & 10.4 min & 10.1 min & 412 & 433 ms & 9 & 15 \\
        & 8 & 0.32 & 257 & 56 & 5.7$\times$10\textsuperscript{4} & 
        2.3$\times$10\textsuperscript{5} & 17.0 min & 7.1 min & 440 & 310 ms & 9 & 21 & 
        1.5$\times$10\textsuperscript{5} & 7.6 min & 6.8 min & 750 & 297 ms & 9 & 17 \\
        & 10 & 0.50 & 401 & 36 & 5.9$\times$10\textsuperscript{4} & 
        3.5$\times$10\textsuperscript{5} & 12.8 min & 5.4 min & 490 & 222 ms & 9 & 21 & 
        2.3$\times$10\textsuperscript{5} & 5.7 min & 5.1 min & 524 & 211 ms & 9 & 17 \\
        \hline
        \hline
    \end{tabular*}

\begin{tabular*}{\textwidth}{|c|c|c|cc|@{\extracolsep{\fill}}ccc ccc@{\extracolsep{\fill}}ccc ccc}
    \multicolumn{17}{c}{}\\
    \multicolumn{17}{l}{\normalsize (b) Fault-Tolerant Architecture}\\
    \multicolumn{17}{c}{}\\
    \multicolumn{5}{c}{} & \multicolumn{6}{c}{\bf Time-Optimal} & \multicolumn{6}{c}{\bf Space-Optimal} \\
    \cmidrule{6-11}\cmidrule{12-17}
    \multicolumn{3}{c}{} &
    \multicolumn{2}{c}{\bf Logical Resources} &
    \multicolumn{3}{c}{\bf Target Hardware} &
    \multicolumn{3}{c}{\bf Desired Hardware} &
    \multicolumn{3}{c}{\bf Target Hardware} &
    \multicolumn{3}{c}{\bf Desired Hardware} \\
    \cmidrule{4-5}
    \cmidrule{6-8}\cmidrule{9-11}
    \cmidrule{12-14}\cmidrule{15-17}
    \multicolumn{1}{c}{\multirow{2}{*}{\shortstack[c]{$U/\tau$}}} &
    \multicolumn{1}{c}{\multirow{2}{*}{\shortstack[c]{$L$}}} &
    \multicolumn{1}{c}{\multirow{2}{*}{\shortstack[c]{$\epsilon$}}} &
    \multicolumn{1}{c}{\multirow{2}{*}{\shortstack[c]{\# Log.\\ qubits}}} &
    \multicolumn{1}{c}{\multirow{2}{*}{\shortstack[c]{\# $T$\\ gates}}} &
    \multirow{2}{*}{\shortstack[c]{\# Phys.\\ qubits}} &
    \multirow{2}{*}{\shortstack[c]{Phys.\\ time}} &
    \multirow{2}{*}{\shortstack[c]{QEC code\\ distances}} &
    \multirow{2}{*}{\shortstack[c]{\# Phys.\\ qubits}} &
    \multirow{2}{*}{\shortstack[c]{Phys.\\ time}} &
    \multirow{2}{*}{\shortstack[c]{QEC code\\ distances}} &
    \multirow{2}{*}{\shortstack[c]{\# Phys.\\ qubits}} &
    \multirow{2}{*}{\shortstack[c]{Phys.\\ time}} &
    \multirow{2}{*}{\shortstack[c]{QEC code\\ distances}} &
    \multirow{2}{*}{\shortstack[c]{\# Phys.\\ qubits}} &
    \multirow{2}{*}{\shortstack[c]{Phys.\\ time}} &
    \multirow{2}{*}{\shortstack[c]{QEC code\\ distances}}\\
    \multicolumn{5}{c}{}&&&&&\\
    \hline\hline

    &&&&&&&&\\
    \multirow{4}{*}{4}
    & 8  & 0.32 & 162  & 2.4$\times$10\textsuperscript{6}
      & 8.9$\times$10\textsuperscript{5} & 21.2 sec & 21 | 25
      & 7.0$\times$10\textsuperscript{5} & 21.2 sec & 17 | 25
      & 5.7$\times$10\textsuperscript{5} & 1.9 min & 23 | 25
      & 5.1$\times$10\textsuperscript{5} & 2.9 min & 17 | 25 \\
    & 10 & 0.50 & 252  & 2.0$\times$10\textsuperscript{6}
      & 1.1$\times$10\textsuperscript{6} & 17.7 sec & 21 | 25
      & 9.4$\times$10\textsuperscript{5} & 17.7 sec & 17 | 25
      & 8.3$\times$10\textsuperscript{5} & 59.4 sec & 21 | 25
      & 7.4$\times$10\textsuperscript{5} & 2.4 min & 17 | 25 \\
    & 16 & 1.28 & 642  & 1.7$\times$10\textsuperscript{6}
      & 2.1$\times$10\textsuperscript{6} & 15.0 sec & 21 | 25
      & 1.9$\times$10\textsuperscript{6} & 15.0 sec & 17 | 25
      & 1.9$\times$10\textsuperscript{6} & 33.1 sec & 23 | 25
      & 1.8$\times$10\textsuperscript{6} & 2.0 min & 17 | 25 \\
    & 20 & 2.00 & 1002 & 1.6$\times$10\textsuperscript{6}
      & 3.1$\times$10\textsuperscript{6} & 14.4 sec & 21 | 25
      & 2.9$\times$10\textsuperscript{6} & 14.4 sec & 17 | 25
      & 3.1$\times$10\textsuperscript{6} & 14.4 sec & 21 | 25
      & 2.7$\times$10\textsuperscript{6} & 1.9 min & 17 | 25 \\
    &&&&&&&&\\
    \hline
    &&&&&&&&\\

    \multirow{4}{*}{8}
    & 8  & 0.32 & 162  & 3.4$\times$10\textsuperscript{6}
      & 8.9$\times$10\textsuperscript{5} & 29.9 sec & 21 | 25
      & 7.0$\times$10\textsuperscript{5} & 29.9 sec & 17 | 25
      & 6.0$\times$10\textsuperscript{5} & 5.0 min & 21 | 27
      & 5.1$\times$10\textsuperscript{5} & 4.1 min & 17 | 25 \\
    & 10 & 0.50 & 252  & 3.1$\times$10\textsuperscript{6}
      & 1.1$\times$10\textsuperscript{6} & 26.8 sec & 21 | 25
      & 9.4$\times$10\textsuperscript{5} & 26.8 sec & 17 | 25
      & 8.8$\times$10\textsuperscript{5} & 4.5 min & 21 | 27
      & 7.4$\times$10\textsuperscript{5} & 3.6 min & 17 | 25 \\
    & 16 & 1.28 & 642  & 2.8$\times$10\textsuperscript{6}
      & 2.3$\times$10\textsuperscript{6} & 24.4 sec & 23 | 25
      & 1.9$\times$10\textsuperscript{6} & 24.4 sec & 17 | 25
      & 2.1$\times$10\textsuperscript{6} & 4.5 min & 23 | 27
      & 1.8$\times$10\textsuperscript{6} & 3.3 min & 17 | 25 \\
    & 20 & 2.00 & 1002 & 2.7$\times$10\textsuperscript{6}
      & 3.5$\times$10\textsuperscript{6} & 25.7 sec & 21 | 27
      & 2.9$\times$10\textsuperscript{6} & 23.8 sec & 17 | 25
      & 3.2$\times$10\textsuperscript{6} & 2.0 min & 21 | 27
      & 2.7$\times$10\textsuperscript{6} & 3.2 min & 17 | 25 \\
    &&&&&&&&\\
    \hline\hline
\end{tabular*}
    
}
\caption{Resource requirements of (a) partially fault-tolerant and (b) fully fault-tolerant implementations of GSEE for a 2D Fermi--Hubbard (FH) model of lattice size $L$$\times$$L$ within a target accuracy 
$\epsilon = 0.005L^2$. The energy parameters of the FH model are specified by the ratio $U/\tau$, where $U$ represents the on-site Coulomb interaction, and $\tau$ represents the kinetic energy of the hopping term. 
In both cases, we report estimates for the physical wall-clock times and the number of physical qubits required for two  hardware specifications, namely, \lq\lq{}target\rq\rq{} and \lq\lq{}desired\rq\rq{} hardware, as summarized in \Cref{tab:physical_params}. 
Furthermore, for both the STAR architecture and the fully fault-tolerant architecture, 
we completely ignore the impacts of real-time decoder latencies, that is, in both cases the classical reaction time associated with real-time decoding tasks is set to zero.  (a) Logical and physical resource requirements for executing the QCELS algorithm on a partially fault-tolerant STAR architecture. The reported logical qubit counts include bus patches of the core processor.
Circuit execution metrics include "Phys. time", which represents the total runtime including the PEC sampling overhead,  "PEC-free time", which represents the circuit execution time including algorithmic sampling but excluding the PEC sampling overhead, and 
"Max.\ single-shot time", which 
refers to the execution time for a single, deepest circuit in the algorithm suite. \lq\lq{}1-Trotter clocks\rq\rq{} denotes the number of average clocks needed to perform a single Trotter step, estimated by simulating the parallel RUS protocol 100 times. \lq\lq{}Max.\ Trotter\rq\rq{} denotes the maximal number of Trotter steps taken to execute the deepest circuit of the QCELS algorithm once. Additionally, we report the initial and final code distances during the rotation resource state production process: $d_i$ represents the code distance of the patches in which the resource states are initially prepared, and $d_f$ represents the target final code distance to which the prepared resource state patches are subsequently grown.
(b) Resource estimates generated using TopQAD~\cite{1qbit2024topqad} for GSEE implementation on a fully fault-tolerant architecture. Our estimates pertain to the logical circuits of the quantum phase estimation scheme based on plaquette Trotterization with Hamming weight phasing from Ref.~\cite{campbell_early_2022}. 
This quantum phase estimation scheme assumes access to the true ground state of the Hamiltonian. 
Logical qubit counts refer solely to the number of qubits in the logical circuit, that is, they exclude architecture considerations. In addition to physical qubit counts and wall-clock execution times, we also report the QEC code distances that are required for running the corresponding circuits fault-tolerantly. For example,  \mbox{[17 | 25]} means that the code distance $d=17$ is  required for the magic state distillation, while the QEC code distance $d=25$ is needed to encode the logical qubits of the core processor. These choices are determined by the architecture's assembler~\cite{1qbit2024topqad} based on optimizations of the various trade-offs between the space and time costs proposed in Ref.~\cite{silva2024optimizing}.  It is worth emphasizing that the physical runtime for the GSEE does not decrease indefinitely. Rather, it saturates to a constant. This occurs due to the \(T\) count of the plaquette Trotterization saturating to a constant at large \(L\). However, the total spacetime volume of this simulation continues to grow as the number of logical qubits needed (equivalently the number of physical qubits discussed here) increase with the system size. See Ref.~\cite{campbell_early_2022} for more details. }
\label{tab: FH2D 0.005L2 resource estimation}
\end{table*}
\begin{figure}[tb]
    \centering
\includegraphics[width=1.0\linewidth]{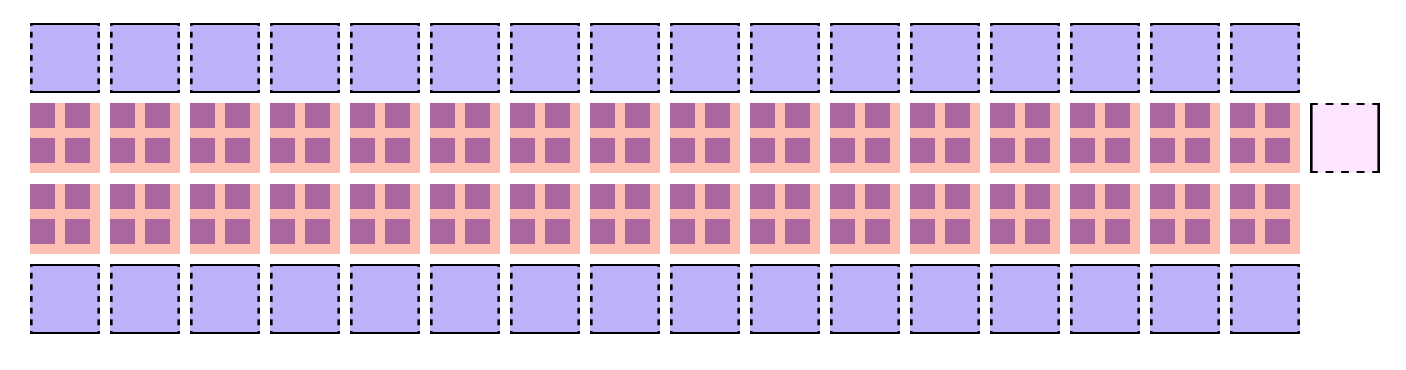}
    \caption{Logical microarchitecture layout used 
    to  execute the QCELS algorithm circuits for GSEE for the 2D Fermi--Hubbard model with a lattice dimension $L$$\times$$L$. 
    The pink-colored code patches in the interior zone form the factory for preparing and growing rotation resource states while also serving as bus qubits. The violet-colored patches in the exterior zone represent the computational data qubits.
    Here, the logical code patch arrangement is shown for $L=4$.}
    \label{fig:2d_fh_4row}
\end{figure}

\begin{table*}[!htbp]
\centering
\setlength{\tabcolsep}{1pt} 
{\scriptsize
\begin{tabular*}{\textwidth}{|c|c|c|ccc|@{\extracolsep{\fill}}ccccccc|ccccccc|}
    \hline
    \hline
    \multicolumn{20}{c}{}\\
    \multicolumn{20}{l}{\normalsize (a) Partially Fault-Tolerant STAR Architecture}\\
    \multicolumn{20}{c}{}\\
    \multicolumn{3}{c}{} & \multicolumn{3}{c}{\bf Logical Resources} & \multicolumn{7}{c}{\bf Target Hardware} & \multicolumn{7}{c}{\bf Desired Hardware} \\
    \cmidrule{4-6} \cmidrule{7-13} \cmidrule{14-20}
    \multicolumn{1}{c}{\multirow{2}{*}{\shortstack[c]{$U$}}} & 
    \multicolumn{1}{c}{\multirow{2}{*}{\shortstack[c]{$L$}}} & 
    \multicolumn{1}{c}{\multirow{2}{*}{\shortstack[c]{$\epsilon$}}} & 
    \multicolumn{1}{c}{\multirow{2}{*}{\shortstack[c]{\# Log.\\ qubits}}} &
    \multicolumn{1}{c}{\multirow{2}{*}{\shortstack[c]{Max.\\ Trotter}}} & 
    \multicolumn{1}{c}{\multirow{2}{*}{\shortstack[c]{\# Rot.\\gates}}} & 
    \multicolumn{1}{c}{\multirow{2}{*}{\shortstack[c]{\# Phys.\\ qubits}}} & 
    \multicolumn{1}{c}{\multirow{2}{*}{\shortstack[c]{Phys.\\ time}}} & 
    \multicolumn{1}{c}{\multirow{2}{*}{\shortstack[c]{PEC free\\ time}}} & 
    \multicolumn{1}{c}{\multirow{2}{*}{\shortstack[c]{1 Trotter \\ clocks}}} & 
    \multicolumn{1}{c}{\multirow{2}{*}{\shortstack[c]{Max. \\ time}}} & 
    \multicolumn{1}{c}{\multirow{2}{*}{\shortstack[c]{$d_i$}}} & 
    \multicolumn{1}{c}{\multirow{2}{*}{\shortstack[c]{$d_f$}}} & 
    \multicolumn{1}{c}{\multirow{2}{*}{\shortstack[c]{\# Phys.\\ qubits}}} & 
    \multicolumn{1}{c}{\multirow{2}{*}{\shortstack[c]{Phys.\\ time}}} & 
    \multicolumn{1}{c}{\multirow{2}{*}{\shortstack[c]{PEC free\\ time}}} & 
    \multicolumn{1}{c}{\multirow{2}{*}{\shortstack[c]{1 Trotter \\ clocks}}} & 
    \multicolumn{1}{c}{\multirow{2}{*}{\shortstack[c]{Max. \\ time}}} & 
    \multicolumn{1}{c}{\multirow{2}{*}{\shortstack[c]{$d_i$}}} & 
    \multicolumn{1}{c}{\multirow{2}{*}{\shortstack[c]{$d_f$}}} \\
    \multicolumn{6}{c}{}&&&&&&&&&&&\\
    \hline
    \hline
    \multirow{4}{*}{4} 
    & 4 & 0.01 & 65 & 3397 & 7.6$\times$10\textsuperscript{5} 
    & 6.9$\times$10\textsuperscript{4} & 5.2 days & 7.1 hrs  & 429.9 & 20.4 sec & 11 & 23 &
    3.8$\times$10\textsuperscript{4} & 12.5 hr & 4.4 hrs & 324.3 & 12.3 sec & 9 & 17 \\
    & 6 & 0.01 & 145 & 5051 & 2.8$\times$10\textsuperscript{6} &
    2.1$\times$10\textsuperscript{5} & 16.4 yr & 1.3 days & 1072.3 & 1.5 min & 13 & 27 & 
    1.0$\times$10\textsuperscript{5} & 4.1 day & 11.1 hrs & 499.3 & 31.8 sec & 11 & 19 \\
    & 8 & 0.01 & 257 & 6750 & 6.9$\times$10\textsuperscript{6} &
    3.7$\times$10\textsuperscript{5} & $10^7$ days & 1.7 days & 1120.0 & 2.0 min & 13 & 27 & 
    1.9$\times$10\textsuperscript{5} & 78.2 days & 16.3 hrs & 560.8 & 47.7 sec & 11 & 19 \\
    & 10 & 0.01 & 401 & 8538 & 1.4$\times$10\textsuperscript{7} 
    & 5.8$\times$10\textsuperscript{5} & $10^{13}$ days & 2.2 days & 1173.0 & 2.7 min & 13 & 27 & 
    3.5$\times$10\textsuperscript{5} & 16.2 yrs & 23.8 hrs & 607.6 & 1.2 min & 11 & 21 \\
    \hline
    \multirow{4}{*}{8} 
    & 4 & 0.01 & 65 & 5111 & 1.1$\times$10\textsuperscript{6} 
    & 6.9$\times$10\textsuperscript{4} & 29.6 days & 10.3 hrs & 417.9 & 29.9 sec & 11 & 23 & 
    4.7$\times$10\textsuperscript{4} & 1.1 days & 9.5 hrs & 424.5 & 27.3 sec & 11 & 19 \\
    & 6 & 0.01 & 145 & 7636 & 4.2$\times$10\textsuperscript{6} & 
    2.1$\times$10\textsuperscript{5} & 277.4 yrs & 1.8 days & 1015.6 & 2.1 min & 13 & 27 & 
    1.0$\times$10\textsuperscript{5} & 12.1 days & 16.5 hrs & 494.6 & 47.6 sec & 11 & 19 \\
    & 8 & 0.01 & 257 & 10162 & 1.0$\times$10\textsuperscript{7} & 
    3.7$\times$10\textsuperscript{5} & $10^{10}$ days & 2.5 days & 1071.8 & 2.9 min & 13 & 27 & 
    2.3$\times$10\textsuperscript{5} & 1.6 yrs & 1.1 days & 552.2 & 1.2 min & 11 & 21 \\
    & 10 & 0.01 & 401 & 12807 & 2.1$\times$10\textsuperscript{7} 
    & 5.8$\times$10\textsuperscript{5} & $10^{16}$ days & 3.2 days & 1106.4 & 3.8 min & 13 & 27 & 3.5$\times$10\textsuperscript{5} & 249.2 yrs & 1.5 days & 600.4 & 1.7 min & 11 & 21 \\
    \hline
    \hline
\end{tabular*}

\begin{tabular*}{\textwidth}{|c|c|c|cc|@{\extracolsep{\fill}}ccc ccc@{\extracolsep{\fill}}ccc ccc}
    \multicolumn{17}{c}{}\\
    \multicolumn{17}{l}{\normalsize (b) Fault-Tolerant Architecture}\\
    \multicolumn{17}{c}{}\\
    \multicolumn{5}{c}{} & \multicolumn{6}{c}{\bf Time-Optimal} & \multicolumn{6}{c}{\bf Space-Optimal} \\
    \cmidrule{6-11}\cmidrule{12-17}
    \multicolumn{3}{c}{} &
    \multicolumn{2}{c}{\bf Logical Resources} &
    \multicolumn{3}{c}{\bf Target Hardware} &
    \multicolumn{3}{c}{\bf Desired Hardware} &
    \multicolumn{3}{c}{\bf Target Hardware} &
    \multicolumn{3}{c}{\bf Desired Hardware} \\
    \cmidrule{4-5}
    \cmidrule{6-8}\cmidrule{9-11}
    \cmidrule{12-14}\cmidrule{15-17}
    \multicolumn{1}{c}{\multirow{2}{*}{\shortstack[c]{$U/\tau$}}} &
    \multicolumn{1}{c}{\multirow{2}{*}{\shortstack[c]{$L$}}} &
    \multicolumn{1}{c}{\multirow{2}{*}{\shortstack[c]{$\epsilon$}}} &
    \multicolumn{1}{c}{\multirow{2}{*}{\shortstack[c]{\# Log.\\ qubits}}} &
    \multicolumn{1}{c}{\multirow{2}{*}{\shortstack[c]{\# $T$\\ gates}}} &
    \multirow{2}{*}{\shortstack[c]{\# Phys.\\ qubits}} &
    \multirow{2}{*}{\shortstack[c]{Phys.\\ time}} &
    \multirow{2}{*}{\shortstack[c]{QEC code\\ distances}} &
    \multirow{2}{*}{\shortstack[c]{\# Phys.\\ qubits}} &
    \multirow{2}{*}{\shortstack[c]{Phys.\\ time}} &
    \multirow{2}{*}{\shortstack[c]{QEC code\\ distances}} &
    \multirow{2}{*}{\shortstack[c]{\# Phys.\\ qubits}} &
    \multirow{2}{*}{\shortstack[c]{Phys.\\ time}} &
    \multirow{2}{*}{\shortstack[c]{QEC code\\ distances}} &
    \multirow{2}{*}{\shortstack[c]{\# Phys.\\ qubits}} &
    \multirow{2}{*}{\shortstack[c]{Phys.\\ time}} &
    \multirow{2}{*}{\shortstack[c]{QEC code\\ distances}}\\
    \multicolumn{5}{c}{}&&&&&\\
    \hline\hline

    &&&&&&&&\\
    \multirow{4}{*}{4}
    & 8  & 0.01 & 148 & 3.0$\times$10\textsuperscript{7}
      & 1.0$\times$10\textsuperscript{6} & 4.7 min & 23 | 27
      & 7.9$\times$10\textsuperscript{5} & 4.7 min & 19 | 27
      & 6.5$\times$10\textsuperscript{5} & 48.5 min & 23 | 29
      & 5.5$\times$10\textsuperscript{5} & 40.0 min & 19 | 27 \\
    & 10 & 0.01 & 224 & 9.7$\times$10\textsuperscript{7}
      & 1.5$\times$10\textsuperscript{6} & 16.5 min & 25 | 29
      & 1.1$\times$10\textsuperscript{6} & 15.3 min & 21 | 27
      & 9.7$\times$10\textsuperscript{5} & 1.4 hrs & 25 | 29
      & 7.9$\times$10\textsuperscript{5} & 2.4 hrs & 21 | 27 \\
    & 16 & 0.01 & 536 & 6.2$\times$10\textsuperscript{8}
      & 3.5$\times$10\textsuperscript{6} & 1.9 hrs & 11, 27 | 31
      & 2.6$\times$10\textsuperscript{6} & 1.9 hrs & 21 | 31
      & 2.6$\times$10\textsuperscript{6} & 9.3 hrs & 11, 31 | 31
      & 2.3$\times$10\textsuperscript{6} & 15.1 hrs & 21 | 31 \\
    & 20 & 0.01 & 828 & 1.9$\times$10\textsuperscript{9}
      & 5.1$\times$10\textsuperscript{6} & 5.8 hrs & 11, 31 | 31
      & 3.8$\times$10\textsuperscript{6} & 5.8 hrs & 23 | 31
      & 4.0$\times$10\textsuperscript{6} & 1.8 days & 11, 29 | 33
      & 3.4$\times$10\textsuperscript{6} & 2.1 days & 23 | 31 \\
    &&&&&&&&\\
    \hline\hline
\end{tabular*}
}
\caption{Resource requirements of (a) partially fault-tolerant and (b) fully fault-tolerant implementations of GSEE for a 2D Fermi--Hubbard (FH) model of lattice size $L$$\times$$L$ within a target accuracy 
 $\epsilon = 0.01$. 
The energy parameters of the FH model are specified by the ratio $U/\tau$, where $U$ represents the on-site Coulomb interaction, and $\tau$ represents the kinetic energy of the hopping term. 
In both cases, we report estimates for the physical wall-clock times and the number of physical qubits required for two  hardware specifications, namely, \lq\lq{}target\rq\rq{} and \lq\lq{}desired\rq\rq{} hardware, as summarized in \Cref{tab:physical_params}. 
Furthermore, for both the STAR architecture and the fully fault-tolerant architecture, 
we completely ignore the impacts of real-time decoder latencies. (a) Logical and physical resource requirements for executing the QCELS algorithm on a partially fault-tolerant STAR architecture. We used $\delta = 0.06$ as in Ref.~\cite{Akahoshi:2024yme}.
The reported logical qubit counts include bus patches of the core processor.
 We used $\delta = 0.06$ as in the choice in Ref.~\cite{Akahoshi:2024yme}. 
 (b) Resource estimates generated using TopQAD~\cite{1qbit2024topqad} for GSEE implementation on a fully fault-tolerant architecture. Here, the resource estimates pertain to the logical circuits of the quantum phase estimation scheme based on 
qubitization techniques of Ref.~\cite{babbush2018encoding}. Here, logical qubit counts refer solely to the number of qubits in the logical circuit, that is, they exclude architecture considerations. In addition to physical qubit counts and wall-clock execution times, we also report the QEC code distances that are required for running the corresponding circuits fault tolerantly. For example,  \mbox{[11, 27 | 31]} 
means that the code distances $d=11$ and $27$ are required for the first and second magic state distillation levels, respectively, while the QEC code distance $d=31$ is needed to encode the logical qubits of the core processor.}
\label{tab: FH2D 0.01 resource estimation}
\end{table*}

\begin{table*}[tbp]
\centering
\scriptsize
\setlength{\tabcolsep}{2pt} 

\end{table*}

\begin{figure*}[t]
\centering
{\small
\centerline{
\Qcircuit @C=0.5em @R=1.5em {
    \lstick{\ket{+}} & \qw & 
    \ctrl{1} & \qw & \ctrl{1} & \qw & 
    \ctrl{1} & \qw & \ctrl{1} & \qw & 
    \ctrl{1} & \qw & \ctrl{1} & \qw & 
    \ctrl{1} & \qw & \ctrl{1} & \qw & 
    \gate{W}\qw &  \meter & \cw & \\
    \lstick{\ket{\psi}} & \qw & 
    \gate{K_i} & \gate{U(H_i, t/4)} & \gate{K_i}& \qw & 
    \gate{K_h} & \gate{U(H_h, t/4)} & \gate{K_h} & \qw &
    \gate{K_h} & \gate{U^{\dagger}(H_h, -t/4)} & \gate{K_h} & \qw &
    \gate{K_i} & \gate{U^{\dagger}(H_i, -t/4)} & \gate{K_i}& \qw & 
}
}
}
\caption{Circuit for estimating $\langle e^{-iHt}\rangle$ for the 2D Fermi--Hubbard Hamiltonian proposed in Ref.~\cite{Akahoshi:2024yme}. Here, $H = H_h + H_i$, with the $H_h$ representing the hopping term,  and the $H_i$ representing the interaction term. The controlled $K$ gates are multi-target X gates or multi-target Z gates  satisfying $K_l H_l =- H_l K_l$ and $U(H, t)\equiv \exp(-iHt)$. The fact that only the $K_l$ gates are controlled is what allows the rotations in the $U(H_l, t)$ to be executed in parallel on the architecture in Fig.\ref{fig:2d_fh_4row}. $W = H$ for estimating $\text{Re}(\langle e^{-iHt}\rangle)$, and $W = SH$ for estimating $\text{Im}(\langle e^{-iHt}\rangle)$.}
\label{fig: FH2D hadamard}
\end{figure*}
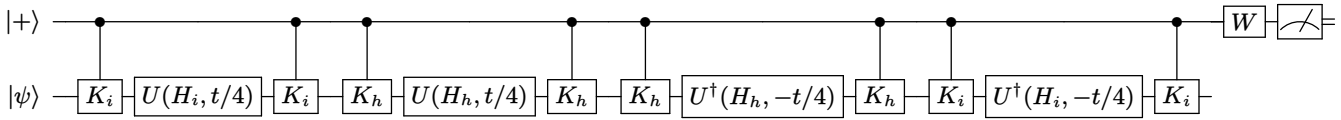

\begin{table}[t!]
    \centering
    \begin{tabular}{|c|c|c|c|c|c|}
    \hline
        &Sites $(L \times L)$ & $4 \times 4$ & $6 \times 6$ & $8 \times 8$ & $10 \times 10$ \\
        \hline\hline
        \multirow{3}{*}{$U = 4$} & $w$ & $1.9 \times 10^3$ & $4.2 \times 10^3$ & $7.5 \times 10^3$ & $1.2 \times 10^4$\\
        \cline{2-6}
        &
        $\lambda$ & 64 & 156 & 288 & 460\\
        \cline{2-6}
        &
        $N_{\text{terms}}$ & 112 & 276 & 512 & 820\\
    \hline
    \hline
        \multirow{3}{*}{$U = 8$} & $w$ & $4.3 \times 10^3$ & $9.6 \times 10^3$ & $1.7 \times 10^4$ & $2.7 \times 10^4$\\
        \cline{2-6}
        &
        $\lambda$ & 80 & 192 & 352 & 560\\
        \cline{2-6}
        &
        $N_{\text{terms}}$ & 112 & 276 & 512 & 820\\
        \hline
    \end{tabular}
    \caption{Trotter error coefficient, Hamiltonian 1-norm, and number of terms in the Hamiltonian. The Trotter error coefficients are computed in Table 6 of Ref.~\cite{kivlichan2020improved}.}
    \label{tab: FH param}
\end{table}

\begin{figure*}[th]
    \centering
    \includegraphics[width=0.73\linewidth]{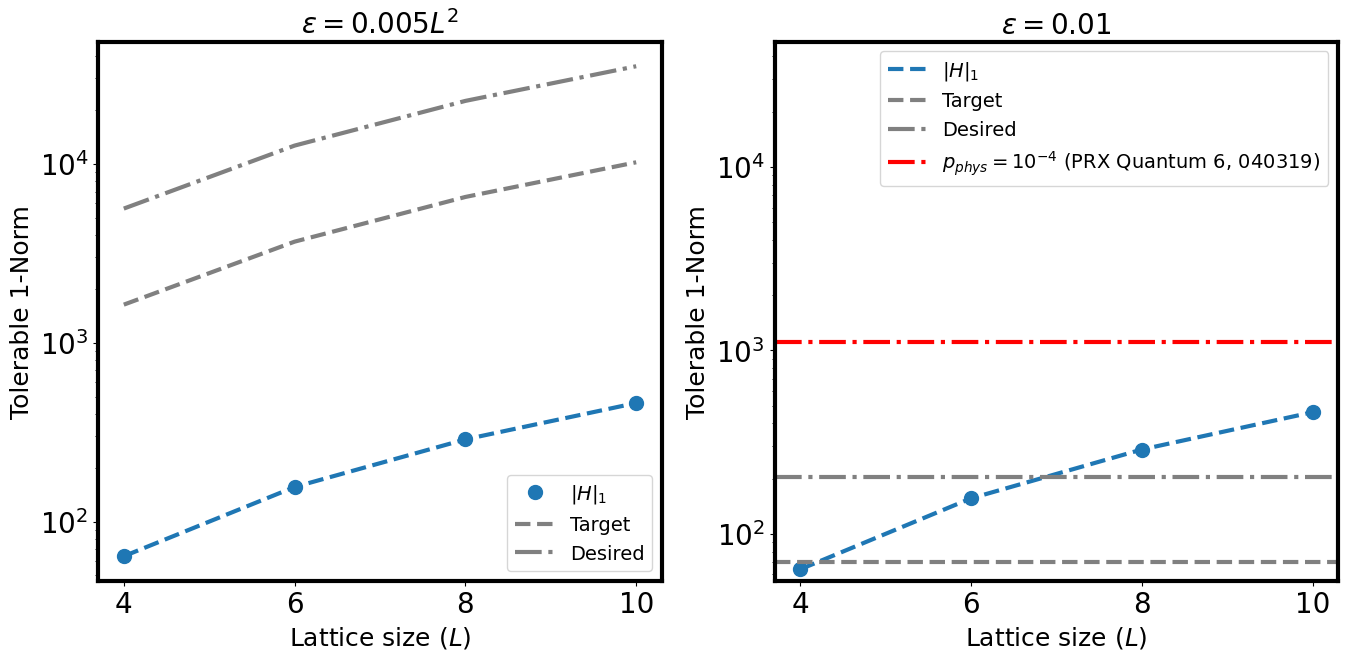}
    \caption{Maximum tolerable Hamiltonian 1-norm estimated using \Cref{eq: h_norm_bound} for target and desired hardware specifications, for two different accuracies in implementing GSEE for the Fermi--Hubbard model, along with actual 1-norm values from~\Cref{tab: FH param}.}
    \label{fig:2DFH_tol_hnorm}
\end{figure*}

Next, we discuss our QRE study for the 2D Fermi--Hubbard model, a local model with a linear relationship between the number of lattice sites and rotation gates, to investigate the potential for quantum advantage in large-scale systems. Previous studies~\cite{toshio2024practical, Akahoshi:2024yme} have demonstrated that executing the QCELS algorithm \cite{ding2023even} on the STAR architecture can yield substantial advantage over classical tensor network methods at a target accuracy of $\epsilon_{\text{target}} = 10^{-2}$ when properly leveraging the locality of the Hamiltonian. 

The major advancement made in Ref.~\cite{Akahoshi:2024yme} was to perform RUS trials and resource state preparation in parallel on the core processor. Their "adaptive injection region updating" technique dynamically reassigns ancilla patches that were measured to give the proper rotation angle to those data patches that have not completed the RUS process. As none of the ancilla qubits are idling during the procedure, it gives much better clock count than na\"ive compilation strategies. This is only possible due to the fact that the Hamiltonian is local and contains a large number of same-angle rotations. As the prepare and measurement are happening in the same zone of the processor, it eliminates the need for additional state preparation factory, leaving the total physical qubit count to simulate the Hamiltonian of an $L\times L$ square grid to be $(4L^2+1)\times 2d_f^2$. The logical microarchitecture layout for this patch arrangement is illustrated in \Cref{fig:2d_fh_4row}. 

Interestingly, this long, narrow microarchitecture layout is also favorable from an experimental perspective. We  briefly outline the reasons here. 
For superconducting hardware, one of the most difficult engineering challenges when scaling to a large number of qubits is connecting the qubits to their analog control signals and readout. One promising approach to scalable wiring is wafer-scale integrated superconducting circuitry, where scalability is achieved through tiling. A promising scheme for a modular design to scale beyond a single wafer was proposed in Ref.~\cite{mohseni2025buildquantumsupercomputer}, in which individual modules, each supporting 20k physical qubits, are capacitively coupled along their edges and arranged in a linear array. This linear design configuration arises from the constraint that one side of each module must remain accessible for control wiring extending from the wafer. Because the 20k-qubit tiles are capacitively coupled at their edges, they can be horizontally stacked, allowing the linear array to be extended to arbitrary length.

To perform QRE for the Fermi--Hubbard model, we again simulate lattice surgery procedures, although in a different setting. As we need to use patch growth to maintain sufficiently high supply rate of resource states, we introduce two major changes. First, the protocol for preparing a resource state through two-step growth takes in total \mbox{$d_i+2+d_f+4$} code cycles. Thus, we have that the depth of a single clock cycle is \mbox{$\mathcal{C} = d_i+2 + d_f+4$} for the condition stated in~\Cref{eq:df_det}.  This is in contrast to the direct-growth protocol, where the resource-state patch can be prepared within a single cycle of $d_f$ rounds consisting of 4 rounds for the analog rotation and $d_f - 4$ rounds for the growth protocol. 
Second, 
while the state preparation is only temporally parallelized by performing $\lfloor d/4 \rfloor$ injections per clock cycle in Ref.~\cite{Akahoshi:2024yme}, in our approach we also exploit spatial parallelization, using $\lfloor d_f/d_i \rfloor^2$ injections in space. With these two changes, we can simulate the lattice surgery using the supplementary code provided in Refs.~\cite{Akahoshi:2024yme, STAR_compilation}. These simulations yield an estimate for  $N_{\text{clock}}$ that depends on all parameters $d_i$, $d_f$, and $\theta_*$, all of which also change the success rate.  We find $d_f$ through  iterative search until the condition \Cref{eq:df_det} is satisfied.

We have conducted resource estimations for two different accuracies, namely, $\epsilon=0.005 L^2$ and  $\epsilon=0.01$. We used $\delta = 0.06$ in these resource estimations as in Ref.~\cite{Akahoshi:2024yme}. For the lower accuracy of $\epsilon = 0.005L^2$, our QRE results are summarized in~\Cref{tab: FH2D 0.005L2 resource estimation}. We observe that  the execution times are merely minutes for both target and desired hardware specifications, which shows that the STAR architecture can indeed be used for physically relevant problems. However, in contrast to Ref.~\cite{akahoshi2024partially}, which showed that the $\epsilon=0.01$ computation can be achieved in a day, our estimations are orders of magnitude larger than theirs due to large PEC overhead. We can estimate the maximal lattice size for which computations are feasible with our qubit specifications in the same manner as in the last paragraph of section~\ref{sec: p-benzyne estimate}. With \Cref{eq: h_norm_bound}, we plot the Hamiltonian 1-norm upper bound tolerable by different device specification in \Cref{fig:2DFH_tol_hnorm}. We can see from the plot that the upper bound for $\epsilon=0.01$ is one to two orders of magnitude higher than the actual 1-norm. However, for $\epsilon = 0.01$, the target device can execute up to $L = 4$ and the desired device can execute up to $L=6$ while that of Ref.~\cite{Akahoshi:2024yme} can cover all sizes from $L=4$ to $L=10$.

\subsection{Factory size requirements, and viability assessment}
\label{sec:factory_size_requirements}

We have also studied the physical cost of generating resource states in an application-agnostic manner by examining the factory footprint required to execute circuits of varying sizes. In particular, it is instructive to compare the factory-size requirements of the partially fault-tolerant STAR architecture with those of fully fault-tolerant implementations on EFTQC platforms, especially also in view of recent developments in designing more-efficient $T$ state preparation methods, such as magic state cultivation~\cite{gidney2024magicstatecultivationgrowing}. 
In fully fault-tolerant schemes, magic state distillation can be used to prepare states with arbitrarily low error rates; however, even a single distillation unit involving code patches of sufficiently large distance requires a substantial footprint. Magic state cultivation, on the other hand, can prepare magic $T$ states relatively cheaply, but only up to a limited error rate; see Ref.~\cite{gidney2024magicstatecultivationgrowing} for more details. For sufficiently large circuits, cultivation alone is therefore insufficient to produce resource states with adequately low error rates.

The unique characteristics of the STAR architecture can be inferred from the simulation results presented in \Cref{fig:rot_gate_error}. In \Cref{fig:rot_gate_error}(a), we observe that the overall gate error increases with the preparation distance, consistent with simulations reported in prior work. As the circuit size grows, the code distance must be increased to protect the Clifford operations in the circuit. However, preparing and injecting resource states at larger code distances leads to a higher overall per-gate error, thereby limiting the size of circuits that can be reliably executed. To some extent, this error can be mitigated using PEC; however, once the cumulative error across all gates in a circuit reaches a value significantly above the allowable taget value, the exponential overhead associated with PEC becomes prohibitive.

We have shown that, by leveraging patch growth, the factory footprint required to execute a circuit can be significantly reduced. In this approach, resource states are first prepared at a lower initial code distance and subsequently grown to a larger final distance. In \Cref{fig:rot_gate_error}(b), the associated error-rate curves scale with the initial distance. Because patch growth introduces only a small error that is independent of the rotation angle, it can reduce the overall error in regimes where the gate error dominates over the growth error. In such cases, patch growth enables both a reduction in the overall error rate and a smaller factory footprint. Furthermore, \Cref{fig:rot_gate_error}(b) shows that for any given rotation angle there exists an optimal initial distance that minimizes the resulting gate error rate.

\begin{figure}[th]
    \centering
\includegraphics[width=0.97\linewidth]{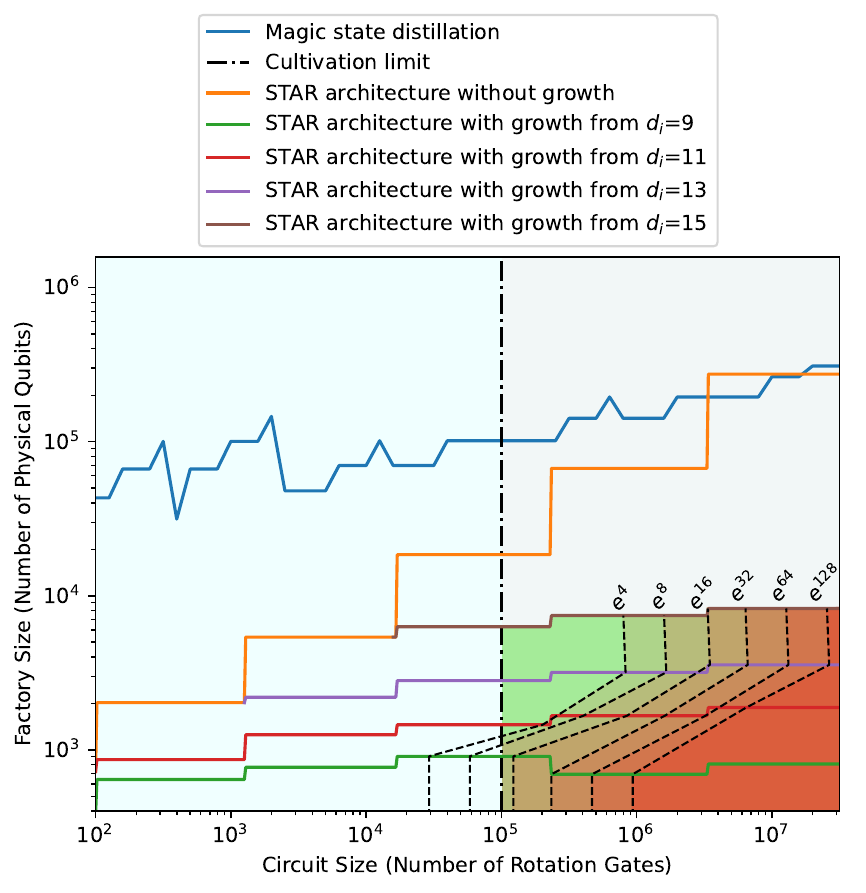}
    \caption{
    Scaling of factory size with increasing circuit size, for a STAR architecture based on rotation resource state production with and without code growth, in comparison with fully fault-tolerant schemes based on magic state distillation or cultivation. Factory sizes are quantified by the number of physical qubits required. Circuit sizes are characterized by the number of small-angle rotation gates to be executed, with angles of rotation $\theta_*=10^{-4}$ or smaller, while we assume 100 logical qubits for the width of the logical circuit. 
    The reported estimates are based on a target logical error rate $\epsilon_{\text{log err}}=0.01$ and the \lq\lq{}target\rq\rq{} hardware specifications 
    defined in \Cref{tab:physical_params}.
    The solid blue line represents the factory-size requirements for magic state distillation inferred using  TopQAD~\cite{1qbit2024topqad}.
    The solid orange line represents the factory-size requirements for the STAR architecture without code growth. The green, red, violet and brown solid lines  represent the factory-size requirements for a STAR architecture leveraging code growth from a specified initial code distance $d_i$ (as explained in the legend). The thick vertical dashed-dotted line at a circuit size $\approx $$10^5$ represents the limitation of the magic state cultivation approach. 
    The multiple dashed lines connecting the lines pertaining to code growth in the bottom-right corner represent the "contour lines" connecting discrete data points 
    of constant PEC overheads for values $e^4, e^8, \dots, e^{128}$ (as indicated).
    The light-blue-shaded  area represents the regime in which magic state cultivation is superior to other approaches. The heat map to the right indicates the increasing magnitude of PEC overheads, with the green color representing a low PEC overhead and red color representing a formidable PEC overhead. 
    The gray-shaded region represents factory-size scales too large to be advantageous for STAR architectures leveraging code growth.
    We observe that STAR architectures are particularly effective for executing logical quantum circuits in the slightly sub-megaquop regime corresponding to roughly $10^5$--$10^6$ small-angle rotation gates, but rapidly become impractical beyond this scale. 
    }
\label{fig:PFTQC-Viability-Regime}
\end{figure}

In \Cref{fig:PFTQC-Viability-Regime}, we plot the total physical qubit cost of a factory that produces resource states. For protocols such as magic state distillation and cultivation, arbitrary-angle rotations are synthesized using $T$ gates. The cost of magic state distillation is estimated by considering the time-optimal magic state factory for a given circuit size, where arbitrary-angle rotations are first decomposed into the Clifford+$T$ gate set, resulting in an increased number of $T$ gates. 
To achieve an overall target logical error rate $\epsilon_{\text{log err}}=0.01$, we allocate half of the error budget to circuit synthesis, and infer the number of $T$ gates needed to implement the circuit fault tolerantly.
For example, for  megaquop circuits  or larger, the number of $T$ gates required to synthesize an arbitrary-angle  rotation to a sufficiently low infidelity is on the order of approximately 100. 
We then optimize the magic state factory using tools in TopQAD~\cite{1qbit2024topqad}, finding the ideal number of magic state distillation levels and the code distances used in them. For a time-optimal configuration, we determine the number of distillation units required to match the production and consumption rates for magic states to ensure that the computation proceeds without delay.

Cultivation (not shown) is significantly cheaper than both the STAR scheme and the full FTQC  approach based on magic state distillation. Consequently, for circuits small enough to be executed using cultivation alone, cultivation provides the lowest-cost method for preparing resource states. 
As shown in Ref.~\cite{gidney2024magicstatecultivationgrowing}, magic state cultivation enables the preparation of magic $T$ states with logical error rates as low as approximately $10^{-9}$. Since synthesizing an arbitrary-angle rotation typically requires about 100 $T$ gates, achieving a target logical error rate of $\epsilon_{\text{log err}} = 0.01$ is feasible for circuits containing up to roughly $10^5$ rotation gates, but is not viable at larger scales. 

For a STAR architecture without patch growth, if the success probability $p$ of preparing a rotation gate is much less than 1\%, the overhead associated with allocating enough preparation patches to compensate for discarded failures, which scale as $1/p$, drastically reduces the advantage of producing arbitrary-angle resource states. For the target hardware parameters, we find that the success probability falls below 1\% at code distances of $d=15$ or higher. Without patch growth, factory sizes remain relatively small compared to those required for magic state distillation, even at larger code distances of $d=21$ or higher. As expected, the STAR architecture can outperform full FTQC architectures in terms of physical qubit count; however, the resulting PEC overheads become overwhelmingly large, making the STAR architecture impractical for code distances larger than $d=15$.

To fully assess the viability of the STAR architecture, we must also ensure that the error rate associated with rotation gates is not so large that the resulting PEC overhead becomes prohibitive. For a range of initial code distances, we compute the required factory size when growing the resource state to the target final distance for a given circuit size. Although the factory is effectively fixed at the initial distance, the total factory size still increases with circuit size. This increase compensates for the longer state-preparation process caused by the additional QEC cycles required by the growth protocol. Using these curves, we evaluate the PEC overhead as defined in \Cref{PEC overhead}. The resulting contours represent combinations of initial distance and circuit size that yield the same PEC overhead. For a circuit of fixed size and a chosen tolerance for slowdown due to PEC, \Cref{fig:rot_gate_error} therefore allows us to identify the smallest factory capable of supporting the computation. \Cref{fig:rot_gate_error} also shows that, for a given tolerable PEC overhead, there exists a global upper bound on the circuit size that can be executed using the STAR architecture. Note, however, that this bound depends on the specific rotation angles present in the circuit.

\section{Conclusion and perspectives}
\label{sec:Conclusion}
In this work, we carefully and systematically analyze the partially fault-tolerant resources required for simulating quantum chemistry and strongly-interacting electrons. In particular, we demonstrate that partial-FTQC algorithms for ground-state energy estimation of $p$-benzyne and the Fermi--Hubbard model require significantly more hardware resources than previously reported \cite{Akahoshi:2024yme}. Very large code distances (beyond $d=20$) are required for the target and desired hardware specifications summarized in \Cref{tab:physical_params}. Consequently, the success probability of the transversal multi-rotation protocol becomes very low, unless resource states are initially prepared in low-distance code patches and then grown to larger-distance code
patches. We find that na\"ive growth protocols for increasing the code distance of rotation resource states dramatically raise the associated logical error rates, leading to substantial probabilistic error cancellation (PEC) overheads. In our simulations of the growth protocol, with distances $d_i=11$ or $d_i=13$ chosen for the initial code patches, the resulting logical error rates of the final code patches entail that the PEC overhead is astronomically large for the target hardware specifications. For the desired hardware parameter set, both the logical infidelity and the success probability improve by roughly an order of magnitude; however, the resulting PEC overhead remains formidable. 

We have also analyzed how the overhead of generating resource states in a STAR architecture scales with increasing circuit size and compared its factory-size requirements with those of fully fault-tolerant implementations in an  
application-agnostic setting. 
Building upon these results, we have analyzed the trade-offs between resource state factory size and error-mitigation overhead, and evaluated the practical viability of STAR schemes for medium-sized circuits in the megaquop regime and beyond.  We conclude that, for quantum hardware
 of near-term quality,  STAR architectures are particularly effective for executing logical circuits in the slightly sub-megaquop regime, corresponding to roughly $10^5$--$10^6$ small-angle rotation gates, 
 but rapidly become impractical beyond this ``Goldilocks scale'', unless substantially more-efficient error-mitigation techniques are developed. 

We anticipate that improved growth protocols could substantially reduce a STAR architecture's resource requirements.
To this end, ideas may be drawn from magic state cultivation, a technique in which a magic $T$ state is rapidly grown to a large code distance with only a small increase in infidelity by using post-selection based on soft-information decoding~\cite{gidney2024cultivation}. Adapting a similar approach to preparing analog rotation resource states  in a STAR architecture would likely further reduce the already low preparation success probabilities, potentially  either requiring larger factories or resulting in slower per-shot execution. However, the resulting improvement in the logical fidelity of the post-growth states could yield significant reductions in PEC overhead, and thus in the overall runtime.

It is worth  highlighting several other opportunities for advancing near-term research on EFTQC algorithms and partially fault-tolerant architectures. Our resource estimates for executing logical quantum circuits in the megaquop regime indicate that the overall algorithm runtimes are dominated by the overhead arising from PEC. These findings suggest that improved error mitigation techniques will be necessary to render the STAR scheme a viable approach for utility-scale applications. 

From an algorithmic perspective, it is important to quantitatively assess the robustness of the EFTQC algorithms that are designed to be inherently noise resilient. Several recent works have already begun to explore this avenue of research. For example, Ref.~\cite{kshirsagar2024proving} presents a study of the  robustness against noise of the randomized Fourier estimation (RFE) algorithm under simple noise models. Similarly, Ref.~\cite{dutkiewicz2024error} presents an analysis of the sensitivity of QPE algorithms based on the quantum Fourier transform to noise, and proposes a phase-estimation-specific error mitigation technique. Another recent work~\cite{Tsubouchi_2025} presents an analysis of symmetric Clifford twirling for cost-optimal quantum error mitigation in the EFTQC regime, showing how certain Pauli noise channels can be scrambled into global depolarizing noise. While the QCELS algorithm is not inherently robust against global depolarizing noise, a recent study in Ref.~\cite{Ding:2023xuw} shows that QCELS can be modified to be more resilient to such noise. This modification, however, comes at the cost of worsening the scaling of $T_{\text{total}}$ with respect to the target energy-estimation accuracy $\epsilon$, from  $\mathcal{O}(\epsilon^{-1})$ to $\mathcal{O}(\epsilon^{-(2 + \Theta(\gamma/\Delta))})$, where $\Delta$ is the gap between the lowest-energy eigenstates of the Hamiltonian and $\gamma$ is a depolarizing noise constant. 

In addition to phase estimation algorithms, it is also important to explore alternatives such as QSCI~\cite{kanno2023quantum}, which has been extended into various forms and applied to obtain qualitatively and, in some cases, quantitatively accurate quantum chemical results. QSCI-like algorithms typically employ short-depth circuits, allowing for reduced code distances and a lower physical qubit overhead. However, it remains an open question whether an accurate error budget model exists for predicting the performance of QSCI methods for larger systems.

From an architectural perspective, it is worth investigating whether alternative logical patch arrangements---beyond those shown in \Cref{fig:layout_partial-FTQC_p_benzyne} and \Cref{fig:2d_fh_4row}---can improve the scheduling of analog rotation gates without requiring large dedicated preparation regions. For instance, it is demonstrated in Ref.~\cite{Akahoshi:2024yme} that modifying the patch layout can reduce the clock depth of a single Trotter step for the 2D Fermi--Hubbard model, eliminating the need for a separate state-preparation area. However, this design assumes qubit quality beyond the desired parameter regime. Further simulations and design optimizations are needed to identify better patch layouts and scheduling strategies for lattice surgery that can enhance the performance of EFTQC algorithms on a STAR architecture.

Finally, we outline the challenges and opportunities for 
EFTQC implementations that are specific to 
quantum simulations of the Fermi--Hubbard model as a high-value application. In recent years, significant advances in algorithmic methods have been made in simulating Fermi--Hubbard models on fault-tolerant quantum computers. These approaches can be broadly divided into three categories: (i) standard Trotterization; (ii) plaquette Trotterization and its generalizations; and (iii) qubitization and quantum signal processing.
Standard Trotterization refers to the use of Trotter--Suzuki product formulae to perform Hamiltonian simulation. In practice, this typically involves the second-order product formulae, although there is evidence that higher-order product formulae can also be advantageous. The resulting quantum circuits generally consist of repeated small-angle rotations, with the scaling dependent on the required precision, making them well-suited for STAR architectures.
Plaquette Trotterization refers to ideas first introduced in Ref.~\cite{campbell_early_2022} and later generalized to variants of the Fermi--Hubbard model. The key idea behind plaquette Trotterization is to improve upon standard split-operator techniques and diagonalize the hopping and the interaction Hamiltonian separately. However, this approach requires the use of the fermionic fast Fourier transform (FFFT), which employs large-angle rotations (namely, many \(T\) gates). However, 
given that the cost of synthesizing \(T\) gates in STAR architectures is prohibitively high, these operations would dominate the overall resource requirements. As a result, STAR architectures appear to be largely incompatible with plaquette Trotterization. It would therefore be interesting to explore whether new algorithms based on split-operator techniques could be developed that would avoid these costly operations.
Techniques based on qubitization or quantum signal processing utilize the recently developed framework referred to as block encoding of Hamiltonians to achieve asymptotically optimal \(T\) complexity. However, for relative accuracy requirements, Trotterization-based techniques have shown improved scaling, at least up to a modest number of orbitals, in simulations of the Fermi--Hubbard model and quantum chemistry systems~\cite{kivlichan2020improved}.

Beyond the choice of Hamiltonian simulation techniques (e.g., standard Trotterization, plaquette Trotterization, and qubitization), additional methods exist for reducing the overall resource requirements of quantum simulations. For example, Ref.~\cite{campbell_early_2022} also shows how to employ Hamming weight phasing, which enables the efficient synthesis of collections of \(Z\)-axis rotations. This technique can exponentially reduce the number of \(Z\)-axis rotations using additional ancillae and Toffoli gates. However, Toffoli gates, similar to $T$ gates, are costly to realize on a STAR architecture. On the other hand, the same repeated angle allows the parallelization of the growth protocol through the use of a specialized architecture.

On the positive side, STAR architectures are particularly well-suited for implementing small-angle rotations. Exploiting this feature may require redesigning standard quantum algorithms. For example, one could consider using higher-order Trotter product formulae~\cite{PhysRevX.11.011020} instead of the second-order Trotter formulae typically employed for Hamiltonian simulation. Higher-order formulae, such as the third-order Suzuki--Trotter formula, introduce only a constant increase in circuit depth while enabling much smaller rotation angles. 
This, in turn, can reduce the overhead associated with PEC and growth protocols, thereby lowering the overall resource requirements for simulating Fermi--Hubbard models. Determining the optimal Suzuki--Trotter product formula under a realistic error budget is thus an interesting direction for future work.

\textbf{Acknowledgments.} 

The authors from QunaSys thank Keita Kanno and Andreas Thomasen for insightful discussions. The authors from 1QBit thank Marko Bucyk for editorial review of the manuscript. QunaSys acknowledges the support of the Cross-ministerial Strategic Innovation Promotion Program (SIP). HPE, Qolab, and 1QBit acknowledge support from DARPA’s Quantum Benchmarking Initiative (QBI) contract no HR00112590116. 1QBit acknowledges the financial support of Pacific Economic Development Canada (PacifiCan) under project number PC0008525. G.~A.~D. is grateful for the support of Mitacs. P.~R. acknowledges the financial support of Innovation, Science and Economic Development Canada (ISED), the Province of Ontario through the Ministry of Colleges and Universities, and the Perimeter Institute for Theoretical Physics.

\onecolumngrid
\bibliography{all_ref_article,all_ref}
\appendix

\section{Partial FTQC protocols}
\label{sec:partial-FTQC_appendix}

\Cref{sec: STAR intro} uses the transversal multi-rotation protocol within the STAR architecture to provide resource estimates for various quantum algorithms. This protocol creates the resource state
\begin{equation}
    \ket{m_{\theta_*}}_{L} \equiv \cos\theta_*\ket{+}_{L} + i\sin\theta_*\ket{-}_{L}.
\end{equation}
In this section, we provide further details on the implementation of the protocol as well as simulation results for this protocol using the target and desired hardware parameter specifications presented in ~\Cref{tab:physical_params}.

The section is organized as follows. \Cref{sec:rotation} provides a high-level description of how the resource state is prepared, and \cref{sec:infidelity} presents a theoretical analysis of the protocol. \Cref{sec:simulation_methodology} discusses the methods used to simulate this protocol, while \cref{sec:simulation_results} presents numerical simulation results, highlighting a noticeable decline in success rate as the distance increases. To address this issue, we propose the growth protocol for expanding the initial patches to larger distances. This is discussed in \cref{sec:growth_protocol}.

\subsection{Transversal multi-rotation protocol}
\label{sec:rotation}

In this section, we explain how the protocol works, focusing on how the different parts lead to the creation of the resource state. The protocol has four steps. First, the logical initial state $\ket{+}_L$ is prepared in a surface code patch. This is done by initializing all data qubits across the entire patch in the $\ket{+}$ state, followed by a round of syndrome measurements. For the next steps to be successful, it is necessary that the $\ket{+}_L$ state be in the canonical logical space, that is, the stabilizer subspace in which the logical state is in the $+1$ eigenstate of all stabilizers. Therefore, if the first round of syndrome measurements yields any non-zero syndromes, the protocol is restarted.

\begin{figure}[tb]
    \centering
    \begin{minipage}[h!]{0.48\textwidth}
        \centering
        \includegraphics[width=0.95\linewidth]{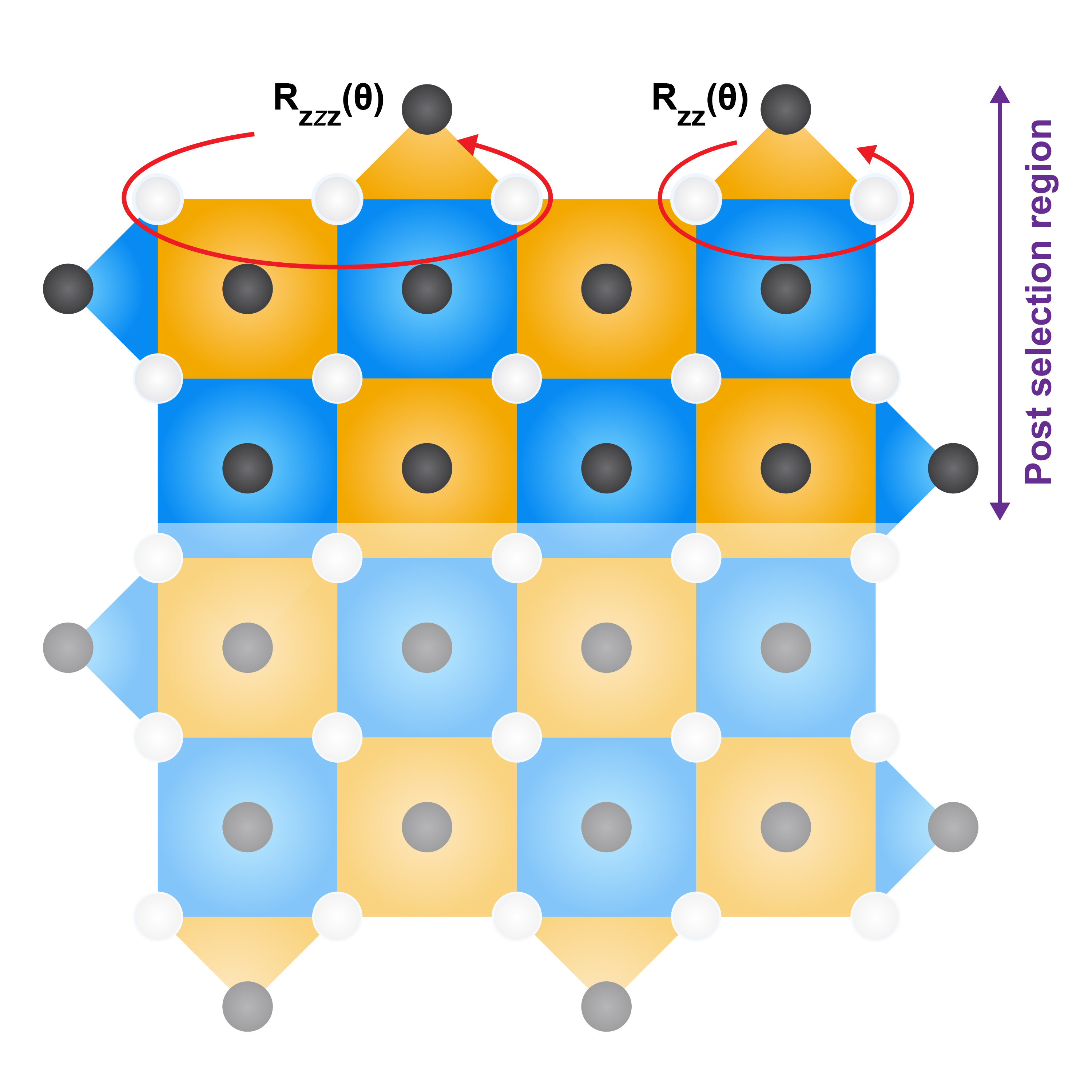}
        \caption*{(a)}
        \label{m2m3}
    \end{minipage}
    \begin{minipage}[h!]{0.48\textwidth}
        \centering
        \includegraphics[width=0.95\linewidth]{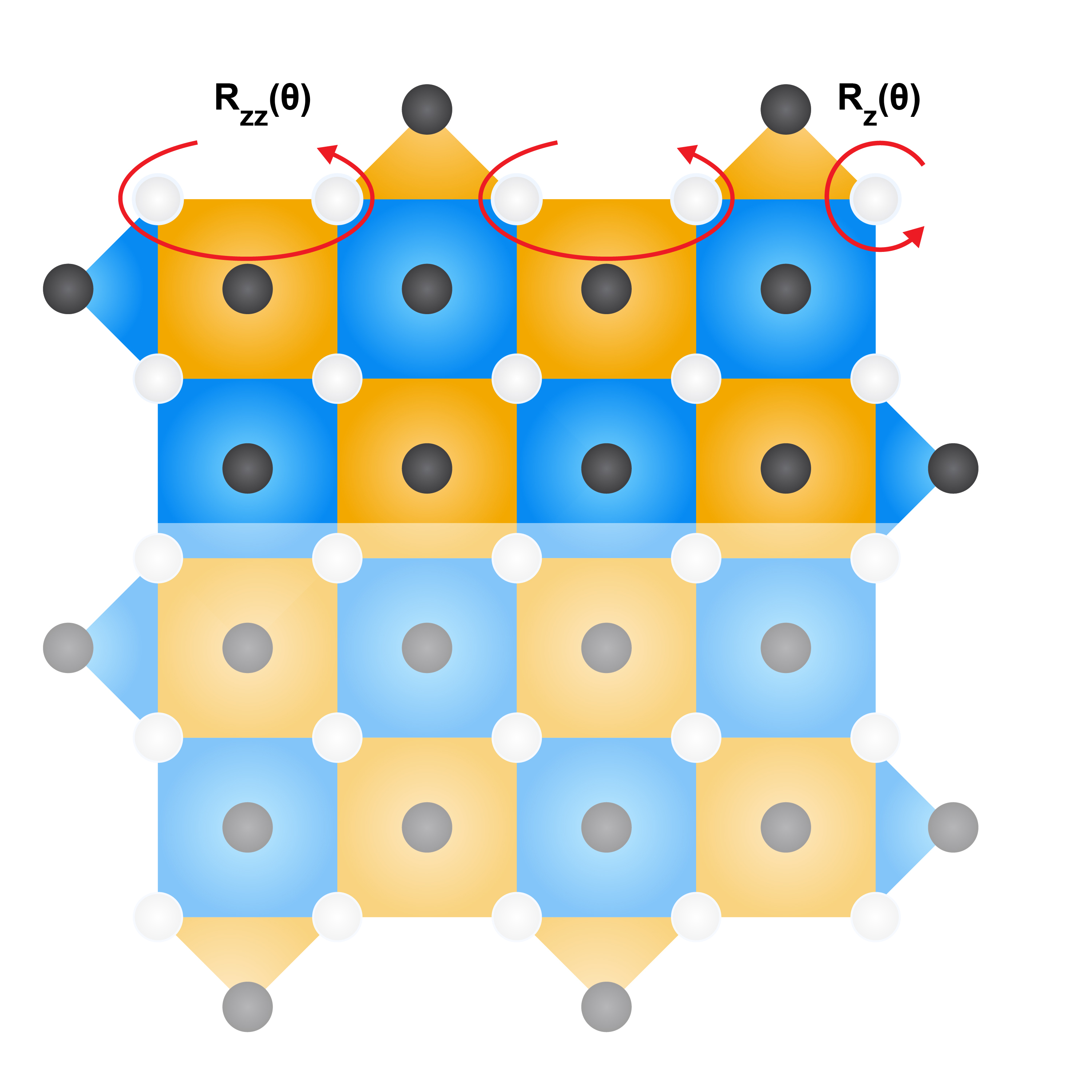}
        \caption*{(b)}
    \end{minipage}
    \caption{Two distance $d=5$ surface code patches are shown, where blue squares represent $Z$ stabilizers, yellow squares represent $X$ stabilizers, white circles are data qubits, and gray circles denote syndrome qubits. In the transversal multi-rotation protocol, physical rotation gates are applied to all the data qubits in the top row. Any combination of physical rotations works for creating the resource state. The patches here show two such examples, which are also used in our simulations. In the protocol, after the physical rotations are applied, syndrome measurements are used to post-select. In the  optimal post-selection scheme, only states for which the top three rows of syndrome measurements are zero are kept, while others are discarded~\cite{Toshio2025PFTQC}. }
    \label{fig:post-select-regime}
\end{figure}

Next, physical analog $Z$-rotation gates of angles $\theta$ are applied as illustrated in \Cref{fig:post-select-regime}. The angle $\theta$ is chosen according to the expressions presented in the next section. After this step, the logical state of the patch is in a superposition of the desired logical rotated state $\ket{m_{\theta_*}}_L$ and undesired other states. Finally, two rounds of syndrome measurements are performed. We post-select states for which all syndromes in the top three rows of the surface patch are zero, and otherwise restart the protocol. This is called the optimal post-selection method~\cite{Toshio2025PFTQC}, and is shown in~\Cref{fig:post-select-regime}. After this post-selection the  logical state is left in the state $\ket{m_{\theta_*}}_L$, if there is no noise in the system. Even if this protocol is executed on a noisy system, there is high probability of preparing the desired logical state. The reason for performing more than one round of syndrome measurements is to protect against faulty syndrome measurement. If we performed only one round of error detection, some faulty states would not be post-selected out. Additional rounds increase the probability that we obtain the correct state, but also increase the rejection rate. Hence, only two rounds are performed.

\begin{figure*}[tb]
    \centering
    \begin{minipage}[h!]{0.43\textwidth}
        \centering
        \includegraphics[width=\linewidth]{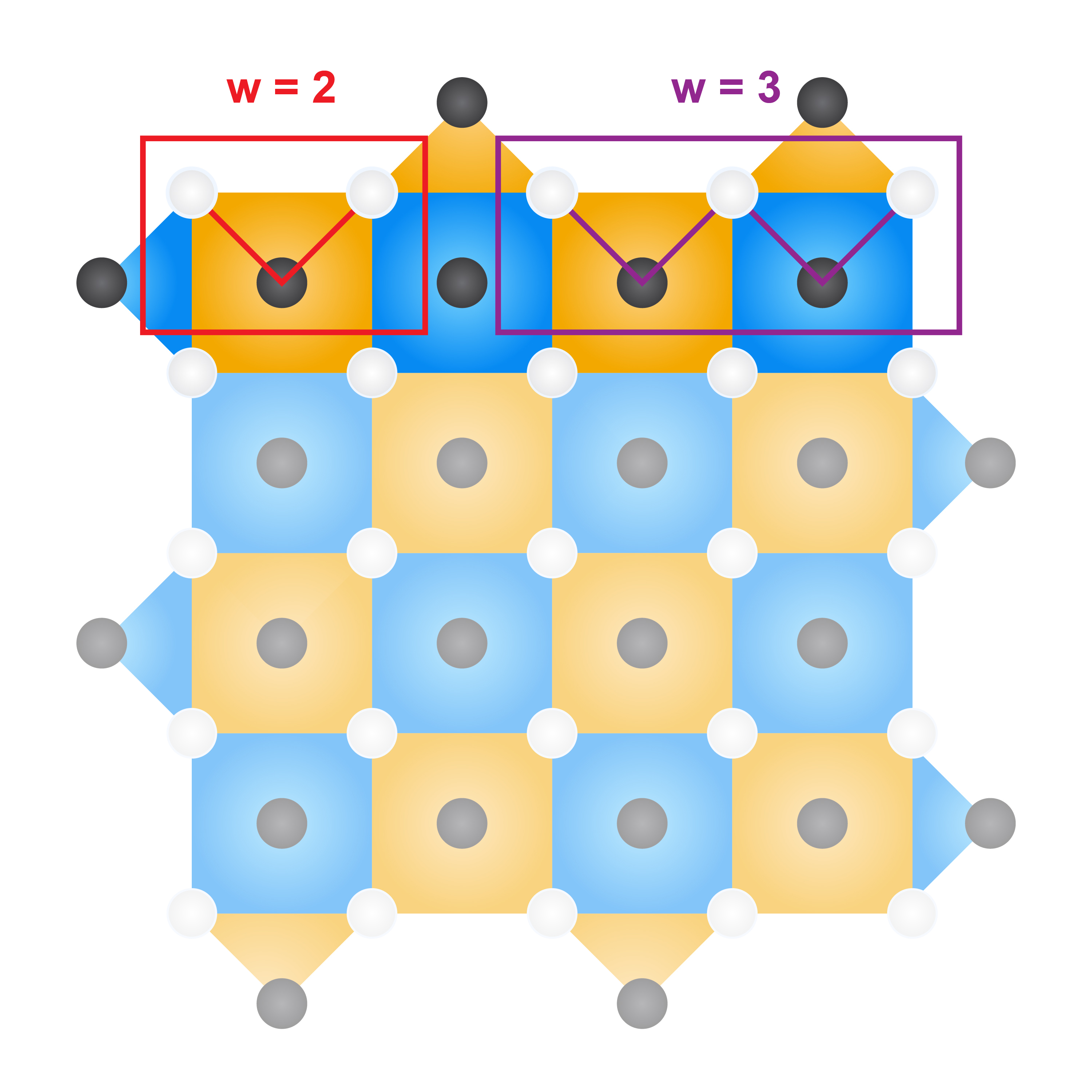}
        \vspace{-7mm}
        \caption*{(a)}
        \label{m2m3-1}
    \end{minipage}
    \begin{minipage}[h!]{0.48\textwidth}
        \centering
        \includegraphics[width=0.63\linewidth]{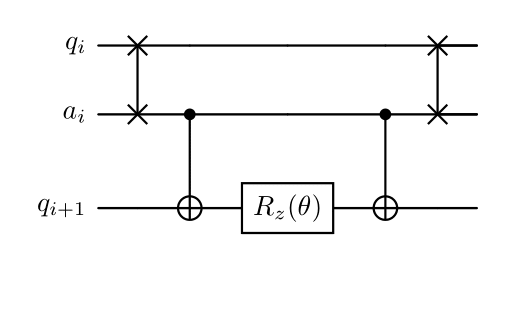}
        \vspace{-10mm}
        \caption*{(b)}
        \vspace{2mm}
        \includegraphics[width=0.73\linewidth]{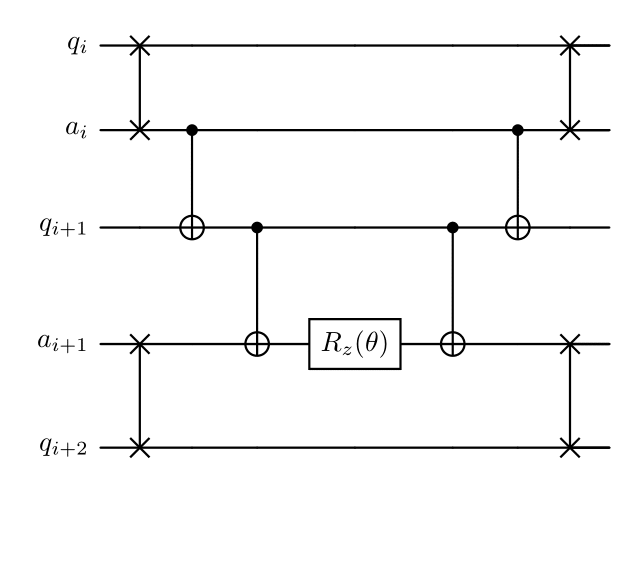}
         \vspace{-8mm}
        \caption*{(c)}
    \end{minipage}
    
    \caption{In superconducting quantum chips, typically only data qubit to syndrome qubit couplers are available. Therefore, two- and three-qubit rotation gates on adjacent data qubits can be implemented by swapping the data qubits with their nearest syndrome qubits. (a) Schematic illustration of the coupler used for implementing the rotation gates: the solid red box represents a $\hat{R}_{zz}(\theta)$ rotation, while the solid magenta box represents a $\hat{R}_{zzz}(\theta)$ rotation. (b) Circuit for $\hat{R}_{zz}(\theta)$ rotation ($m=2$). Here, $q$ denotes a data qubit, $a$ denotes a syndrome qubit, and the indices denote positions in the row. (c) Circuit for $\hat{R}_{zzz}(\theta)$ rotation ($m=3$).}
    \label{fig:circuits}
\end{figure*}

The physical multi-qubit $Z$-rotation gates mentioned above can be executed on a superconducting quantum chip, which only has data qubit to syndrome qubit couplers. It's relatively simple to achieve this for a two-qubit $R_{zz}(\theta)$ rotation (with weight $w=2$) and for three-qubit $R_{zzz}(\theta)$ rotation ($w=3$), with the aid of swap gates. In \Cref{fig:circuits} explicit circuits for these rotations are provided~\cite{Toshio2025PFTQC}.

\subsection{Theory of the transversal multi-rotation protocol}
\label{sec:infidelity}
We now discuss the mathematical details of this protocol. First, we find an equation that determines the angle $\theta$ for any choice of $\theta_*$. Next, we determine theoretical expressions for the infidelity~\cite{Akahoshi2024PFTQC} and success rate. These expressions are in terms of $\theta$ and the physical error rate, and can be used to validate the simulation results we present in the next section. 

For clarity of analysis we make two assumptions. First, we analyze rotated surface code patches of even distance. Second, we restrict the derivations to physical two-qubit $Z$-Pauli rotations ($w=2$ case). In contrast, in our simulations whose results are used for the resource estimates, we employ rotated surface code patches of odd distances, which require the use of a mix of rotations of different weights. However, this more complicated case does not significantly change the expressions for the infidelity or success rate.

Our goal is to demonstrate that the protocol creates the logical state
\begin{align}
    \ket{m_{\theta_*}} &= \exp(i\theta_* Z_L)\ket{+}_L \nonumber \\ &= \cos\theta_* \ket{+}_L + i\sin\theta_* \ket{-}_L,
\end{align}
with high fidelity. Recall that, in the protocol, after the first syndrome measurement round and post-selection, the surface code patch is in the state $\ket{+}_L$ state with all stabilizers in the $+1$ state. Acting with a sequence of $d/2$ two-qubit rotation gates of angle $\theta$ on the top row of data qubits results in the state
\begin{align}
\prod_{i=0}^{k}\hat{R}_{zz, i}(\theta)\ket{+}_L &=\prod_{i=0}^{k}\left(\cos\theta\, \hat{I}_{2i}\hat{I}_{2i+1} + i\sin\theta\, \hat{Z}_{2i}\hat{Z}_{2i+1}\right)\ket{+}_L, \nonumber \\
&= \Big[\big(\cos^k\theta I_0I_1\cdots I_{d-1}I_d + (i\sin\theta)^{k}Z_0Z_1\cdots Z_{d-1}Z_d\big)\nonumber \\
&\qquad + \big(\cos^{k-1}\theta\, (i\sin\theta)\, Z_{0}Z_{1}I_{2}I_{3}\cdots I_{d-1}I_{d} \nonumber \\ &\qquad + (i\sin\theta)^{k-1}\cos\theta\, I_{0}I_{1}Z_{2}Z_{3}\cdots Z_{d-1}Z_{d}\big) + \cdots\Big]\ket{+}_L.
\label{eq:zz-rotation}
\end{align}
The indices of the Pauli operators above refer to the positions of the data qubits in the top row. Note that $Z_0Z_1\cdots Z_{d-1}Z_d$ in the first parenthesis corresponds to the logical $Z$ operator, thus, this operator will transforms the $\ket{+}_L$ state into $\ket{-}_L$ state. Furthermore, the terms in this expression appear in pairs, with each pair mapping the $\ket{+}_L$ state into a distinct stabilizer subspace. This can be seen because the second term in each pair can be obtained from the first by applying the logical $Z$ operator. This means that the subsequent two syndrome measurement rounds project the state onto one of the stabilizer subspaces. Only if all syndrome measurements are $0$ is the state projected onto 
\begin{equation}
    \cos^{k}\theta\, \ket{+}_L 
    + (i\sin\theta)^{k}\, \ket{-}_L.
\end{equation}
Therefore, we post-select on all syndrome measurements being $0$ to ensure the patch is in this state. If the angle $\theta$ is appropriately chosen, then this post-selected state will be equal to $\ket{m_{\theta_*}}$. By solving 
\begin{equation*}
\cos\theta_* \ket{+}_L + i\sin\theta_*\ket{-}_{L} = \frac{\cos^{k}\theta }{p_{\text{ideal}}}\ket{+}_L + \frac{(i\sin\theta)^{k}}{p_{\text{ideal}}}\ket{-}_L.
\end{equation*}
we can determine the rotation angle,
\begin{equation}
    \theta_* = \arcsin\left(\frac{1}{\sqrt{p_{\text{ideal}}}} \sin\theta^{k}\right),
\end{equation}
where
\begin{equation}
    p_{\text{ideal}} = \sqrt{\cos^{2k}\theta + \sin^{2k}\theta},
\end{equation}
is the success rate of the protocol in the noise-free regime. When $\theta \ll 1$, the rotation angle can be approximated as $\theta_* \approx \theta^{k} + O(\theta^{k + 2})$, and the success probability simplifies to $p_{\text{ideal}} \approx 1 - k\theta^{2}$, which is close to one.

We can also determine the leading order probability of error. The most likely source of error are correlated $ZZ$-errors that act in the same way as the rotation operators. These cannot be detected by post-selection. The action of such errors is to swap the first pair of terms with the second pair of terms in \Cref{eq:zz-rotation}. If this happens, the patch will instead be left in the state $\ket{m_{\theta_{\text{error}}}}$, where
\begin{align}
\theta_{\text{error}} &= -\arcsin\left(\frac{1}{\sqrt{p_{\text{error}}}}\sin^{k - 1}\theta\cos\theta\right), \nonumber\\
&\simeq -\theta^{k -2} + O(\theta^k),
\label{eq:thetaerror}
\end{align}
where again we have taken the $\theta \to 0$ limit. Given a $ZZ$ error does occur, the probability of this happening is
\begin{align}
p_{\text{error}} &= (\cos^{k -1}\theta\sin\theta)^{2} + (\sin^{k - 1}\theta\cos\theta)^{2}, \nonumber\\
&=\sin^2\theta\cos^2\theta(\sin^{2k - 4}\theta + \cos^{2k - 4}\theta), \nonumber\\
&\simeq \theta^{2} + O(\theta^4).
\label{eq:perror}
\end{align}
This is of order $\theta^2$ as was the asymptotic limit of $p_\text{ideal}$.

To estimate the fidelity of the output state, one first quantifies the probability of $ZZ$ errors. Assuming that every operation in the circuit fails with probability $p_\text{ph}$ and this leads to correlated $ZZ$ error with probability $P_\text{ud}$, the output density matrix is
\begin{align}
    \hat{\rho}_{\text{out}} &= \frac{1}{p_{\text{succ}}} \Big[ p_{\text{ideal}}(1 - Q) \ket{m_{\theta_{*}}} \bra{m_{\theta_{*}}}_L + p_{\text{error}} P_{ud} \ket{m_{\theta_{\text{error}}}} \bra{m_{\theta_{\text{error}}}}_L \Big] + O(p_{\text{ph}}^2),
    \label{eq:pftqc_simple_density_matrix}
\end{align}
where $Q \propto p_{\text{ph}}$ is the probability of discarding the output state even when it's actually correct, and the normalization factor $p_\text{succ} = p_\text{ideal}(1-Q) + p_\text{error}P_\text{ud}$ is the probability of success of the protocol if high order errors are ignored. Using these expressions, the infidelity of the analog protocol can be estimated up to the leading-order in $p_{\text{ph}}$ as
\begin{align}
1 - F_\text{resource} &=  1 - \langle m_{\theta_{*}}|\hat{\rho}|m_{\theta_{*}}\rangle\nonumber\\
&=  P_{\text{ud}}\left(\frac{p_{\text{error}}}{p_{\text{ideal}}}\right)\sin^2(\theta_{\text{error}} - \theta_{*}).
\label{eq:infidelity_theory}
\end{align}
This linear relationship of the infidelity on $P_\text{ud}$ means that as the physical error rate $p_\text{ph}$ goes to zero, so does the infidelity. The asymptotic behavior of the infidelity is $P_\text{ud}\theta_*^{2(1-1/k)}$. Hence the infidelity also goes to zero as $\theta_*$ becomes smaller. However, as the distance $d=kw$ increases, the infidelity goes to a constant. 

This shows that this is not a fault-tolerant protocol. With this theoretical expression for infidelity in hand, we now present the framework for simulating this protocol on noisy hardware.

\subsection{Simulation methodology for the transversal multi-rotation protocol}
\label{sec:simulation_methodology}

In this subsection, we discuss how simulations of the protocol on noisy hardware are performed, and how the raw data from the simulation is post-processed to extract the infidelity and success rate of the protocol. The challenge of simulating the transversal rotation protocol is that it consists of non-Clifford gates. Faithful simulations of such quantum circuits typically require full state vector simulation. This is not realistically possible for surface codes which contain a lot of qubits. Here, we present an alternate approach which leverages cheap Clifford simulations to assess the infidelity and success rate with high accuracy~\cite{choi2023pftqc, Toshio2025PFTQC}.

In this method, the application of physical analog rotation gates, as well as the occurrence of errors is treated as a stochastic process, consisting of either $I$ or $Z$ gates. This can be inferred in  the following way. Let $\hat{Z}^b \equiv \prod_{i: b_{i}=1}\hat{Z}_{2i}\hat{Z}_{2i + 1}$, where $b=b_1b_2b_3\dots b_k\in\{0,1\}^k$ is a bit-string and $|b|$ is its Hamming weight. This operator allows the re-expression of the rotation operator as
\begin{equation}
    \prod_{i=1}^{k}\hat{R}_{zz, i}(\theta) = \sum_{b = 0}^{2^k}u_{|b|}\hat{Z}^{b}, \quad \text{where } u_{|b|} \equiv i^{|b|}\sin^{|b|}\theta\cos^{k - |b|}\theta.
    \label{eq:z_b}
\end{equation}
Recall that, in \Cref{eq:zz-rotation} terms come in pairs, where the second term in each pair is obtained from the first by applying the logical $Z$ operator. In the notation used here, this means that the term labeled by $b$ is paired with the term labeled by $\bar{b}$, which is the bit-wise negation of $b$. Hence, summation on the right hand side of \Cref{eq:z_b} can be rewritten as
\begin{equation}
    \prod_{i=1}^{k}\hat{R}_{zz, i}(\theta) = \sum_{b = 0}^{2^{k-1}}\big(u_{|b|}\hat{Z}^{b} + u_{|\bar{b}|}\hat{Z}^{\bar{b}}\big).
    \label{eq:zbzbbar}
\end{equation}
Both terms in a pair project the logical state of the surface code into the same stabilizer subspace, that is, $Z^b\ket{\psi}_L$ and $Z^{\bar{b}}\ket{\psi}_L$ both trigger the same set of stabilizers. On the other hand different pairs project logical states onto different stabilizer subspaces. This means that after doing the physical rotations in the protocol, each distinct set of syndromes measurements in the subsequent rounds projects the logical state onto a distinct stabilizer subspace, and the probability of projecting onto stabilizer subspace labeled by $b$ is $|u_{b}|^2 + |u_{\bar{b}}|^2$. A $Z^b$ operator is Clifford and so the action of each $Z^b$ operator can be determined by an independent Clifford simulation. Moreover, since the action of $Z^{\bar{b}}$ is similar to $Z^b$ we only need to simulate half the bit-strings. Hence, a state vector simulation with physical analog rotations can be replaced by a stochastic Clifford simulation as follows. For each $b \in \{0, \dots, 2^{k-1}\}$, perform the following noisy Clifford simulation with $N_\text{shots}$ shots. 
\begin{enumerate}
    \item Prepare the initial state $\ket{+}_L$ of a rotated surface code of odd distance $d$ by initializing each data qubit in the $\ket{+}$ state and performing a round of syndrome measurements. Post-select if all measurements are zero.
    \item  Apply the operator $Z^b$ to the top row of data qubits. This requires utilizing the circuits shown in \Cref{fig:circuits} with $R_Z(\theta)$ replaced by $Z$.
    \item Perform two rounds of syndrome measurements. Perform post-selection on the syndrome measurements.
\end{enumerate}
Let $N_{\text{pass}, b}$ be the number of shots that are not discarded for bit-string $b$.

The stochastic simplification also holds if errors occur during the protocol as in these noisy simulations. Any error $E$, whose weight is less than $d$, moves the logical state to a different stabilizer subspace. The errors that are not captured by the post-selection are correlated $Z$ errors that are equivalent to one of the $Z^b$ operators with $|b| \ne 0$. So, in a noisy simulation with $|b|=0$, the states that pass the post-selection correspond to the desired logical state, while the states that pass the post-selection in the simulations with $|b| \ne 0$ result in undesired logical states.

We use the raw statistics from the simulation, $N_\text{shots}$ and $\{N_\text{pass, b}\}$ to estimate the success rate and the infidelity as follows. After post-selection, the logical state of the patch will either be in the desired state $\ket{m_{\theta_{*}}}$ or in an incorrect state. The angle of these desired and undesired logical states will be 
\begin{equation}
\theta_{n}(\theta, k) = \arcsin\left(\sqrt{\frac{|u_{k - n}|^2}{|u_{k}|^2 + |u_{k - n}|^2}}\right),
\end{equation}
where $\theta_0 = \theta_*$. Each such state will pass post-selection with probability 
\begin{equation}
q_{n} = \frac{\big(|u_{n}^2| + |u_{|k-n|}|^2\big)}{{}^kC_{n}}\sum_{b: |b| = n} \frac{N_{\text{pass}, b}}{N_{\text{shots}}}.
\end{equation}
This allows us to estimate that the probability of successful post-selection as 
\begin{equation}
    p_{\text{succ}} = \sum_{n=0}^{k-1} q_{n}.
\end{equation}
The output density matrix \Cref{eq:pftqc_simple_density_matrix} with all possible terms is
\begin{equation}
    \rho_{\text{out}} = \frac{1}{p_{\text{succ}}}\sum_{n=0}^{k-1}q_{n} \ket{m_{\theta_{n}}}\bra{m_{\theta_{n}}}.
\end{equation}
This yields the infidelity,
\begin{align}
1 - F_\text{resource} &= 1 - \langle m_{\theta_{*}}|\rho_{\text{out}}|m_{\theta_{*}}\rangle, \nonumber\\
&= \frac{1}{p_{\text{succ}}}\sum_{n = 0}^{k - 1}q_n \sin^2(\theta_n - \theta_{*}).
\end{align}

Given the success rate and the infidelity from the numerical simulation, we can assess the overall performance of this approach.

\section{Improved Trotter error factor for $p$-benzyne Hamiltonian}
\label{app:improved trotter}
\begin{figure}[tbph]
    \centering
    \includegraphics[width=1\linewidth]{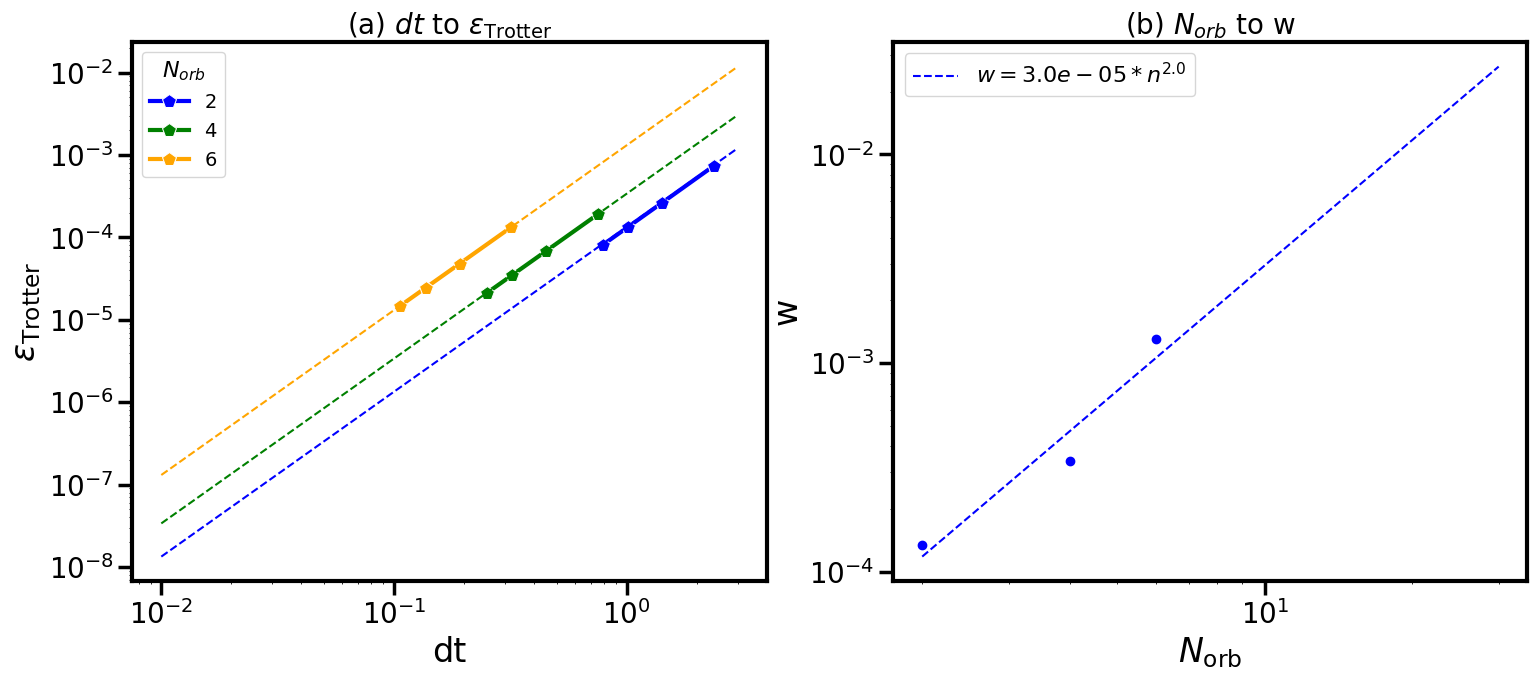}
    \caption{(a) The Trotter error of first order product formula at different $dt$s. $\epsilon_{\text{Trotter}}$ is computed by diagonalizing the both the Trotter effective Hamiltonian and the exact Hamiltonian. (b) The $w$ factor and the $N_{\text{Trotter}}$.}
    \label{fig:Trotter eval}
\end{figure}

The Trotter error bound coefficient used in the $p$-benzyne resource estimation in \Cref{sec: p-benzyne estimate} is smaller the empirical bound used in \cite{mohseni2025buildquantumsupercomputer}, leading to a tighter bound and an even more optimistic Trotter step estimation. Here, we discuss how we obtained the Trotter error coefficient $w$.  Intuitively, we numerically estimate this value instead of using loose theoretical bounds.

Given the Hamiltonian written in the Pauli string basis: $H = \sum_{i=0}^{L-1}c_iP_i$, previous estimations of the Trotter error bound utilizes the theoretical bound of:
\begin{equation}
    w = \frac{1}{12}\sum_{a=1}^L \left( \left\lVert \left[ \sum_{c=a+1}^LH_c,\left[ \sum_{b=a+1}^{L}H_b, H_a\right] \right]\right\rVert + \frac{1}{2}\left\lVert \left[ H_a,\left[ H_a, \sum_{b=a+1}^LH_b\right] \right]\right\rVert\right),
\end{equation}
or the empirical evaluation of the Hamiltonian 2-norm. Here, we directly diagonalize the exact Hamiltonian and the Trotter effective Hamiltonian:
\begin{equation}
    H_{\text{eff}} \equiv \frac{i}{dt}\log\left(\prod_{i=0}^{L-1}\exp\left(-ic_i P_i dt\right)\right)
\end{equation}
at different $dt$s and fit against Trotter error model:
\begin{equation}
    \epsilon_{\text{Trotter}} = |E_{\text{exact}}-E_{\text{Trotter}}| = w(dt)^2.
    \label{eq: Trotter to dt fit}
\end{equation}
We can only perform the diagonalization for $\text{HL} \pm 0, 1, 2$, but it is sufficient for us to perform a fit to obtain $w$ as a function of $N_{\text{orb}}$. With the fitting result, we obtained the relation $w = (3\times 10^{-5}) N_{\text{orb}}^2$ where we plot the data in \Cref{fig:Trotter eval}~(b).

It should also be noted that we are using the first Trotter order formula and the fitting model \Cref{eq: Trotter to dt fit}. As noted in \cite{reiher2017elucidating}, the first order product formula also has the same scaling the the second order product formula when the Hamiltonian is real-valued. It is indeed the case for our Hamiltonian, as it only contains $Z\cdots Z$, $XZ\cdots ZX$ and $YZ\cdots ZY$ Pauli strings with real coefficients. We explicitly checked this relation in \Cref{fig:Trotter eval}~(a), where the dashed lines all scale as $\mathcal{O}(dt^2)$.

\section{A review of the QCELS algorithm and its error budget equation}\label{app: QCELS scaling}
\label{app:appendix-QCELS}

\begin{algorithm}

\caption{\textbf{multi-level QCELS algorithm} from  Ref.~\cite{ding2023even}, p.8}
\label{alg: QCELS pseudo}

\textbf{Preparation:} Number of data pairs: \( N \);
number of samples: \( N_s \); 
number of iterations: \( J \); 
sequence of time steps: \( \{\tau_j\}_{j=1}^{J} \); 
Quantum oracle: \( \{\exp(-i \tau_j H)\}_{j=1}^{J} \);

\vspace{0.5em}
\textbf{Running:}

\SetKwInOut{Input}{Input}
\SetKwInOut{Output}{Output}

Set \( \lambda_{\min} \leftarrow -\pi \), \( \lambda_{\max} \leftarrow \pi \) \tcp*[r]{\( [\lambda_{\min}, \lambda_{\max}] \) contains \( \lambda_0 \)}

\For(\tcp*[f]{Loop over time steps}){\( j = 1 \) \KwTo \( J \)}{
    \For(){\(n = 0\) \KwTo \(N-1\)}{
        Estimate \(Z_n = \langle e^{-in\tau_j H}\rangle\) using \(N_s\) samples \;
    }
    \( \mathcal{D}_H \leftarrow \{(n \tau_j, Z_n)\}_{n=0}^{N-1} \)\;
    Minimize the loss function \Cref{eq:qcels loss} over \(\mathcal{D}_H\)\;
    \(
    (r_j^*, \theta_j^*) \leftarrow \arg\min_{\substack{r \in \mathbb{C} \\ \theta \in [ \lambda_{\min}, \lambda_{\max} ]}} L(r, \theta)
    \)\;
    Update search interval\;
    \( \lambda_{\min} \leftarrow \theta_j^* - \frac{\pi}{2 \tau_j} \), \quad
    \( \lambda_{\max} \leftarrow \theta_j^* + \frac{\pi}{2 \tau_j} \) \tcp*[r]{Shrink interval by 1/2. See \Cref{eq:tau_def}}
}

\Output{\( \theta^* \)}

\end{algorithm}

Here we review the quantum complex exponential least squares (QCELS) algorithm introduced in Ref.~\cite{ding2023even}. It is an EFTQC phase estimation algorithm that keeps the total runtime scaling satisfying the Heisenberg limit with a short circuit depth. The algorithm is summarized in Algorithm \ref{alg: QCELS pseudo}. In each iteration, it chooses a specific $\tau$ and collects data
\begin{eqnarray}
    Z_n = \frac{1}{N_s}\sum_{i=1}^{N_s}(X_i + i Y_i),
\end{eqnarray}
where $X_i + i Y_i \in \{\pm 1 \pm i\}$ are measurement results of a Hadamard-test-like circuit such as \Cref{fig:short_hadamard}

that estimates the expectation value $\langle e^{-it_n H}\rangle$ for $t_n = n\tau$.  (Note that the real part and imaginary part are computed separately.)  Given this data, we then optimize over a set of parameters $(r, \;\theta)$ to find a minimum of the loss function $L(r, \theta)$:
\begin{equation}
    L(r, \theta) = \frac{1}{N}\sum_{n=0}^{N-1} \left|Z_n-r e^{-i t_n \theta}\right|^2.
    \label{eq:qcels loss}
\end{equation}
The optimization is performed over the interval $\theta\in [\lambda_{\min}, \lambda_{\max}]$, which contains the target phase $\theta^*$. After each iteration, the search interval is halved.  This repeats until the specified number of iterations have occurred.

In the algorithm, the hyperparameters $J$, $N$, $N_s$, $\tau_j$ are chosen according to the conditions:
\begin{subequations}
\begin{align}
  \delta &= \Theta(\sqrt{1 - p_0}) ,
  \label{eq:delta_def} \\
  J &= \lceil \log_2 \epsilon^{-1} \rceil + 1 ,
  \label{eq:J_def} \\
  \tau_j &= 2^{j-J} \frac{\delta}{N\epsilon} ,
  \label{eq:tau_def} \\
  NN_s &= \widetilde{\Theta}(\delta^{-2}) ,
  \label{eq:Ns_def}
\end{align}
\end{subequations}
with $p_0$ defined as the overlap between the ground state and the algorithm input state. $N_s$ is taken to be the empirical value of 100, which is consistent with the simulation in Ref.~cite{ding2023even}. $N$ is then determined by $N = \max\{5, \delta^{-2}N_s^{-1}\}$, where the minimum of 5 data points is chosen based on empirical values used in the simulation of Ref.~\cite{ding2023even}. We can see from the pseudo-code and the above parameter choice that the $T_{\text{max}}$ and $T_{\text{total}}$ scale as 
\begin{subequations}
\begin{align}
\small
T_{\text{max}} &= N\tau_J = \frac{\delta}{\epsilon}, \\
\small
T_{\text{total}} &= 
\sum_{j=1}^J\sum_{n=0}^{N}n N_s \tau_j =\sum_{j=1}^J\frac{N(N+1)}{2}N_s \tau_j = \widetilde{\Theta}(\delta^{-1}\epsilon^{-1})
\label{eq:t_total}
\end{align}
\end{subequations}

Following the algorithm above, we assume that the time evolution operator used for estimating $\langle e^{-iHt}\rangle$ is implemented by second-order Trotter--Suzuki decomposition. We can then rewrite the terms in the error budget equation 
\begin{equation}
    \epsilon_{\text{Trotter}} + \epsilon_{\text{QCELS}} \leq \epsilon_{\text{target}}
\end{equation}
following \cite{toshio2024practical, Akahoshi:2024yme}:
\begin{equation}
    \epsilon_{\text{Trotter}} = w\left(\frac{T_{\text{max}}}{2N_{\max}}\right)^2 =w \left(\frac{\delta}{2\epsilon_{\text{QCELS}} N_{\text{Trotter}}}\right)^2,
\end{equation}
where $N_{\max}$ is the maximum number of Trotter steps used to evolve to $T_{\max} = N\tau_J$. This gives:
\begin{equation}
    N_{\max} 
    = \frac{\delta}{2\epsilon_{\text{QCELS}}}\sqrt{\frac{w}{\epsilon_{\text{Trotter}}}} .
\end{equation}
To reduce the runtime, we minimize the total number of Trotter steps $N_{\text{total}}$ throughout the entire QCELS algorithm, which can be accomplished by numerically solving the Lagrange multiplier equation:
\begin{equation}
\small
\begin{split}
    L(\epsilon_{\text{Trotter}}, \epsilon_{\text{QCELS}}) = N_{\text{total}} - \lambda (\epsilon_{\text{Trotter}} + \epsilon_{\text{QCELS}} -\epsilon_{\text{target}}),
\end{split}
\label{eq: error budget lagrangian}
\end{equation}
with $N_{\text{total}}$ given by:
\begin{equation}
\begin{split}
    N_{\text{total}} 
    &= N_s\sum_{j=1}^J\sum_{n=0}^{N} \left\lceil n \frac{\tau_j}{T_{\max}} N_{\max} \right\rceil \\
    &= (N+1)N_s\sum_{j=1}^J \left\lceil\frac{2^{j-J(\epsilon_{\text{QCELS}})-1}\delta}{\epsilon_{\text{QCELS}}} \sqrt{\frac{w}{\epsilon_{\text{Trotter}}}}\;\right\rceil.
\end{split}
\end{equation}
The total execution time can be estimated with:
\begin{equation}
    T_\text{total} = N_s\sum_{j=1}^J\sum_{n=0}^{N} \gamma_{n, j}^2 \left\lceil \frac{n \tau_j}{T_{\max}} N_{\max} \right\rceil \times T_{1-\text{trotter}}
\end{equation}
with $T_{1-\text{trotter}}$ being the execution time of 1-Trotter step and $\gamma_{n, j}^2$ is the execution time of the time evolution circuit up to evolution time $n\tau_j$.  Note that, in Ref.~\cite{ding2023even}, the summation over $n$ is only up to $N-1$ so the overall factor becomes $N-1$ instead of $N+1$ here. We perform our resource estimation by summing to $N$.

\subsection{Choice of the $\delta$ parameter}
\label{app: delta choice}

The $\delta$ factor in the QCELS algorithm is a tunable parameter that determines the maximal and total evolution times via $T_{\text{max}} = \delta/\epsilon$ and $T_{\text{total}} = \Theta(\delta^{-1}\epsilon^{-1})$. It depends on the overlap $p_0 = |\langle \psi_0| \psi \rangle|^2$ between the QCELS input state $|\psi\rangle$ and the true ground state $|\psi_0
\rangle$ through $\delta = \Theta(\sqrt{1-p_0})$. This means that we can choose $\delta$ to be arbitrarily small if we can prepare a state with overlap $p_0 = 1$ at the cost of more fitting data point $N$ or shots $N_s$ determined by \Cref{eq:Ns_def}. While the theoretical bound of $\delta$ is given by $\sqrt{1-p_0}$, Fig. 4 of Ref.~\cite{ding2023even} explored the empirical choice of $\delta$ via concrete numerical simulation of the QCELS algorithm assuming different $p_0$. They showed that $\delta$ can be chosen to be $0.06$ even when the overlap is as low as $p_0 = 0.6$ ($\delta = \sqrt{1-p_0} \approx 0.63$) or $p_0 = 0.8$ ($\delta = \sqrt{1-p_0} \approx 0.45$). As we assume the overlap of the prepared ground state is $p_0\approx 1$, we can adopt $\delta = 0.06$ for the empirical resource estimation tables \Cref{tab: p-benzyne qre}, \Cref{tab: FH2D 0.005L2 resource estimation}, and  \Cref{tab: FH2D 0.01 resource estimation}.

On the other hand, when $\delta = 0.06$ yields a deep circuit that requires large PEC overhead, we leverage our assumption of $p_0 \approx 1$ and choose $\delta = 0.001$, which corresponds to $p_0 = 0.999999$. We refer to this choice as "theoretical" choice of the $\delta$ parameter in the resource estimation tables above. We need to pay the price of needing more data points $N$ per iteration in QCELS with this choice of $\delta$. Using the relation $N N_s = \delta^{-2}$ and the choice of $N_s = 100$, we need $N=10^4$ instead of $N = 5$ as in the empirical choice of $\delta = 0.06$ parameter set. This choice is validated by simulated the QCELS algorithm with $p$-benzyne of active space size (14e, 14o), assuming the input state is the FCI vector.

In \Cref{tab:qcels_side_by_side_nested}, we list the explicit QCELS parameters used in generating the resource estimation \Cref{tab: p-benzyne qre}, \Cref{tab: FH2D 0.005L2 resource estimation}, and \Cref{tab: FH2D 0.01 resource estimation}. They are derived from the solution of \Cref{eq: error budget lagrangian}.

\begin{table*}[tbp]
    \centering
    \begin{subtable}[b]{1\textwidth} 
        \centering
        {\scriptsize
        \begin{tabular}{|cc|cccccc|cccccc|}
            \hline 
            & & 
            \multicolumn{6}{c|}{\textbf{Empirical}} & \multicolumn{6}{c|}{\textbf{Theoretical}} \\
            $N_{orb}$ & $\epsilon$ & 
            $J$ & $N_s$ & $N$ & $\epsilon_{\text{QCELS}}$& 
            $\epsilon_{\text{Trotter}}$& 
            $T_{\text{max}}$ &
            $J$ & $N_s$ & $N$ &
            $\epsilon_{\text{QCELS}}$& 
            $\epsilon_{\text{Trotter}}$& 
            $T_{\text{max}}$\\ 
            \hline 
            $6$  & 1.6 mHa & 
            12 & 100 & 5 & 1.067 mHa & 0.533 mHa & 56.25 &
            12 & 100 & $10^4$ & 1.067 mHa & 0.533 mHa & 0.9375 \\
            $14$ & 1.6 mHa & 
            15 & 100 & 5 & 1.067 mHa & 0.533 mHa & 56.25 &
            15 & 100 & $10^4$ & 1.067 mHa & 0.533 mHa & 0.9375 \\
            $18$ & 1.6 mHa & 
            16 & 100 & 5 & 1.067 mHa & 0.533 mHa & 56.25 &
            16 & 100 & $10^4$ & 1.067 mHa & 0.533 mHa & 0.9375\\
            $26$ & 1.6 mHa & 
            18 & 100 & 5 & 1.067 mHa & 0.533 mHa & 56.25 &
            18 & 100 & $10^4$ & 1.067 mHa & 0.533 mHa & 0.9375 \\
            \hline 
        \end{tabular}
        }
        \caption{$p$-benzyne $\epsilon=1.6 \text{mHa}$}
        \label{tab:sub_molecules}
    \end{subtable}
    \hfill 
    \begin{subtable}[b]{0.5\textwidth}
        \centering
        {\scriptsize
        \begin{tabular}{|c|c|c|cccccc|} 
            \hline
            & & & \multicolumn{6}{c|}{\textbf{Empirical}} \\
            $\epsilon$ & $U$ & $L$ & $J$ & $N_s$ & $N$ &$\epsilon_{\text{QCELS}}$& 
            $\epsilon_{\text{Trotter}}$& 
            $T_{\text{max}}$\\ 
            \hline
            \multirow{8}{*}{$0.005L^2$} 
               & \multirow{4}{*}{4} 
                 & 4  & 10 & 100 & 5 & 0.0533& 0.0267& 1.1250\\
               & & 6  & 10 & 100 & 5 & 0.1200 & 0.0600 & 0.5000 \\
               & & 8  & 10 & 100 & 5 & 0.2133& 0.1067 & 0.2813\\
               & & 10 & 10 & 100 & 5 & 0.3333 & 0.1667 & 0.1800\\
               \cline{2-9} 
               & \multirow{4}{*}{8} 
                 & 4  & 10 & 100 & 5 & 0.0533 & 0.0267 & 1.1250 \\
               & & 6  & 10 & 100 & 5 & 0.1200 & 0.0600 & 0.5000 \\
               & & 8  & 11 & 100 & 5 & 0.2133& 0.1067 & 0.2813\\
               & & 10 & 11 & 100 & 5 & 0.3333 & 0.1667 & 0.1800\\ 
            \hline
        \end{tabular}
        }
        \caption{2D Fermi--Hubbard $\epsilon=0.005L^2$}
        \label{tab:sub_fh_nested}
    \end{subtable}%
    \hfill 
    \begin{subtable}[b]{0.5\textwidth}
        \centering
        {\scriptsize
        \begin{tabular}{|c|c|c|cccccc|}
            \hline
            & & & \multicolumn{6}{c|}{\textbf{Empirical}} \\
            $\epsilon$ & $U$ & $L$ & $J$ & $N_s$ & $N$ & $\epsilon_{\text{QCELS}}$& 
            $\epsilon_{\text{Trotter}}$& 
            $T_{\text{max}}$\\ 
            \hline
            \multirow{8}{*}{$0.01$} 
               & \multirow{4}{*}{4} 
                 & 4  & 13 & 100 & 5 & 2/3$\times$10\textsuperscript{-2} & 1/3$\times$10\textsuperscript{-2} & 9.0\\
               & & 6  & 14 & 100 & 5 & 2/3$\times$10\textsuperscript{-2} & 1/3$\times$10\textsuperscript{-2} & 9.0\\
               & & 8  & 15 & 100 & 5 & 2/3$\times$10\textsuperscript{-2} & 1/3$\times$10\textsuperscript{-2} & 9.0\\
               & & 10 & 16 & 100 & 5 & 2/3$\times$10\textsuperscript{-2} & 1/3$\times$10\textsuperscript{-2} & 9.0\\
               \cline{2-9}
               & \multirow{4}{*}{8} 
                 & 4  & 13 & 100 & 5 & 2/3$\times$10\textsuperscript{-2} & 1/3$\times$10\textsuperscript{-2} & 9.0\\
               & & 6  & 15 & 100 & 5 & 2/3$\times$10\textsuperscript{-2} & 1/3$\times$10\textsuperscript{-2} & 9.0\\
               & & 8  & 15 & 100 & 5 & 2/3$\times$10\textsuperscript{-2} & 1/3$\times$10\textsuperscript{-2} & 9.0\\
               & & 10 & 16 & 100 & 5 & 2/3$\times$10\textsuperscript{-2} & 1/3$\times$10\textsuperscript{-2} & 9.0\\
            \hline
        \end{tabular}
        }
        \caption{2D Fermi--Hubbard $\epsilon = 0.01$}
        \label{tab:sub_fh_nested_2}
    \end{subtable}

    \caption{Summary of QCELS algorithmic parameters $J, N_s$, and $N$ for $p$-benzyne and 2D Fermi--Hubbard systems. }
    \label{tab:qcels_side_by_side_nested}
\end{table*}

\section{Clock count estimation of the 2D Fermi--Hubbard simulation}

\begin{table}[htbp!]
\centering
\begin{tabular}{|c|c|c|c|}
\hline
\textbf{Operation Type} & \textbf{Clocks per Occurrence} & \textbf{Occurrences} & \textbf{Total Clocks} \\ \hline
{$ZZ$ Rotations} & $T_{\text{RUS}}(V, ZZ)$ & $2$ & $2T_{\text{RUS}}(V, ZZ)$ \\ \hline
{Patch moving} & $3$ & $2$ & $6$ \\ \hline
{fSWAP network} & $7(N-1)$ & $2$ & $14N - 14$ \\ \hline
{$XX+YY$ Rotations} & $9 + T_{\text{RUS}}(V-N, ZZ) + T_{\text{RUS}}(V-N, Z)$ & $7$ & $63 + 7T_{\text{RUS}}(V-N, ZZ) + 7T_{\text{RUS}}(V-N, Z)$ \\ \hline
\end{tabular}
\caption{Summary of operations and their execution clocks per Trotter step.}
\end{table}
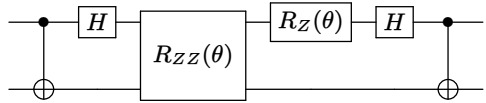
\begin{figure}
    \centering
    \leavevmode
    \Qcircuit @C=1.0em @R=1.5em {
    & \ctrl{1} & \gate{H} & \multigate{1}{R_{ZZ}(\theta)} & \gate{R_Z(\theta)} & \gate{H} & \ctrl{1} & \qw \\
    & \targ    & \qw      & \ghost{R_{ZZ}(\theta)}        & \qw                & \qw      & \targ    & \qw
}
    \caption{Circuit for implementing the $XX+YY$ rotations. Same as the one in Fig. 11 of Ref.~\cite{Akahoshi:2024yme}.}
    \label{fig:XX YY Rotation}
\end{figure}

Here, we review the clock count estimation of the circuit \cref{fig: FH2D hadamard} in \Cref{sec: 2D FH}. The estimation scheme was first presented in Ref.~\cite{Akahoshi:2024yme}. The Hamiltonian simulated by this circuit is written as
\begin{equation}
    H = -\frac{t}{2}\sum_{\langle i, j\rangle,\sigma} \left(X_{i,\sigma}X_{j,\sigma}+Y_{i,\sigma}Y_{j,\sigma}\right)Z_{i,j,\sigma}^{\leftrightarrow} +\frac{U}{4} \sum_{i}Z_{i,\uparrow}Z_{i,\downarrow}.
    \label{eq: FH2D Hamiltonian}
\end{equation}
So the Trotter time evolution includes $ZZ$ rotation for the interaction term, $XX+YY$ rotations for the hopping term and other lattice surgery operations to smoothly connect these two types of operations.

\paragraph{Interaction term $ZZ$ rotations}
The $ZZ$ rotation terms correspond to the interactions between spin-up and spin-down components on the same lattice site, and thus these rotations can be executed in parallel across the $V = N^2$ lattice sites. A single layer of this operation takes $T_{\text{RUS}}(V, ZZ)$ lattice surgery clock cycles. In a second-order Trotter decomposition, this operation occurs twice, once at the beginning and once at the end of the step, resulting in a total of $2T_{\text{RUS}}(V, ZZ)$ clock cycles.

\paragraph{Patch moving}
Before executing the hopping interactions, the logical patches must be rearranged to separate the spin-up and spin-down components into the upper and lower rows of the architecture. This moving operation requires $3$ lattice surgery clock cycles. This rearrangement must happen twice per Trotter step, once to split the spins before the hopping terms, and once to reverse the arrangement afterward, consuming $6$ clock cycles in total.

\paragraph{Hopping terms}
After the patch movement, the next step is the hopping term. The hopping terms are decomposed into four sets, namely, two different Jordan--Wigner orderings, each with their own horizontal and vertical interaction terms. The two Jordan--Wigner orderings relate to each other via layers of fSWAP gates. Thus, the hopping term is divided into four $XX+YY$ rotation layers and one fSWAP network term. We discuss these two types of operations separately:
\begin{itemize}
    \item fSWAP network: The fSWAP network contains $N-1$ fSWAP layers, each of which takes 7 clock cycles to execute. So the clock count for fSWAP sums up to $7(N-1)$.
    \item $XX+YY$ rotations: Each $XX+YY$ rotation layer contains $N(N-1)$ parallel rotations. And the $XX+YY$ rotation is implemented by the circuit shown in \Cref{fig:XX YY Rotation}. The Clifford operation takes 9 clocks, $R_{Z}$ rotation takes $T_{\text{RUS}}(V-N, Z)$ clocks and ZZ rotation takes $T_{\text{RUS}}(V-N, ZZ)$ clocks, where $V= N^2$. Thus, one hopping layer takes $9 + T_{\text{RUS}}(V-N, ZZ) + T_{\text{RUS}}(V-N, Z)$ clocks to execute.
\end{itemize}
Na\"ively, the hopping operation in the second-order Trotter formula takes $8$ $XX+YY$ rotations and $2$ fSWAP networks, but we can eliminate one $XX+YY$ rotation layer by merging the middle terms in the product formula. This gives us:
\begin{equation}
    7 \left(9 + T_{\text{RUS}}(V-N, ZZ) + T_{\text{RUS}}(V-N, Z)\right) + 2\times 7(N-1)
    = 14N + 49 + 7T_{\text{RUS}}(V-N, ZZ) + 7T_{\text{RUS}}(V-N, Z)
\end{equation}
clock cycles for the hopping terms.

\paragraph{Overhead due to control operations}
The above discussions only contains the operations for Trotter time evolution circuit. However, there are multi-target CNOTs and multi-target CZs interleaved between the hopping and interaction terms. They are responsible for the control operations that couples the data and measurement qubits. The controlled-$K_i$s are multi-target CNOTs and controlled-$K_h$s are multi-target CZs. It is estimated in Appendix A of Ref.~\cite{Akahoshi:2024yme} that controlled-$K_i$s take five clock cycles and  the controlled-$K_h$s takes four clock cycles. Looking at the circuit diagram \Cref{fig: FH2D hadamard}, there are four controlled-$K_i$s and four controlled-$K_h$s in a single Trotter step. However, some optimization can be applied to the combined circuit. The controlled-$K_h$s in the middle of the circuit should cancel with each other so that the middle $XX+YY$ rotation layers can be merged. In addition, the controlled-$K_i$s at the end of one Trotter step can cancel with those at the beginning of the next Trotter step. Thus, only two additional controlled-$K_i$s and two additional controlled-$K_h$s are needed for each new Trotter step.  Note that to implement the controlled-$K_i$s, the first row of the patch must be filled with spin-up data patches and the last row must be filled with spin-down data patches. This arrangement is different from that of the interaction term $ZZ$ rotations. However, one can apply the patch movement discussed above, which takes three steps each time. Adding this to the beginning and the end of the circuit, the controlled-$K$ operations takes
\begin{equation}
\begin{split}
    T_{\text{ctrl}}(N_{\text{Trotter}}) 
    &= 2\times 5 + 2\times 3 + (2\times 4 + 2 \times 5)N_{\text{Trotter}} \\
    &= 16 + 18 N_{\text{Trotter}}.
\end{split}
\end{equation}

In summary, we already see that the interaction terms take $2T_{\text{RUS}}(ZZ, V)$ clock cycles, patch movement takes $6$ clock cycles, and hopping term takes $14N + 49 + 7T_{\text{RUS}}(V-N, ZZ) + 7T_{\text{RUS}}(V-N, Z)$ clock cycles.  Combining these estimates, we find that a single Trotter step takes
\begin{equation}
  T_{\text{1-Trotter}} = 2T_{\text{RUS}}(V, ZZ) + 7T_{\text{RUS}}(V-N, ZZ) + 7T_{\text{RUS}}(V-N, Z) + 55 + 14N
\end{equation}
clock cycles, which reproduces eq.~(17) of \cite{Akahoshi:2024yme}. Finally, for $N$ Trotter steps, the total clock count becomes 
\begin{equation}
    T(N_{\text{Trotter}}) = T_{1-\text{Trotter}}N_{\text{Trotter}} + 16+18N_{\text{Trotter}}.
\end{equation}

\end{document}